\documentclass[pra,twocolumn,showpacs,superscriptaddress,notitlepage,amsmath,amssymb]{revtex4-1}
\usepackage[dvipsnames]{xcolor}
\definecolor{dodgerblue}{rgb}{0.12, 0.56, 1.0}
\usepackage[colorlinks=true,citecolor=dodgerblue,linkcolor=dodgerblue,urlcolor=dodgerblue,pdftitle={surf}]{hyperref}
\usepackage{dcolumn,graphicx,amsfonts,amsthm,bm,color,appendix,float}
\usepackage{booktabs}
\usepackage{braket}
\usepackage[normalem]{ulem}
\usepackage[version=4]{mhchem}
\usepackage{siunitx}
\usepackage{amsmath}
\usepackage{amssymb}
\usepackage{CJKutf8}
\usepackage{mathtools}
\usepackage{enumitem}
\usepackage{booktabs}
\usepackage{bbold}
\usepackage{wasysym}
\usepackage{mathrsfs}
\usepackage{calligra}
\usepackage{soul}
\usepackage{xfrac}
\usepackage{quantikz}
\usepackage{wasysym}
\usepackage{url}
\usepackage[export]{adjustbox}

\usepackage{tikz}
\usetikzlibrary{calc}

\DeclareMathAlphabet{\mathcalligra}{T1}{calligra}{m}{n}
\DeclareFontShape{T1}{calligra}{m}{n}{<->s*[2.2]callig15}{}

\newcommand{\Dfour}{D_4}
\newcommand{\CZ}{\widetilde{\mathrm{CZ}}}
\newcommand{\tsigma}{\widetilde{\sigma}}

\definecolor{crimson}{rgb}{0.86, 0.08, 0.24}
\definecolor{royalblue}{rgb}{0.25, 0.41, 0.88}

\newcommand{\red}[1]{{ \color{crimson} #1}}
\newcommand{\blue}[1]{{ \color{royalblue} #1}}
\newcommand{\green}[1]{{ \color{ForestGreen} #1}}

\newcommand{\R}{\red{R}}
\newcommand{\G}{\green{G}}
\newcommand{\B}{\blue{B}}
\newcommand{\Z}{\mathbb{Z}}

\newcommand{\tr}{\textrm{tr}}

\renewcommand{\v}[1]{\boldsymbol{#1}}

\newcommand{\borromeansymbol}{
\begin{tikzpicture}[scale=0.3, baseline=-0.6ex]
  \draw (0,0) circle (0.5);
  \draw (0.6,0) circle (0.5);
  \draw (0.3,0.5) circle (0.5);
\end{tikzpicture}
}
\newcommand{\borromean}{\mathrel{\raisebox{-0.2em}{\borromeansymbol}}}

\newcommand{\borromeansymbolcolor}{
\begin{tikzpicture}[scale=0.3, baseline=-0.6ex]
  \draw[color=crimson] (0,0) circle (0.5);
  \draw[color=ForestGreen] (0.6,0) circle (0.5);
  \draw[color=royalblue] (0.3,0.5) circle (0.5);
\end{tikzpicture}
}
\newcommand{\borromeancolor}{\mathrel{\raisebox{-0.2em}{\borromeansymbolcolor}}}

\newcommand{\trianglesymbol}{
\begin{tikzpicture}[scale=0.3, baseline=-0.6ex]
  \draw (0,0) circle (0.6);
  \draw (240:0.6) -- (0:0.6) -- (120:0.6) -- cycle;
  \draw (60:0.3) -- (180:0.3) -- (300:0.3) -- cycle;
\end{tikzpicture}
}

\begin{document}

\title{Statistical Mechanics and Symmetries of Non-Abelian Anyon Proliferation:\\ From Deformation to Decoherence}
\author{Avi Vadali}
\email{avadali@mit.edu}
\affiliation{California Institute of Technology, Pasadena, CA 91125, USA}
\author{Robijn Vanhove}
\affiliation{
Department of Physics and Institute for Quantum Information and Matter, \\ California Institute of Technology, Pasadena, California 91125, USA}
\affiliation{Department of Physics and Astronomy, Ghent university, Krijgslaan 281, S9, B-9000, Ghent, Belgium}
\author{Ruben Verresen}
\affiliation{Pritzker School of Molecular Engineering, University of Chicago, Chicago, IL 60637, USA}
\author{Jason Alicea}
\affiliation{
Department of Physics and Institute for Quantum Information and Matter, \\ California Institute of Technology, Pasadena, California 91125, USA}
\affiliation{Walter Burke Institute for Theoretical Physics, California Institute of Technology,\\ Pasadena, California 91125, USA}
\author{Pablo Sala}
\email{psala@berkeley.edu}
\affiliation{
Department of Physics and Institute for Quantum Information and Matter, \\ California Institute of Technology, Pasadena, California 91125, USA}
\affiliation{Walter Burke Institute for Theoretical Physics, California Institute of Technology,\\ Pasadena, California 91125, USA}
\affiliation{Department of Physics, University of California, Berkeley, CA 94720, USA}
\affiliation{Simons Institute for the Theory of Computing, University of California at Berkeley}
\date{\today}

\begin{abstract}

Topological quantum computation relies on braiding non-Abelian anyons, but requires the underlying topological order to survive imperfect state preparation and environmental noise. 
We show that the instability of topological order to wavefunction deformations and to decoherence, with the latter probed by syndrome distributions, is generically captured by stat-mech models whose $G\times \textrm{Rep}(G)$ symmetry naturally exposes the corrupting anyonic excitations.
As an example, we combine this framework with Monte-Carlo simulations to resolve the stability of $D_4$ topological order under deformations and quantum channels that proliferate multiple non-Abelian anyon species that are individually unable to condense. We show that beyond a finite threshold, proliferation of two non-Abelian anyon species parasitically condenses a shared Abelian-anyon fusion outcome---destroying the topological order.  Our symmetry-based approach suggests the resulting trivial phase is distinct from that obtained by condensing all Abelian anyons; in other words, the trivial phase ``remembers’' which anyons condensed. This framework provides a first step into identifying the relevant symmetry for optimal decoders, conditioned on syndrome measurements, of non-Abelian topological order.
\end{abstract}

\maketitle

\textbf{Introduction.---} Quantum error correction for non-Abelian codes remains in its early stages, with a growing body of work yielding increasingly encouraging results~\cite{Wootton_14,Brell_14,Wootton_16,Burton_2017,Dauphinais_2017,Schotte_22a,Schotte_22b, davydova2025, Rossoneri_25,lyons2026quantumcomputinganyonsfault,Julio_26}. While the \emph{optimal error threshold} of Abelian codes connects to thermal phase transitions in disordered statistical-mechanics (stat-mech) models~\cite{Dennis_2002}, recently revisited from a variety of perspectives~\cite{fan2023diagnostics,bao2023mixedstate,LeeYouXu2022,chen2023separability, Renorm_QECC_23,wang2023intrinsic,Mong_24,ellison2024classificationmixedstatetopologicalorders,sohal_24,Mong_24, tapestry_24,chen2024unconventional,Hauser_24,lyons24,2024_sala_SSSB,Markov_length_24,TshungCheng_24,lee2024exactcalculationscoherentinformation}, analogous predictions for non-Abelian codes~\cite{Kitaev_2003,Mochon03,Mochon04,Freedman_2000gwh,Freedman_2006,Nayak_08} remains a rather open question due to the extensive coherence injected by non-Abelian anyons. Recent work has identified a remarkable stability to processes that proliferate a single non-Abelian anyon type \cite{sala_D4,loops_25}, leading to optimal decoding schemes when conditioned upon measuring local syndromes~\cite{Rossoneri_25}.  
Yet, the generalization of these results to multiple non-Abelian anyon species remains open. 
\begin{figure}[t!]
     \centering
    \includegraphics[width=\linewidth]{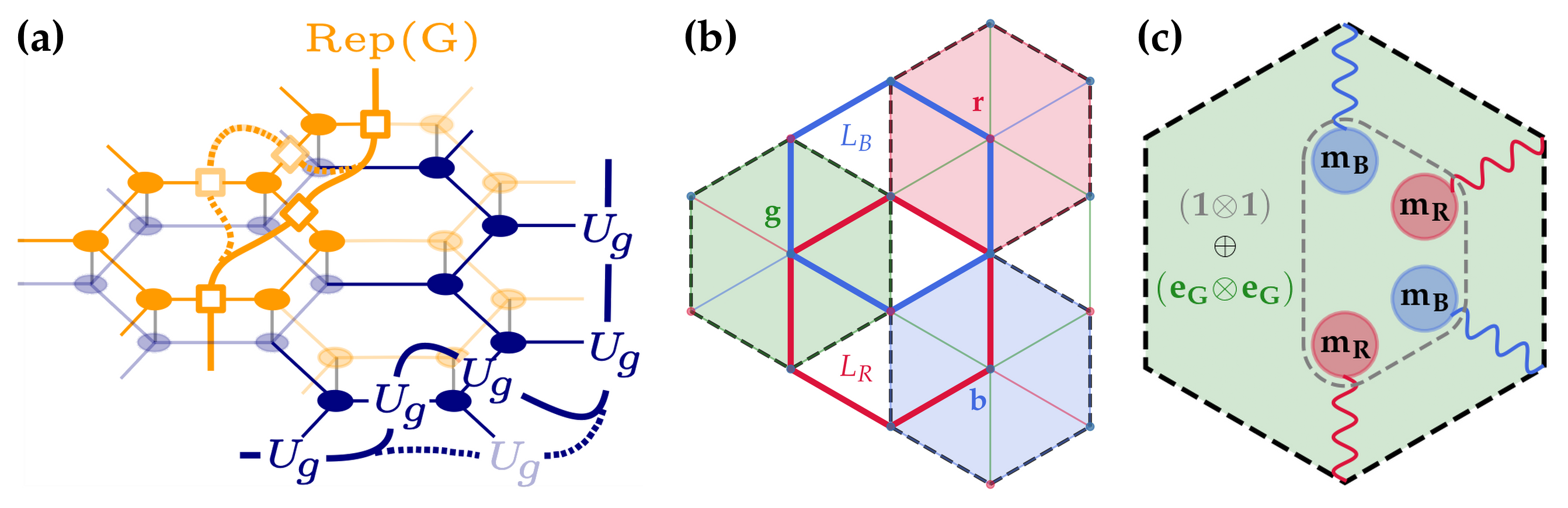}
    \caption{ \textbf{Fusion processes and symmetries.} (a) Norm of the deformed wavefunction as a tensor network with the bra and ket in different PEPS representations. The stat-mech model has a $G\times \textrm{Rep}(G)$ symmetry inherited from the virtual symmetries of the double layer. (b) Lattice for the stat-mech model in Eq.~\eqref{eq:Z_psi}, with $r, g, b$ respectively labeling centers of red, green, blue hexagons. Shown loop configuration has cyclomatic number $C_{L_\R \cup L_\B}=3$. (c) Allowed fusions between $m_R$ and $m_B$ anyons: $(m_B\times m_B) \times (m_R \times m_R)\to 1\times 1=1$, or $(m_B\times m_B) \times (m_R \times m_R)\to e_G\times e_G=1$. }
    \label{fig:composite_figure}
\end{figure}

Here we extend these results to noise processes proliferating, either via finite-depth non-unitary wavefunction deformations~\cite{Castelnovo_2008,Haegeman_15,GuoYi_19,Mari_n_2017, Xu_2021, Xu_2022, Schotte_2019, Fendley_2008} or local quantum channels, different types of non-Abelian anyons with non-trivial fusion and braiding, specializing to $D_4$ topological order (TO) as an experimentally relevant case study~\cite{Iqbal_24}.  We identify a class of coupled loop models which consistently account for such processes, permitting extensive numerical simulations as well as analytical progress. Specifically, we find that the joint proliferation of two non-Abelian anyon species hosted by the $D_4$ TO ($m_R$ and $m_B$) drives a single transition in which they simultaneously condense together with a shared fusion outcome ($e_G$).  In stat-mech language, the resulting trivial phase corresponds to a unique pattern of spontaneous non-Abelian symmetry breaking. We further show condensing different sets of non-Abelian anyons can yield sharply distinct trivial phases, distinguished by unbroken symmetries of the stat-mech model.
More generally, symmetries of the stat-mech models follow from the TO under consideration and the anyons nucleated by deformation/quantum channels ~\cite{haegeman2015shadows, schuch2013topological, schuch2010peps, Xu_2021,Xu_2022}. We use tensor network techniques to identify a $G\times \textrm{Rep}(G)$ symmetry~\cite{vanhove2018mapping,lootens2021matrix,lootens2023dualities, lootens2024dualities} when considering wavefunction deformations as well as syndrome distributions for the quantum double of a group $G$. The existence of a stat-mech model with such a symmetry can be demonstrated by choosing different suitable Projected Entangled Pair State (PEPS) representations of the bra and ket in the norm of the deformed wavefunction (see Fig.~\ref{fig:composite_figure}(c)). We also present an alternative formulation that naturally exposes the proliferated anyons.

\textbf{Model and anyon proliferation.---} We consider the exactly solvable qubit model for $D_4$ TO realized on three intertwined honeycomb sublattices (${\R}$, ${\G}$, ${\B}$)~\cite{Yoshida_2016,Iqbal_24}. All anyons can be generated by the fusion of three Abelian anyons $e_{R,G,B}$ and three self-bosonic non-Abelian anyons $m_{R,G,B}$ with quantum dimension $d = 2$. We can locally generate adjacent Abelian anyon pairs via $Z$ operators acting on the corresponding colored sublattices. Similarly, $X$ operators generate minimally separated non-Abelian anyon pairs.  
We will primarily focus on generation of $m_R$ and $m_B$, which exhibit self-fusion channels $m_R \times m_R = 1 + e_B + e_G + e_{B}e_{G}$ and $m_B \times m_B = 1 + e_R + e_G + e_{R}e_{G}$ that yield only Abelian anyons with shared $e_G$ outcome. Moreover, $m_R$ and $m_B$ braid non-trivially with each other, toggling the fusion outcome of pairs of anyons from vacuum to $e_G$~\cite{Iqbal_24}. 

Consider a pristine ground state $\ket{D_4}$ non-unitarily deformed by
\begin{equation}
\label{eq:psi_bR_bB}
    \ket{\beta^z_\alpha,\beta_\alpha} = e^{\sum_\alpha \frac{\beta^z_\alpha}{2} \sum_{j \in {\alpha}} Z_j^{\alpha}} e^{\sum_{\alpha}\frac{\beta_\alpha}{2} \sum_{j \in {\alpha}} X_j^{\alpha}}  \ket{D_4},
\end{equation}
with $\alpha = \R, \G,\B$, and where finite $\beta_\alpha, \beta^z_\alpha\in [0,\infty)$ quantify the strengths of deformations nucleating anyon pairs.  Generation of only pairs of $m_R$ and $m_B$ anyons is achieved by acting with on-site $X$ on the respective sublattice, i.e., taking $\beta^z_\alpha=\beta_\G=0$. Phase transitions as a function of $\beta_\R, \beta_\B$ can be diagnosed by examining the decay of correlation functions as well as the norm of $\ket{\beta_\R, \beta_\B}$. The norm maps to coupled O$(2)$ loop models in $(2+0)$ dimensions~\cite{sala_D4}: $ \langle \beta_\B, \beta_\R|\beta_\B,\beta_\R\rangle= \sum_{L_\R, L_\B} t_\R^{|L_\R|} t_\B^{|L_\B|} W(L_\R,L_\B)$ with
\begin{equation} \label{eq:Z_psi}
    W(L_\R,L_\B)\equiv  \frac{2^{C_{L_\R \cup L_\B}}}{\sqrt{2}^{|L_\R| + |L_\B|}} \geq 0
\end{equation}
and tensions $t_\alpha = \tanh(\beta_\alpha)$ for $\alpha=\R,\B$. When $\ket{D_4}$ is a $+1$ eigenstate of all $\mathcal{Z}$ logical operators, the sum runs over contractible loop configurations; otherwise non-contractible configurations appear as well. Loops of different color, ${L_\R}$ and ${L_\B}$, interact through the shared green sublattice, reflecting the common $e_G$ fusion outcome of $m_R$ and $m_B$ pairs (see Fig.~\ref{fig:composite_figure}). This interaction is encoded in the coupling $C_{L_\R \cup L_\B}$, which for a connected component, denotes the cyclomatic number~\cite{Essam70} in $L_\R \cup L_\B$, namely, the minimum number of edges that must be removed to break all its cycles, making it into a tree (e.g., $C_{L_\R \cup L_\B}=5$ in Fig.~\ref{fig:composite_figure}(a)). This coupling acts as an attractive potential between $L_\R$ and $L_\B$, increasing their probability to cross.

Understanding the effect of decoherence on non-Abelian TO---specifically, when noise creates multiple anyon types---is more challenging. We characterize such processes using local quantum channels 
\begin{equation} \label{eq:E_channel}
    \mathcal{E}^\alpha_j(\rho_0) = (1-p_\alpha) \rho_0 + p_\alpha P_j^\alpha \rho_0 P_j^\alpha
\end{equation}
acting on the pure density matrix $\rho_0 = \ket{D_4}\bra{D_4}$. Here $P_j^{\alpha}$ are local Pauli operators acting with probability $p_\alpha \in [0,\sfrac{1}{2}]$ for $\alpha = \R, \B$. The resulting mixed density matrix $\rho$ follows from composing all such (commuting~\footnote{Unlike $X$ versus $Z$ wavefunction deformations in Eq.~\eqref{eq:psi_bR_bB}, now local quantum channels $\mathcal{E}^X_j$ and $\mathcal{E}^Z_j$ commute on every site $j$.}) local channels. We will consider certain limiting regimes ($p_\R=p_\B=\sfrac{1}{2}$) and R\'{e}nyi quantities due to their computational tractability. 
In particular, following a similar calculation as for the deformed wavefunction, one finds that the purity of $\rho$ with only $X$ noise on $\R$ and $\B$ lattices leads to coupled $O(4)$ loop models~\cite{sala_D4}
\begin{equation} \label{eq:purity} 
\tr(\rho^2)= \sum_{L_\R, L_\B} r_\R^{|L_\R|} r_\B^{|L_\B|}W(L_\R,L_\B)^2
\end{equation}
with $r_\alpha = \frac{2p_\alpha (1-p_\alpha)}{(1-p_\alpha)^2 + p_\alpha^2}$. Similar expressions follow from including other types of errors as given in Eq.~\eqref{eq:E_channel}. 

\textbf{Symmetries.---} 
Although the preceding loop model formulations nicely expose properties of the nucleated anyons, their symmetries are not manifest. We thus turn to a rotor formulation---which both enables clean symmetry identification and Monte Carlo simulations---by rewriting Eq.~\eqref{eq:Z_psi} as the exact high-temperature expansion of the local stat-mech model
\begin{equation} \label{eq:H_local_rotor}
     H= - \sqrt{2} \beta_\R \! \! \! \! \sum_{\langle b, g \rangle \in \red{\hexagon}} \v{n}^{\R}_b\cdot \v{n}^{\R}_g  - \sqrt{2} \beta_\B \! \! \! \! \sum_{\langle r, g \rangle \in \blue{\hexagon}} \v{n}^{\B}_r\cdot \v{n}^{\B}_g, 
\end{equation} 
defined on the AB-stacked honeycomb lattices in Fig.~\ref{fig:composite_figure}(a). A detailed calculation appears in App.~\ref{sec:high_T_expansion}. The $\v{n}$'s are discrete two-component vectors $\v{n}=(\cos(\theta),\sin(\theta))$, where the angles are constrained such that $\theta^{\R}_g, \theta^{\B}_g = \frac{(2n + 1) \pi}{4}$ and $\theta^{\R}_b, \theta^{\B}_r = \frac{n \pi}{2}$, together with the constraint $(\theta^{\B}_g - \theta^{\R}_g) \bmod \pi \equiv 0$.

The unconstrained model is $D_4 \times D_4$ symmetric for $\beta_\R \neq \beta_\B$ and $(D_4 \times D_4) \rtimes \Z^{\textrm{swap}}_2$ along $\beta_\R = \beta_\B$, with each $D_4$ factor acting only on $\v{n}^{\R}$ or $\v{n}^{\B}$. Imposing the constraint reduces the symmetry to $G_{m_R, m_B} \cong \Z_2^3 \rtimes \Z_2^2$ for $\beta_\R \neq \beta_\B$ and to $G_{m_R, m_B} \rtimes \Z^{\textrm{swap}}_2$ along $\beta_\R = \beta_\B$.
Here $\v{n}^{\R}$ and $\v{n}^{\B}$ transform in two distinct $2$-dimensional irreducible representations (irreps) of $G_{m_R, m_B}$ (see App.~\ref{sec:symmetry_group}). The action of $G_{m_R, m_B}$ on rotor variables is given by
\begin{center}
\scalebox{0.85}{
\begin{tikzpicture}
\node at (-4.7,0) {\Huge$($};
\begin{scope}[shift={(-4,0)}]
    \draw[crimson, thick, opacity=0.5] (0,0) circle (0.5);
    \draw[dashed,opacity=0.4] (-0.6,0) -- (0.6,0);
    \draw[->] (0.6,0) arc[start angle=0,end angle=180,radius=0.6];
    \node at (0.0,0.85) {$V_R$};
    \node at (0.0,-0.8) {$\{\theta^{\R} \}$};
\end{scope}
\node at (-3.2,0) {\Large$\times$};
\begin{scope}[shift={(-2.35,0)}]
    \draw[royalblue, thick, opacity=0.5] (0,0) circle (0.5);
    \draw[dashed,opacity=0.4] (-0.6,0) -- (0.6,0);
    \draw[->] (0.6,0) arc[start angle=0,end angle=180,radius=0.6];
    \node at (0.0,0.85) {$V_B$};
    \node at (0.0,-0.8) {$\{\theta^{\B} \}$};
\end{scope}
\node at (-1.52,0) {\Large$\times$};
\begin{scope}[shift={(-0.7,0)}]
    \draw[thick, opacity=0.5] (0,0) circle (0.5);
    \draw[<->] (0,-0.4) -- (0,0.4);
    \draw[dashed,opacity=0.4] (-0.6,0) -- (0.6,0);
    \node at (0.0,0.82) {$T_G$};
    \node at (0.0,-0.8) {$\{\theta^{\R}, \theta^{\B} \}$};
\end{scope}
\node at (-0.06,0) {\Huge$)$};
\node at (0.23,0) {\Large$\rtimes$};
\node at (0.55,0) {\Huge$($};
\begin{scope}[shift={(1.15,0)}]
    \draw[thick, crimson, opacity=0.5] (0,0) circle (0.5);
    \draw[<->] (0.3,-0.3) -- (-0.3,0.3);
    \draw[dashed,opacity=0.4] (-0.39,-0.39) -- (0.39,0.39);
    \node at (0.0,0.82) {$S_R$};
    \node at (0.0,-0.8) {$\{\theta^{\R} \}$};
\end{scope}
\node at (1.9,0) {\Large$\times$};
\begin{scope}[shift={(2.65,0)}]
    \draw[thick, royalblue, opacity=0.5] (0,0) circle (0.5);
    \draw[<->] (0.3,-0.3) -- (-0.3,0.3);
    \draw[dashed,opacity=0.4] (-0.39,-0.39) -- (0.39,0.39);
    \node at (0.0, 0.82) {$S_B$};
    \node at (0.0,-0.8) {$\{\theta^{\B} \}$};
\end{scope}
\node at (3.25,0) {\Huge$)$};
\end{tikzpicture}
}
\end{center}
where $\{\theta^\alpha\}$ indicates that the corresponding symmetry acts on all $\theta^\alpha_j$ rotors.

\begin{figure}[t]
    \centering
    \includegraphics[width=\linewidth]{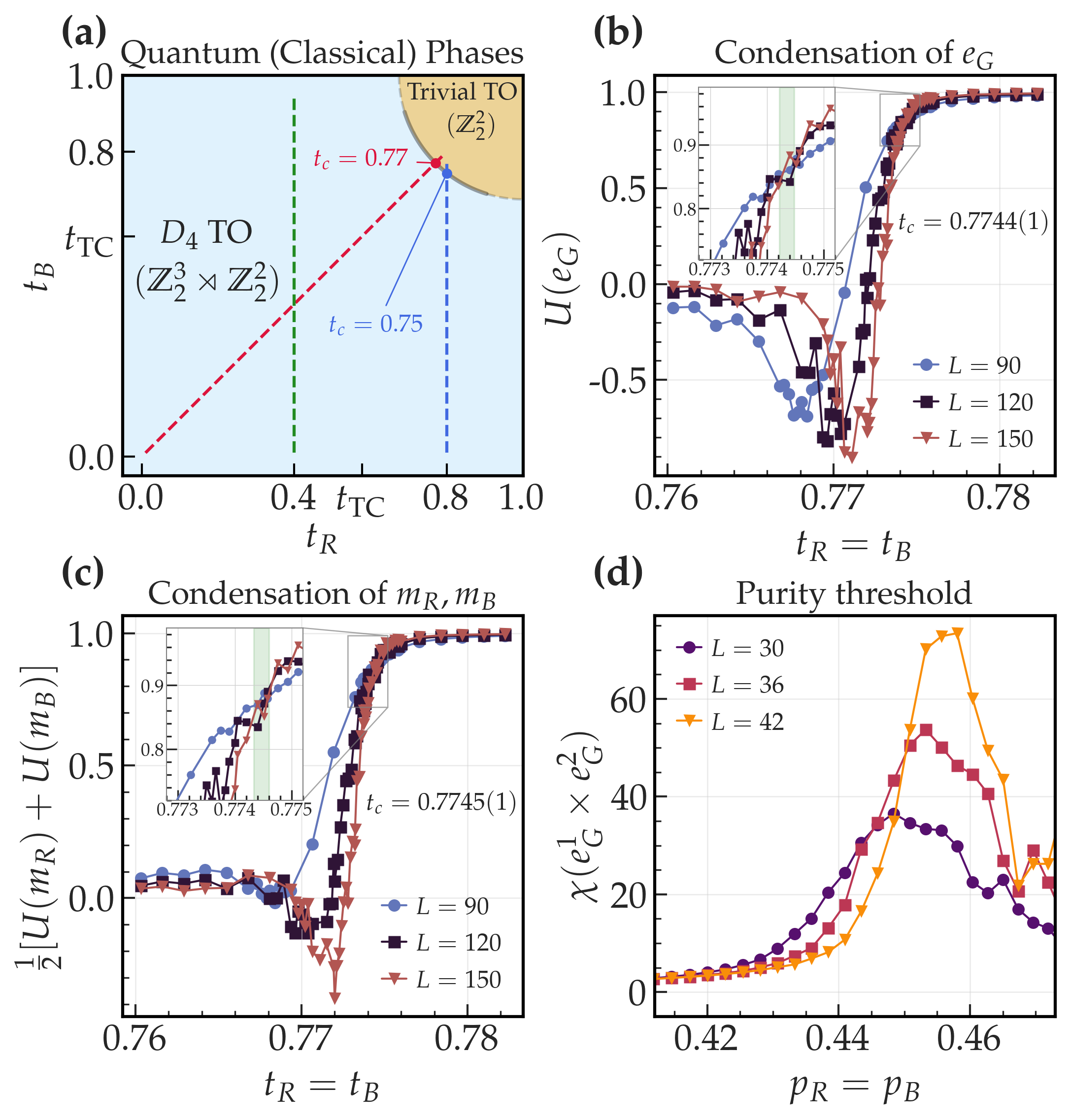}
    \caption{\textbf{Deformation and quantum channel simulations.} (a) Phase diagram capturing $m_R, m_B, e_G$ condensation with the associated $G_{m_R,m_B} \rightarrow \mathbb{Z}_2^2$ classical transition. Red line corresponds to data from panels (b,c), blue and green lines indicate traces from App.~\ref{app:2_NA_numerics}, and $t_\mathrm{TC} = 1/\sqrt{3}$ is the toric code threshold on the honeycomb lattice~\cite{HOUTAPPEL1950425}. (b, c) Binder cumulants $\frac{1}{2} (U_{m_R} + U_{m_B})$ and $U(e_G)$ along the diagonal $t_\R = t_\B$ showcasing condensation of $m_R, m_B$ and $e_G$ anyons. \textit{Inset:} Critical tension $t_c$ given by the crossing point among different linear system sizes $L$. (d) Susceptibility $\chi(e_G^{1} \times e_G^2)$ for $p_R = p_B$. The maximum in $\chi(e_G^{1} \times e_G^2)$ evidences a phase transition for $p_R=p_B \leq 1/2$. See App.~\ref{sec:numerical_purity_instability} for details.}
    \label{fig:Fig_2}
\end{figure}

\textbf{Induced instability.---} Previous work~\cite{sala_D4} has shown that for $\beta_\B=0$ ($\beta_\R=0$), no transition exists at \emph{any} finite $\beta_\R$ ($\beta_\B$).  To diagnose possible instabilities when both parameters are finite, we first focus on the limit $\beta_\R = \beta_\B= \infty$ (upper right corner in Fig.~\ref{fig:Fig_2}(a)). Upon ignoring the constraint discussed above, the coupled loop model corresponding to Eq.~\eqref{eq:Z_psi} maps to two decoupled O$(2)$ loop models at the Berezinskii-Kosterlitz-Thouless point, which is characterized by two decoupled Luttinger liquids with \emph{coarse-grained} fields $\theta^{\R}, \theta^{\B}$ (see App.~\ref{sec:field_theory}). The constraint can be energetically imposed via the $G_{m_R, m_B}$-symmetric local term\footnote{For $\theta^{\R}_g, \theta^{\B}_g = \frac{(2n+1) \pi}{2}$, $\cos^2(\theta^{\R}_g - \theta^{\B}_g) = \sin^2(\theta^{\R}_g + \theta^{\B}_g)$.} $-|J|(\v{n}^{\R}\cdot \v{n}^{\B})^2 =-|J| \sin(2 \theta^{\R})\sin( 2\theta^{\B})$, which is highly relevant at this point. This coupling leads to an $8$-fold degenerate manifold with either $\theta^{\R},\theta^{\B} \in \{\frac{\pi}{4}, \frac{5\pi}{4}\}$ or $\theta^{\R},\theta^{\B} \in \{-\frac{\pi}{4}, \frac{3\pi}{4}\}$, all of which correspond to long-range-ordered $\theta^{\R}_g, \theta^{\B}_g$. Each minimum explicitly breaks the symmetry $G_{m_R,m_B}$ down to a $\mathbb{Z}_2\times \mathbb{Z}_2$ subgroup ($ \langle S_R \rangle \times \langle S_B \rangle$ for the former configurations). The remaining $\Z_2^2$ symmetry reflects a lack of long-range order in $\theta_b^{\R}, \theta_r^{\B}$; such order would respectively indicate condensation of $e_R, e_B$ anyons. The resulting quantum phase is topologically trivial and becomes $\ket{\beta_\R=\beta_\B=\infty}=\ket{+}^{\otimes |{\R}|}\ket{+}^{\otimes |{\B}|}\ket{\uparrow}^{\otimes |{\G}|}$ (see App.~\ref{sec:sb_phase}). As we show next, condensation of $e_G, m_R, m_B$ anyons underlies the onset of this trivial state.

\textbf{Condensation order parameters.---}
We confirm the predicted instability by numerically simulating the thermal phase diagram of  Eq.~\eqref{eq:H_local_rotor} using Monte-Carlo methods with single-site updates (see App.~\ref{sec:monte_carlo_sims}). 
We diagnose anyon condensation in the deformed state $\ket{\beta_{\B},\beta_{\R}}$ by considering the (unnormalized) matrix elements ${W}_a(x, y) = \bra{\beta_\B, \beta_\R} M_\beta  \hat{W}_a(x, y)M_\beta^{-1} \ket{\beta_\B, \beta_\R}$ for any finite deformation with  $M_\beta$ representing the invertible transformation~\footnote{Notice that for any anyon $a$, the deformed Wilson operator $M_\beta \hat{W}_a M_\beta^{-1}$ on any contractible loop is a (non-unitary) symmetry of the deformed wavefunction $M_\beta \ket{D_4}$. This being a similarity transform, the deformed operators obey the same algebraic relations as the original $\hat{W}_a$.} defined in Eq.~\eqref{eq:psi_bR_bB}, and where $\hat{W}_a(x,y)$ is the Wilson operator creating a pair of $a$ anyons at $x$ and $y$ on top of $\ket{D_4}$~\cite{fan2023diagnostics}.  In the limit $|x-y|\to \infty$, the normalized ${W}_a(x,y)$ vanishes within the phase where $a$ anyons are well-defined quasi-particles, and is finite otherwise. 
In the local stat-mech model formulation in Eq.~\eqref{eq:H_local_rotor}, the quantities ${W}_a(x,y)$ for $a=m_R, m_B, e_{G}$ map to two-point functions $W_{m_R}(b,g)=\langle \v{n}_b^{\R} \cdot \v{n}_g^{\R} \rangle$, $W_{m_B}(r,g)=\langle \v{n}_r^{\B} \cdot \v{n}_g^{\B} \rangle$, and $W_{e_G}(g,g')=\langle \widetilde{\sigma}_g \widetilde{\sigma}_{g'} \rangle $ with $\widetilde{\sigma}_g\equiv \v{n}_g^{\R} \cdot \sigma^x\v{n}_{g}^{\R}=\pm 1$ transforming in a $1$-dimensional irrep of $G_{m_R, m_B}$
(see App.~\ref{sec:wilson_line_observables}). We numerically evaluate $W_{a}=\frac{1}{N_xN_y}\sum_{x,y}W_{a}(x,y)$, where $N_{x(y)}$ denotes the number of $x(y)$ vertices, and their corresponding Binder cumulants: $U(e_G)=\frac{3}{2}(1-\frac{\langle (\sum_g \widetilde{\sigma}_g)^4\rangle}{3\langle (\sum_g \widetilde{\sigma}_g)^2\rangle^2})$ and $U(m_R), U(m_B)$. The latter two are generalizations to vector order parameters~\cite{BinderLandauFiniteSize, sandvik_AT}, which are further discussed in the End Matter. Binder cumulants are normalized such that in the thermodynamic limit, $U \rightarrow 0  \ (1)$ in the symmetric (symmetry-broken) phase.

\textbf{Numerical results.---}
We first focus on the higher-symmetry line $t_\R = t_\B$ that features the larger symmetry $G_{m_R, m_B} \rtimes \Z_2^{\mathrm{swap}}$. As shown in Fig.~\ref{fig:Fig_2}(b,c), all three Binder cumulants cross at a common threshold $t_c \approx 0.7744$---coincident to within one standard deviation.  The cumulants approach $1$ for $t_\R = t_\B > t_c$, signaling simultaneous condensation of not only $m_R$ and $m_B$, but also $e_G$ anyons driven by their production via fusion of the proliferated non-Abelian anyons; $D_4$ TO accordingly enters a topologically trivial phase~\cite{d4_hasse}\footnote{The wavefunction may nevertheless retain subtle nontrivial characteristics, similar to the strongly deformed toric code \cite{CastelnovoChamon2007DefToricCode, HuxfordNguyenKim2023,Sahay2025FunkyIsing,Manoj2026,Sahay2026}.} in agreement with the preceding field-theory analysis (see App.~\ref{sec:field_theory}). The \emph{joint} proliferation of $m_R$ and $m_B$ destabilizes $D_4$ TO, though the threshold $t_c$ still exceeds the toric-code value $t_\mathrm{TC} = 1/\sqrt{3}$ on the honeycomb lattice~\cite{HOUTAPPEL1950425}. The absence of an intermediate phase along this line is consistent with the Ginzburg-Landau (GL) analysis of App.~\ref{sec:ginzburg_landau}, where the relevant $G_{m_R, m_B}$-symmetric couplings between the two-dimensional order parameters $\v{n}^{\R}, \v{n}^{\B}$ and the scalar order parameter for $e_G$ condensation generically lock the three orderings together. At a mean-field level the GL theory admits a continuous transition, though fluctuations potentially drive it weakly first order. Our numerics support but do not conclusively establish the latter scenario, as the negative extrema in the Binder cumulants (Fig.~\ref{fig:Fig_2}(b,c)) are consistent with proximity to a multi-critical point or fluctuations along a projection of a vector order parameter (as in the Ashkin-Teller model~\cite{sandvik_AT}). 

Away from the diagonal $t_\R \neq t_\B$, the mean-field GL analysis allows in principle for an intermediate phase where only a subset of $\{m_R, m_B, e_G\}$ condenses. Numerically, however, sampling a constant $t_\R = 0.8$ cut (blue line in Fig.~\ref{fig:Fig_2}(c); see App.~\ref{sec:no_intermediate}) yields Binder-cumulant crossings for all three diagnostics ($t_c \approx 0.7511$) that coincide within statistical error, indicating that no such intermediate phase is resolved. Although we cannot rule out a narrow intermediate window beyond our resolution, our observations, particularly the bimodal energy histograms in App.~\ref{sec:energy_hist}, are again consistent with a weakly first-order transition that locks the three condensations together~\cite{LeeKosterlitzBimodal, LeeKosterlitzBimodal2}. Numerical simulations for $t_\R > 0.8$ (App.~\ref{sec:gapless_edge}) become inconclusive due to proximity to the zero-temperature limit, but suggest that no additional phase exists close to the $t_\R=1$ or $t_\B=1$ axes. We conjecture that the $t_\R = 1$ and $t_\B = 1$ lines are critical for a finite range of $t_\B$ and $t_\R$, respectively, due to an ``accidental'' U$(1)$ 1-form symmetry exhibited by the deformed wavefunction~\cite{sohal2025obstructionergodicitylocalityu1}, which precludes the appearance of a relevant term for a finite region of the Luttinger parameter. Finally, App.~\ref{sec:mR_mB_eG} presents additional results incorporating direct proliferation of $e_G$ anyons via the $\beta^z_{\G}$ deformation in Eq.~\eqref{eq:psi_bR_bB}.

\textbf{Symmetry-resolved condensates.---}
Of the 22 anyons in $D(D_4)$, $m_R$ and $m_B$ generate 14 anyons via fusion~\cite{Iqbal_24}: $\mathcal{A}_{m_R,m_B} = \{\text{all 8 Abelian anyons}\} \cup \{m_R, m_B, f_R, f_B, m_{RB}, f_{RB}\}$. Notably, $G_{m_R, m_B}$ is the minimal group with exactly 8 one-dimensional and 6 two-dimensional irreps that precisely reproduce the fusion rules of $\mathcal{A}_{m_R,m_B}$ (see App.~\ref{sec:symmetry_group}). In fact, the stat-mech model including finite $\beta^z_{c=\R, \G, \B}$, i.e., anyons already generated by fusing $m_R$ and $m_B$, retains the same $G_{m_R,m_B}$ symmetry.

The mapping between anyons and irreps enables differentiation between ordinarily indistinguishable phases in a conventional $2+1$d ground-state classification. Such distinctions can be made by identifying the irreps needed to support the un-condensed anyons projected to a condensate, revealing the remnant symmetry of the corresponding symmetry broken phase.
For instance, the residual symmetries characterizing the phases for sufficiently large $\beta_\R, \beta_\B$ on one hand and $\beta^z_{\R, \G, \B}$ on the other are distinct $\Z_2^2$ subgroups of $G_{m_R, m_B}$, which identify the un-condensed set of $\langle e_R, e_B\rangle$ and $\langle m_R, m_B\rangle$ anyons, respectively. However, since these groups are isomorphic, one has to tread with care. Indeed, while the wavefunction-deformed toric code has $e$- and $m$-condensates which are separated by a phase transition \cite{Haegeman_15,GuoYi_19}, they can be connected by a finite-depth local unitary using $em$-duality symmetry (see App.~\ref{app:TC_e_m_equiv}).

But unlike the toric code case, allowing for all deformations from Eq.~\eqref{eq:psi_bR_bB} (in particular, including possible $m_G$ generation) makes the two trivial condensates for $D_4$ TO explicitly distinct. In this case, the symmetry of the stat-mech model enlarges as $G_{m_R, m_B}\to G_{m_R, m_B, m_G}$ since now all $22$ anyons of $D_4$ TO are generated by fusing $m_R, m_B, m_G$; see App.~\ref{sec:3_NA_anyons}. The non-isomorphic symmetries characterizing the phases for sufficiently large (but finite) $\beta_\R, \beta_\B$ and $\beta^z_{\R, \G, \B}$ are then $D_4$ and $\Z_2^3$ subgroups of $G_{m_R, m_B, m_G}$, identifying the un-condensed sets $\langle e_R, e_B, m_G\rangle$ and $\langle m_R, m_B, m_G\rangle $ respectively (see App.~\ref{sec:3_NA_A_anyons}). These are again the minimal groups whose irreps consistently fuse as the set of un-condensed anyons, e.g., $m_G\times m_G=1+e_R+e_B+e_Re_B$.

The previous analysis hence suggests that for invertible wavefunction deformations, the proliferating-anyon-label remains meaningful away from criticality for finite $\beta$ (since for $\beta=\infty$ the deformation becomes a projector). As we show below, this additional symmetry label extends to arbitrary finite-depth invertible deformations acting on topologically ordered ground states that admit an exact tensor network representation

\textbf{General approach to identify the symmetry.---}
Fusion in the group $G_{m_R,m_B}$ is consistent with that of anyons in $\mathcal{A}_{m_R,m_B}$, yet the correspondence is not one-to-one: pairs of bosons and fermions share the same fusion multiplicities. Even more drastically, this correspondence explicitly breaks down when trying to identify irreps of $G_{m_R,m_B,m_G}$ with the two semions of $D_4$ TO (see App.~\ref{sec:3_NA_anyons}). Central to this mismatch is the fact that local order parameters cannot faithfully reproduce the full anyon content, including, e.g., non-trivial statistics \cite{kitaev2003fault}. Hence, a general approach should rather expose a full equivalence (Morita equivalence) between the category of anyons of $D(G)$ and the symmetry of the stat-mech model \cite{ostrik2002module, etingof2005fusion, lootens2023dualities, lootens2024dualities}. 

The natural symmetries for a stat-mech model from a deformation that generates the full anyon content are the intrinsic ``double layer'' $G\times G$ and $G\times \textrm{Rep}(G)$ symmetries~\cite{haegeman2015shadows, schuch2013topological, schuch2010peps, Xu_2021,Xu_2022}, corresponding to different PEPS representations of bra and ket (see Fig.~\ref{fig:composite_figure}(a)). In this case, degrees of freedom in the resulting stat-mech models correspond to group elements, rather than to the proliferated anyons. Here, we provide an alternative but direct connection between these two approaches inspired by the strange correlator construction and the recent insights for dualities in stat-mech models~\cite{you2014wave,vanhove2018mapping,etingof2005fusion, lootens2023dualities}. 
As shown in the End Matter, we find that a dual $G\times \mathrm{Rep}(G)$-symmetric~\footnote{$G\times \mathrm{Rep}(G)$ is used instead of the more formal $\mathrm{Vec}_{G} \times \mathrm{Rep}(G)$ to denote the (untwisted) category itself.} model---different from that in Fig.~\ref{fig:composite_figure}(a)---can be obtained by resolving the deformation of a quantum double $D(G)$ wavefunction in the anyon basis and viewing the wavefunction norm as an overlap between a local product state and a string-net ground state with input the category of anyons of $D(G)$ and a virtual $G\times \mathrm{Rep}(G)$ symmetry~\cite{lootens2021matrix}. 
As a result, degrees of freedom of the stat-mech model now directly correspond to proliferated anyons (i.e., conjugacy classes and irreps of their centralizers). In particular, both local and non-local order parameters are labeled by the $G\times \mathrm{Rep}(G)$ topological sectors~\cite{lootens2019cardy}, in this case tuples of anyons $(\alpha,\beta)$ with $\alpha$, $\beta \in D(G)$~\cite{etingof2005fusion}.

For the microscopic model of $D_4$ TO considered in this work, there exists a clear relationship between the group-like symmetry $G_{m_R, m_B}$ and the categorical symmetry $D_4\times \mathrm{Rep}(D_4)$. The loop model $\langle \beta_\B, \beta_\R|\beta_\B, \beta_\R\rangle$ can be equivalently written as a stat-mech model with a categorical symmetry obtained from a $\Z_2$ quotient of $D_4 \times \mathrm{Rep}(D_4)$. The former $G_{m_R, m_B}$-symmetric formulation is then exactly recovered from the latter, causing the non-invertible symmetry to reduce to the group-like one (see End Matter).
Although we currently lack an \emph{equivalent} $D_4\times \mathrm{Rep}(D_4)$-symmetric formulation of the $G_{m_R, m_B, m_G}$-symmetric model~\eqref{eq:3NA_local_main} that reproduces the same partition function, 
the generality of the approach introduced in this section suggests that such a formulation exists.

\textbf{Topological quantum memory.---}
We now turn to the fate of non-Abelian topological order under decoherence.  Given a general completely positive trace-preserving map $\mathcal{E}$ that generates local anyon insertions, along with a projector $P_{a_v}$ onto a fixed anyon type $a$ at vertex $v$, the corresponding anyon syndrome distribution is $p(\{a_v\})=\tr\left[\prod_v P_{a_v}\mathcal{E}(\ket{D(G)}\bra{D(G)})\right]$. Writing $p(\{a_v\})$ as a wavefunction overlap $p(\{a_v\})= \langle D(G)|\mathcal{E}^{\dagger}\left(\prod_v P_{a_v}\right)|D(G)\rangle$ (where $\mathcal{E}^\dagger$ is the adjoint completely positive quantum channel), we can directly apply the discussion from the previous section to conclude that $p(\{a_v\})$ corresponds to a (disordered) stat-mech model with a $G\times \mathrm{Rep}(G)$ (or in the dual picture $G\times G$) symmetry. Hence, the symmetry analysis for deformed wavefunction overlaps, extends naturally to the anyon syndrome configuration. This connection then suggests that a parallel analysis to that of non-isomorphic trivial condensates for deformed wavefunctions could help to distinguish the resulting classical memories~\cite{Mong_24, sohal_24, axio_25}.

Turning to the stability under the noise channel in Eq.~\eqref{eq:E_channel}, our previous analysis suggests that $D_4$ TO breaks down, as diagnosed by Rényi-like quantities, when the error rates associated with proliferating two non-Abelian anyon species exceed a nonzero threshold. In particular, the purity in Eq.~\eqref{eq:purity} maps to a local stat-mech model with $(G_{m_R,m_B})^{\times 2}\rtimes\mathbb{Z}_2$ symmetry (see App.~\ref{sec:numerical_purity_instability}). The corresponding order parameters $W_{a^{1}\times a^{2}}$ diagnose simultaneous anyon proliferation on the two replicated copies of $\rho$. Because the Binder cumulants are noisy in this regime, we instead analyze the susceptibility $\chi(e^1_G\times e^2_G)$ associated with $W_{e^1_G\times e^2_G}$ (see End Matter).

Along the diagonal cut $p_R=p_B$, this susceptibility identifies a critical error rate $p_c^{(2)}$ beyond which the quantum memory is lost according to the purity diagnostic; see Fig.~\ref{fig:Fig_2}(d). This conclusion is bolstered by the exactly solvable point $p_R=p_B=1/2$, where eigenvalues of $\rho$ become $G_{m_R,m_B}$-symmetric partition functions (see App.~\ref{sec:decoherence_eigenvalues}). At this point, the infinite moment $\lim_{n\rightarrow\infty}\tr(\rho^n)^{1/n}$ maps to the partition function in Eq.~\eqref{eq:Z_psi} with $t_\R=t_\B=1$, corresponding to the upper-right corner of the phase diagram, where $e_G$, $m_R$, and $m_B$ anyons proliferate. Hence, $p_c^{(\infty)}<1/2$. These results together provide strong evidence for a finite information-theoretic threshold when multiple non-Abelian anyons proliferate.

\textbf{Outlook.---} 
Our work invites many future directions. We did not conclusively determine the order of the phase transition in Fig.~\ref{fig:Fig_2}. Future work simulating the coupled loop models using more tailored algorithms and/or renormalization group treatments of the Ginzburg-Landau action could settle this issue. It would also be interesting to understand the stability of $D_4$ TO under the proliferation of all three $m_R, m_B, m_G$ non-Abelian anyons~\cite{d4_hasse}, and when local deformations do not commute (e.g., in the quantum double formulation via genuine ribbon operators). 

Our general symmetry identification focused on the quantum double formulation. However, the underlying tensor-network construction applies to general string-nets, including double Fibonacci, where the corresponding stat-mech models have non-invertible symmetries. To connect the optimal error threshold for string-net-based quantum memories (upon measuring local syndromes) to a phase transition of classical stat-mech remains an important open question. Equally interesting would be a rigorous generalization of our work to general ground states which do not admit an exact tensor network representation.

More generally, a long-term goal is to identify the optimal threshold of non-Abelian topological quantum memories. This work provides a further step toward the identification of the relevant symmetries when arbitrary anyons are proliferated, and the corresponding mechanisms leading to its breakdown. We expect that the $G\times \textrm{Rep}(G)$ symmetry predicted for syndrome distributions provides consistency checks for future works and a route to gauge the stability of the involved TO based on symmetry considerations. Overall, characterizing the optimal threshold with both perfect and faulty syndrome measurements in non-Abelian codes remains a long-term endeavor.

\textbf{Acknowledgments.---} We acknowledge helpful discussions and feedback from Ehud Altman, Xie Chen, Arpit Dua, Rossoneri Jing, Vibhu Ravindran, Nat Tantivasadakarn, and Frank Verstraete. We also thank Nandagopal Manoj for feedback on the manuscript. A.V. acknowledges support from the Bill Davis SURF fellowship. R.~Vanhove is supported by a fellowship from the Flemish Research Foundation (FWO), the Simons collaboration on `Ultra-Quantum Matter' (grant number
651438) and the Simons Investigator Award (award ID 828078). P.S. acknowledges the support from the U.S. Department of Energy, Office of Science, Office of High Energy Physics, under QuantISED Award DE-SC0019380; the NSF QLCI program through Grant No. OMA-2016245; and the support from the Caltech Institute for Quantum Information and Matter, an NSF Physics Frontiers Center (NSF Grant No.PHY-1733907), and the Walter Burke Institute for Theoretical Physics at Caltech.  The U.S. Department of Energy, Office of Science, National Quantum Information Science Research Centers, Quantum Science Center supported the Ginzburg-Landau analysis portion of this work.

\let\savedacl\addcontentsline
\renewcommand{\addcontentsline}[3]{}
\section*{End Matter} 
\label{sec:end_matter}

\textbf{Loop model symmetries.---}
The symmetry group $G_{m_R, m_B}$ (2 non-Abelian anyons) can be expressed as $G_{m_R, m_B} \cong \Z_2^3 \rtimes \Z_2^2 \cong \Z_2^4 \rtimes \Z_2$, or in terms of its generators as 
\begin{equation}
    \begin{split}
        G_{m_R, m_B} &\cong ( \langle V_R \rangle \times \langle V_B \rangle \times \langle T_G \rangle ) \rtimes ( \langle S_R \rangle \times \langle S_B \rangle ) \\
        &\cong ( \langle V_R \rangle \times \langle V_B \rangle \times \langle S_R \rangle \times \langle S_B \rangle) \rtimes \langle T_G \rangle.
    \end{split}
\end{equation}
The action of the semidirect product is specified by relations $V_\alpha^2 = T_G^2 = S_\alpha^2 = e$, $S_R S_{B} = S_{B} S_R$, $V_\alpha = [T_G, S_\alpha]$. Here $[g, h] = g h g^{-1} h^{-1}$, $\alpha = R, B$. $G_{m_R, m_B}$ is indexed by the GAP~\cite{GAP4} ID $\mathrm{SmallGroup}(32, 27)$.

\begin{figure}
    \centering
    \includegraphics[width=\linewidth,page=12]{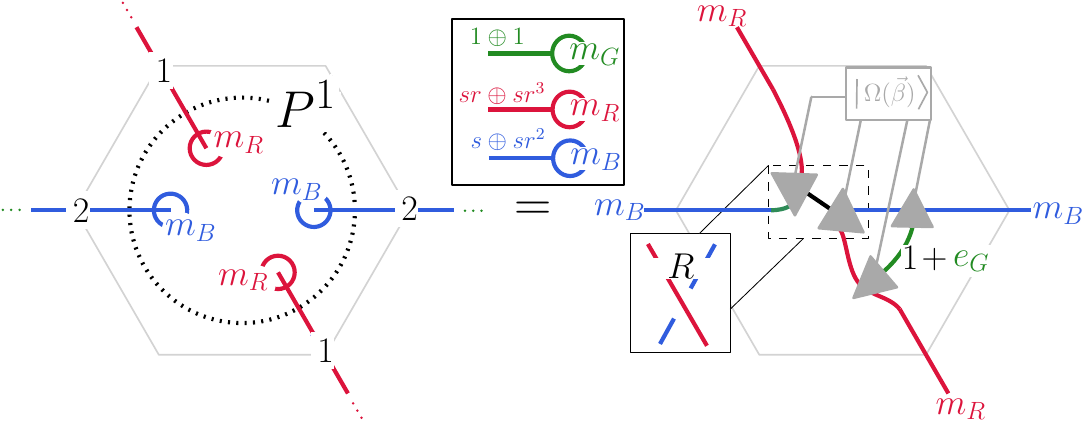}
    \caption{\textbf{MPO anyon network.} One particular anyon configuration created by a deformation on a plaquette of the $D_4$ quantum double groundstate: an $m_R$ and $m_B$-anyon pair is created in the specified order. The overlap forces the final anyon on every plaquette to be the vacuum, indicated by the projector $P^1$, and evaluates to a closed network of fusion tensors, representing the braiding and fusion processes consistent with the given ordering. The network is exactly a string-net strange correlator as in Eq.~\eqref{SCstring-net}.}
    \label{fig:anyon_process}
\end{figure}

When proliferating the all three of $m_\R, m_\G$ and $m_\B$, the coupled-loop model reads
\begin{equation} \label{eq:3NA_main}
   \mathop{\scalebox{2}{$\sum$}}_{L_\R, L_\B, L_\G \ \not\ni \borromeancolor} \prod_{c=\R, \G,\B}\bigg(\frac{t_c}{\sqrt{2}} \bigg)^{|L_c|}  2^{C_{L_\R \cup L_\B \cup L_\G}},
\end{equation}
where $\borromeancolor$ indicates Borromean-like loop configurations in which all three loops of different colors cross, and the corresponding Boltzmann weight vanishes (see App.~\ref{sec:3_NA_anyons}). This partition function appears as the exact high-temperature expansion of the local stat-mech model   
\begin{equation}
\label{eq:3NA_local_main}
    H = - \sum_{c=\R, \G,\B}\left[\beta_c \sum_{\langle i,j\rangle_c} \sigma_i^{c} \sigma_j^{c} \CZ_{i j}\right],
\end{equation}
where $\langle i,j\rangle_c $ denotes nearest-neighbor sites on a honeycomb of color $c$ (see Fig.~\ref{fig:anyon_process}(a)) and $t_c=\tanh(\beta_c)$. Here $\CZ_{i j} = \frac{1}{2} (1 + \widetilde{\sigma}_i + \widetilde{\sigma}_j - \widetilde{\sigma}_i \widetilde{\sigma}_j)$ for $\widetilde{\sigma}_j \in \{-1, 1\}$. The corresponding symmetry group $G_{m_R, m_B, m_G}$ can be expressed as $G_{m_R, m_B, m_G} \cong \Z_2^3 \rtimes D_4 \cong (\Z_2^2 \times D_4 ) \rtimes \Z_2$, and it is indexed by the GAP~\cite{GAP4} ID $\mathrm{SmallGroup}(64, 73)$. Notably, $G_{m_R, m_B, m_G}$ has 2 (out of 22) complex irreps which are dual to each other. In contrast, all 22 anyons of $D(D_4)$ are self-dual~\cite{Iqbal_24}. Hence, as noted in the main text, the fusion rules of the irreps of $G_{m_R, m_B, m_G}$ and those of the anyons of $D(D_4)$ cannot be in one-to-one correspondence.

\textbf{Vector Binder cumulants. ---}
We define the Binder cumulants for the 2-vector order parameters probing the condensation of non-Abelian anyons $m_R, m_B$ as
\begin{equation}
\begin{split}
    U(m_\alpha) &= 2 \bigg( 1  - \frac{\langle [\sum_{g, g'} \v{n}_{g}^{\alpha} \cdot \v{n}_{g'}^{\alpha}]^2 \rangle }{2 \langle \sum_{g, g'} \v{n}_g^{\alpha} \cdot \v{n}_{g'}^{\alpha} \rangle^2} \bigg)
\end{split}
\end{equation}
for $\alpha = \R, \B$. This Binder cumulant is normalized such that $U(m_\alpha)= 0 ~(1)$ in the ordered (disordered) phase. See App.~\ref{sec:binder_derivations} for additional details. 

\textbf{Purity quantities. ---}
The following spin Hamiltonian's partition function recovers the expression for the $\tr(\rho^2)$ in Eq.~\eqref{eq:purity}:
\begin{equation}
H_2= -2\sum_{\alpha = \R, \B} \mu_\alpha \sum_{\langle i, j \rangle_\alpha} \sigma_i^{\alpha} \sigma_j^{\alpha} \CZ^{(1)}_{i j} \CZ^{(2)}_{i j}
\end{equation}
where $\tanh(\mu_\alpha) = p_\alpha / (1 - p_\alpha)$. The hexagons $\red{\hexagon}, \blue{\hexagon}$ are A-B stacked as in Eq.~\eqref{eq:H_local_rotor}. We define the magnetization as $m(e_G^{12}) = \frac{1}{|\R \cap \B|}\sum_{g \in \R \cap \B} \widetilde{\sigma}_g^{(1)} \widetilde{\sigma}_g^{(2)}$. The corresponding susceptibility is then
\begin{equation}
\begin{split}
    \chi(e_G^1 \times e_G^2) = \frac{p |\R \cap \B|}{(1 - p)} [ \langle m(e_G^{12})^2 \rangle - \langle |m(e_G^{12})| \rangle^2],
\end{split}
\end{equation}
where $p = p_R = p_B$, and $|\R \cap \B|$ denotes the total number of sites on the triangular lattice shared by $\red{\hexagon}, \blue{\hexagon}$. See App.~\ref{sec:dec_density_mat} for additional details.

\textbf{Tensor network approach and $D_4\times \mathrm{Rep}(D_4)$ symmetric stat mech model. ---}
The aim of this section is to use the existing Matrix Product Operator (MPO) formalism for TO \cite{haegeman2015shadows, schuch2013topological, schuch2010peps, Xu_2021,Xu_2022, csahinouglu2021characterizing, bultinck2017anyons, williamson2017symmetry} to first, identify a generic $G\times \mathrm{Rep}(G)$ symmetry for general deformed quantum double overlaps $\bra{D(G)} O(\vec{\beta}) \ket{D(G)}$, where the operator $O(\vec{\beta})$ creates the full set of anyons in $D(G)$. Second, we provide an alternative $D_4\times \mathrm{Rep}(D_4)$-symmetric stat-mech model equivalent to that introduced in Eq.~\eqref{eq:Z_psi}. 

We address both points by resolving the operator $O(\vec{\beta})$ in the anyon basis and considering its action on the fixed-point ground state Projected Entangled Pair State (PEPS) representation, which is guaranteed to have exact virtual MPO symmetries. Since no non-trivial excitations survive in the overlap, the effect of $O(\vec{\beta})$ is represented at the virtual level of the groundstate by a closed network of MPOs that can be freely pulled through the lattice. This network can itself be interpreted as a strange correlator \cite{vanhove2018mapping} of a string-net groundstate of the category of anyons of $D(G)$ (the Drinfeld center $\mathcal{Z}(\mathrm{Vec}_{G})$):
\begin{align}
\nonumber \bra{D(G)} O(\vec{\beta}) \ket{D(G)}&= \substack{\text{MPO}\\ \text{network}}(\vec{\beta}) \times \braket{D(G)|D(G)} \\
&= \braket{\Omega(\vec{\beta})|\Psi_{\mathrm{SN}(\mathcal{Z}(\mathrm{Vec}_{G}))}}.
\label{SCstring-net}
\end{align}

The weights $\vec{\beta}$ and the microscopic details of the deformation are contained in the topologically trivial state $\bra{\Omega(\vec{\beta})}$ and the topological aspects of the anyon fusion and braiding processes are captured by the string-net wavefunction, whose virtual symmetry is inherited by the resulting stat-mech model. Note that in writing $\braket{D(G)|D(G)} = 1$, we have effectively assumed that either $O(\vec{\beta})$ does not create non-contractible deformations or that those contributions have been killed by a suitable choice of boundary conditions in the partition function. 

One can intuitively understand the appearance of the string-net wavefunction by folding, i.e. writing the overlap as $\bra{\phi(\beta)}( \ket{D(G)}\otimes \ket{D(G)})$ (an overlap between two copies of $D(G)$ and an adorned Bell-like state $\ket{\phi(\beta)}$). The string-net model for $\mathcal{Z}(\mathrm{Vec}_{G})$ realizes the same topological order as the doubled state $\ket{D(G)}\otimes \ket{D(G)}$, so they are related by a finite-depth unitary quantum circuit that can be absorbed in the state $\bra{\phi(\beta)}$. The strange correlator inspired way of writing the partition function comes with a host of dual stat-mech models that correspond to different choices of the tensor network representation of the string-net groundstate in Eq.~\eqref{SCstring-net}. Every representation comes with its own symmetry, given by a unique fusion category that is Morita equivalent to $\mathcal{Z}(\mathrm{Vec}_{G})$ (and by extension also Morita equivalent to $G\times G$) \cite{lootens2021matrix, lootens2023dualities, lootens2024dualities}. One particular choice yields a stat-mech model with a general $G\times \mathrm{Rep}(G)$ symmetry; such a choice is motivated by the direct correspondence between the stat-mech degrees of freedom and the proliferated anyon strings.

One way of obtaining the explicit stat-mech Boltzmann weights in Eq.~\eqref{SCstring-net} is to construct the fusion and braiding tensors for the anyon configurations created by the deformation on every plaquette of the quantum double ground state (via the recipe in Section IV of Ref.~\cite{williamson2017symmetry}). The method uses the representations of the anyons on the virtual level of the PEPS in terms of idempotents of the $D_4$ tube algebra ~\cite{williamson2017symmetry}. The procedure is shown in Fig.~\ref{fig:anyon_process} for a particular anyon configuration created by an unspecified deformation. When proliferating the two non-Abelian anyons $m_R,m_B$, anyon loops are spanned by an internal two-dimensional space labeled by a combination of the elements in their corresponding $D_4$ conjugacy class. We use the convention that $m_G$ is the pure charge anyon (trivial conjugacy class) while $m_R$ and $m_B$ are dyons.

\begin{figure}[t!]
    \centering
    \includegraphics[width=\linewidth]{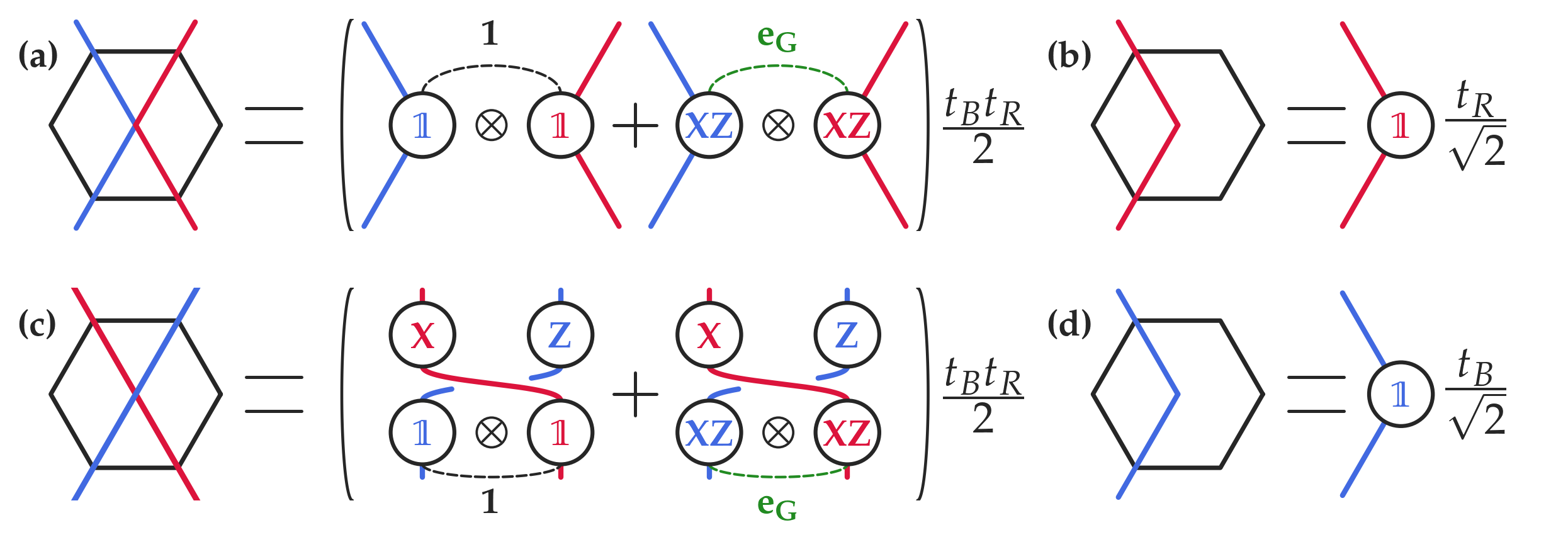}
    \caption{\textbf{Boltzmann weights of the coupled-loop stat-mech model.} Loops interact by exchanging a trivial charge and an Abelian charge ($e_G$ for the red/blue interaction) (a) and in addition braid non-trivially when they intersect (c). (b,d) A non-interacting loop segment acts trivially in its respective 2-dimensional space. Other weights are analogously defined.}
    \label{fig:Boltzmann_weights}
\end{figure}

We now study an equivalent stat-mech model (obtained using this procedure) that reproduces the coupled-loop model in Eq.~\eqref{eq:Z_psi}. This model consists of closed loops on two honeycomb lattices (one for each color) and an inter-color interaction. On every bond, we place a 3-dimensional Hilbert space: one degree of freedom for an empty bond, while red and blue strings span a two-dimensional subspace labeled by the group elements in their respective conjugacy class (see Fig.~\ref{fig:anyon_process}). The explicit Boltzmann weights are shown in Fig.~\ref{fig:Boltzmann_weights}.
Closed loops weights correspond to tracing over the Pauli matrices generated by the loop interactions, and attaching the corresponding tension factors. By contracting the resulting network, the coupled-loop weight in Eq.~\eqref{eq:Z_psi}, is recovered. 
The Boltzmann weights are symmetric under a local $D_4$ symmetry with $D_4=\langle r,s|r^4=s^2=e, srs=r^3\rangle$ generated by $T_s$ and $T_{sr}$ and an additional $\Z_2^2$ symmetry generated by $V_R$ and $V_B$. On the 3-dimensional subspaces, for each color the action of the generators can be written as: $T_s^R = 1\oplus X_R, T_s^B = 1\oplus Z_B$, $T_{sr}^R = 1\oplus Z_R, T_{sr}^B=1\oplus X_B$, $V_{R,B} = 1\oplus(-\mathbb{1}_{R,B})$, with $Z_R=\ket{sr}\bra{sr}-\ket{sr^3}\bra{sr^3}$, $X_R=\ket{sr}\bra{sr^3}+\ket{sr^3}\bra{sr}$; and $Z_B, X_B,-\mathbb{1}_R,-\mathbb{1}_B $ analogously defined. Moreover, one can identify a non-invertible symmetry $T_{\pi}$ with $T_{\pi}^2 = 1 + V_R + V_B + V_R\otimes V_B$. The full categorical symmetry is then obtained from $D_4\times \mathrm{Rep}(D_4)$ by noticing that $T_{r^2}^{R,B}=V_{R,B}$, namely the $r^2$ central element corresponds to a 1 dimensional irrep of $\mathrm{Rep}(D_4)$. The resulting symmetry $(D_4\times \mathrm{Rep}(D_4)) / \langle(r^2,s_1)\rangle$ can then be understood as a $\mathbb{Z}_2$ quotient of $D_4\times \mathrm{Rep}(D_4)$. 
The non-invertible symmetry $T_{\pi}$ appears as an MPO symmetry in the stat-mech model, i.e. the local Boltzmann weights satisfy the \emph{pulling-through} tensor equation
\begin{align}
\label{pulling_through}
\includegraphics[valign=c,page=14,scale=0.8]{figures/figures_Robijn_D4.pdf} = \includegraphics[valign=c,page=15,scale=0.8]{figures/figures_Robijn_D4.pdf},
\end{align}
shown here explicitly for the crossing weights (Fig.\ref{fig:Boltzmann_weights}(c)) (an analogous equation holds for the non-crossing weight in Fig.\ref{fig:Boltzmann_weights}(a)). The symmetry element $T_{\pi}$ is obtained by closing the MPO on periodic boundary conditions $T_{\pi} = \includegraphics[valign=c,page=17,scale=0.35]{figures/figures_Robijn_D4.pdf}$, where the entries of the MPO tensor \includegraphics[valign=c,page=16,scale=0.65]{figures/figures_Robijn_D4.pdf} $= \Gamma^{\pi}(g)_{i,j}$, are given by the 2-dimensional representation matrices $\Gamma^{\pi}(s)=X$ and $\Gamma^{\pi}(sr)=Z$. When the MPO crosses a colored string, this symmetry acts diagonally as $Z_B$ and $Z_R$ on the conjugacy class elements for each color, and inserts the matrix $X(Z)$ for blue (red) on the virtual level of the MPO (dashed lines). The crossing weights in particular require the anti-commutation between $X$ and $Z$ acting at the virtual level to satisfy the pulling-through condition. This is not necessary for the non-crossing weights which are also symmetric under $T_3=Z_R\otimes Z_B$. In fact, since only closed loop configurations contribute to the partition function, and loop intersections come always in pairs, one can also discard the additional $X$ and $Z$ operators for the crossing weights in Fig.~\ref{fig:Boltzmann_weights}, and recover the same partition function. Hence, the resulting stat-mech model becomes symmetric under the $T_3$, which can be understood as a localized version of the MPO symmetry $T_\pi$ in Eq.\ref{pulling_through}. All together, one then recovers the group $G_{m_R, m_B}$.

\nocite{CastelnovoChamon2007DefToricCode,HuxfordNguyenKim2023,Sahay2025FunkyIsing,Manoj2026,Sahay2026}

\bibliographystyle{apsrev4-1}
\bibliography{references}

\newpage

\appendix
\onecolumngrid

\let\addcontentsline\savedacl
\renewcommand{\tocname}{Appendix Contents}
\tableofcontents

\newpage

\section{Loop model derivation for two non-Abelian anyon proliferation}
\label{sec:Loop_model_2NA}

\subsection{High-temperature expansion of spin Hamiltonian}
\label{sec:high_T_expansion}

As shown in Sec.~C in Ref.~\cite{sala_D4}, evaluation of the norm $\braket{\beta_\R, \beta_\B | \beta_\B, \beta_\R}$ yields the loop model
\begin{equation}
    \langle \beta_\B, \beta_\R|\beta_\B, \beta_\R\rangle= Z(\beta_\R, \beta_\B) = \sum_{L_\R, L_\B} t_\R^{|L_\R|} t_\B^{|L_\B|} \frac{2^{C_{L_\R \cup L_\B}}}{\sqrt{2}^{|L_\R| + |L_\B|}} .
    \label{eq:two_color_partition_functino}
\end{equation}
Here we propose a local spin Hamiltonian inspired by the Hamiltonian obtained from applying ungauging maps on $\ket{\beta_\R}$~\cite{sala_D4}. We verify that the exact high-temperature expansion of this Hamiltonian yields a partition function identical (up to a multiplicative constant) to $Z(\beta_\R, \beta_\B)$. Consider the Hamiltonian
\begin{equation}
    \label{eq:2na_ham}
    H[\beta_\R, \beta_\B] = -\beta_\R \sum_{\langle b, g \rangle \in \red{\hexagon}} \sigma_b^{\R} \sigma_g^{\R} \CZ_{bg} - \beta_\B \sum_{\langle r, g \rangle \in \blue{\hexagon}} \sigma_g^{\B} \sigma_r^{\B} \CZ_{gr}
\end{equation}
where $\CZ_{i j} = \frac{1}{2} (1 + \widetilde{\sigma}_i + \widetilde{\sigma}_j - \widetilde{\sigma}_i \widetilde{\sigma}_j)$ for $\widetilde{\sigma}_j \in \{-1, 1\}$. Here the honeycomb lattices $\red{\hexagon}, \blue{\hexagon}$ are A-B stacked with a common $g$-sublattice as in Fig.~\ref{fig:composite_figure}(a). We now carry out the exact high-temperature expansion of the Hamiltonian:
\begin{equation}
    \begin{split}
        Z_H(\beta_\R, \beta_\B) &= \sum_{\{\sigma^{\R}, \sigma^{\B}, \tilde{\sigma}\}} e^{-H[\beta_\R, \beta_\B]} \\
        &= \sum_{\{\sigma^{\R}, \sigma^{\B}, \tilde{\sigma}\}} \bigg( \prod_{\langle b, g \rangle \in \red{\hexagon}} \cosh(\beta_\R) \bigg[1 + \sigma_b^{\R} \sigma_g^{\R} \CZ_{bg}\tanh(\beta_\R) \bigg] \bigg) \bigg( \prod_{\langle g, r \rangle \in \blue{\hexagon}} \cosh(\beta_\B) \bigg[1 + \sigma_g^{\B} \sigma_r^{\B} \CZ_{gr}\tanh(\beta_\B) \bigg] \bigg) \\
        &= \cosh(\beta_\R)^{|\R|} \cosh(\beta_\B)^{|\B|} \sum_{L_\R, L_\B} \tanh(\beta_\R)^{|L_\R|} \tanh(\beta_\B)^{|L_\B|} \sum_{\{\tilde{\sigma}\}} \bigg(\prod_{\langle b, g \rangle \in L_\R} \CZ_{bg} \prod_{\langle g, r \rangle \in L_\B} \CZ_{gr} \bigg) \\
        &\propto \sum_{L_\R, L_\B} \tanh(\beta_\R)^{|L_\R|} \tanh(\beta_\B)^{|L_\B|} \sum_{\{\tilde{\sigma}\}} \bigg(\prod_{\langle b, g \rangle \in L_\R} \CZ_{bg} \prod_{\langle g, r \rangle \in L_\B} \CZ_{gr} \bigg)
    \end{split}
\end{equation}
The quantity $\sum_{\{\tilde{\sigma}\}} \bigg(\prod_{\langle b, g \rangle \in L_\R} \CZ_{bg} \prod_{\langle g, r \rangle \in L_\B} \CZ_{gr} \bigg)$ is exactly the ``topological weight'' computed in Appendix C.2.b of Ref.~\cite{sala_D4}. Thus the partition function evaluates to
\begin{equation}
\begin{split}
    Z_H(\beta_\R, \beta_\B) &\propto \sum_{L_\R, L_\B} \tanh(\beta_\R)^{|L_\R|} \tanh(\beta_\B)^{|L_\B|} \frac{2^{C_{L_\R \cup L_\B}}}{\sqrt{2}^{|L_\R| + |L_\B|}} = Z(\beta_\R, \beta_\B) = \langle \beta_\B , \beta_\R | \beta_\B, \beta_\R \rangle.
\end{split}
\end{equation}
A similar calculation leads to a local microscopic representation of the purity $\textrm{tr}(\rho^2)$.

By repackaging the Ising-like variables introduced in Eq.~\eqref{eq:2na_ham}, we can recover the rotor Hamiltonian in Eq.~\eqref{eq:H_local_rotor}. Consider the following 2-vectors:
\begin{equation}
\label{eq:n_vec_def}
\v{n}^{\R}_g = \frac{1}{\sqrt{2}} \sigma^{\R}_g
\begin{pmatrix}
1 \\
\widetilde{\sigma}_g
\end{pmatrix}, \ 
\hfill
\v{n}^{\B}_g = \frac{1}{\sqrt{2}} \sigma^{\B}_g
\begin{pmatrix}
1 \\
\widetilde{\sigma}_g
\end{pmatrix}, \ 
\hfill
\v{n}^{\R}_b = \frac{1}{2} \sigma^{\R}_b
\begin{pmatrix}
1+\widetilde{\sigma}_b \\
1-\widetilde{\sigma}_b
\end{pmatrix}, \ 
\hfill
\v{n}^{\B}_r = \frac{1}{2} \sigma^{\B}_r
\begin{pmatrix}
1+\widetilde{\sigma}_r \\
1-\widetilde{\sigma}_r
\end{pmatrix}.
\end{equation}
These vectors are constrained such that on the common sublattice $\v{n}_g^{\R} \cdot \v{n}_g^{\B} = \pm 1$. Moreover, the vectors $\v{n}_g^{\R}, \v{n}_b^{\R}$ transform in the irrep $\chi_{11}$ (see Table~\ref{tab:char_tab_GRB}):
\begin{equation}
\label{eq:rho_11_def}
        \rho_{11}(V_R) = -I, \  \rho_{11}(V_B) = I, \ \rho_{11}(T_G) = \sigma^z, \ \rho_{11}(S_R) = \sigma^x, \ \rho_{11}(S_B) = I.
\end{equation}
On the other hand, the vectors $\v{n}_g^{\B}, \v{n}_r^{\B}$ transform in the irrep $\chi_{13}$ of $G_{m_R, m_B}$: 
\begin{equation}
\label{eq:rho_13_def}
    \begin{split}
        \rho_{13}(V_R) = I, \  \rho_{13}(V_B) = -I, \ \rho_{13}(T_G) = \sigma^z, \ \rho_{13}(S_R) = I, \ \rho_{13}(S_B) = \sigma^x.
    \end{split}
\end{equation}

Using these 2-vectors, the Hamiltonian can be written as in Eq.~\eqref{eq:H_local_rotor} in the main text, or parametrizing these unit vectors in terms of the respective angles $\theta_g^{\R}, \theta_g^{\B}, \theta_b^{\R}, \theta_r^{\B}$. Such a parametrization yields
\begin{equation}
    H[\beta_\R, \beta_\B] = - \beta_\R \sqrt{2} \sum_{\langle b, g \rangle \in \red{\hexagon}} \cos(\theta^{\R}_b -\theta^{\R}_g) - \beta_\B \sqrt{2} \sum_{\langle r, g \rangle \in \blue{\hexagon}} \cos(\theta^{\B}_r -\theta^{\B}_g )
\end{equation}
where now the constraint $\v{n}_g^{\R} \cdot \v{n}_g^{\B} = \pm 1$ becomes $(\theta_g^{\R} - \theta_g^{\B}) \bmod \pi \equiv 0$. The vector and rotor expressions of $H$ are a convenient way of understanding the symmetry action on the microscopic degrees of freedom. In particular, the vector formulation explicitly reveals the order parameters that detect the induced instability discussed in the main text.

\subsection{Symmetry group $G_{m_R,m_B}$}
\label{sec:symmetry_group}

The symmetry group $G_{m_R, m_B}$ of the Ising-like Hamiltonian in Eq.~\eqref{eq:2na_ham} can be expressed as $G_{m_R, m_B} \cong \Z_2^3 \rtimes \Z_2^2 \cong \Z_2^4 \rtimes \Z_2$, or in terms of its generators as
\begin{equation}
    G_{m_R, m_B} \cong \bigg ( \langle V_R \rangle \times \langle V_B \rangle \times \langle T_G \rangle \bigg) \rtimes \bigg( \langle S_R \rangle \times \langle S_B \rangle \bigg) \cong \bigg ( \langle V_R \rangle \times \langle V_B \rangle \times \langle S_R \rangle \times \langle S_B \rangle \bigg) \rtimes \langle T_G \rangle.
\end{equation}

The action of the semidirect product is specified by the presentation of this group:
\begin{equation}
G_{m_R, m_B} \;=\; \left\langle\, V_R, V_B, T_G, S_R, S_B \;\middle|\;
\begin{gathered}
V_\alpha^2 = T_G^2 = S_\alpha^2 = e, \quad S_R S_{B} = S_{B} S_R, \quad V_\alpha = [T_G, S_\alpha]
\end{gathered}
\right\rangle
\end{equation}
where $[g, h] = g h g^{-1} h^{-1}$ and $\alpha = R, B$. 

The action of the generators of $G_{m_R, m_B}$ on the Ising variables is given in Table~\ref{tab:mR_mB_symmetries}. 
\begin{table}[htbp]
\centering
\renewcommand{\arraystretch}{1.25}
\setlength{\tabcolsep}{4pt}
\begin{tabular}{c|ccccccc}
 $G_{m_R, m_B}$ & $\widetilde{\sigma}_r$ & $\widetilde{\sigma}_b$ & $\widetilde{\sigma}_g$ & $\sigma_b^{\R}$ & $\sigma_g^{\R}$ & $\sigma_g^{\B}$ & $\sigma_r^{\B}$ \\ \hline
$V_R$ & & & & $-1$ & $-1$ & & \\
$V_B$ & & & & & & $-1$ & $-1$ \\
$T_G$ & & & $-1$ & $\widetilde{\sigma}_b$ & & & $\widetilde{\sigma}_r$ \\
$S_R$ & & $-1$ & & & $\widetilde{\sigma}_g$ & & \\
$S_B$ & $-1$ & & & & & $\widetilde{\sigma}_g$ & \\
\end{tabular}
\caption{Symmetry generators of $H[\beta_\R, \beta_\B]$. An entry $(g, \sigma) = c$ implies the symmetry $g$ acts on $\sigma$ as $g :  \sigma \mapsto c * \sigma$. Blank entries are to be interpreted as $+1$, i.e., not transforming.}
\label{tab:mR_mB_symmetries}
\end{table}
When $\beta_\R = \beta_\B$, the symmetry is enlarged and includes an extra swap factor such that $G_{m_R, m_B} \rightarrow G_{m_R, m_B} \rtimes \Z_2$. This swap acts as $\{\sigma^{\R} \leftrightarrow \sigma^{\B}, \widetilde{\sigma}_b \leftrightarrow \widetilde{\sigma}_r\}$. In the following sections, we will develop a phenomenological Ginzburg-Landau theory as well as an effective quantum field theory to identify potential phase transitions in $Z(\beta_\R, \beta_\B)$. Hence it will prove useful to identify the irreducible representations (irreps) of $G_{m_R, m_B}$ so that we include all relevant order parameters in these  theories. 

The character table of $G_{m_R, m_B}$ is given in Table~\ref{tab:char_tab_GRB}. We have identified each irrep of $G_{m_R, m_B}$ with an anyon generated by the fusion descendants of the proliferated $m_R$ and $m_B$ anyons. Notably, the fusion multiplicities of anyons in $D(D_4)$ are preserved under this identification of anyons with irreps. Moreover, there is a freedom in assigning irreps while preserving the fusion products, since the pairs of irreps $\{9,10\}$, $\{11,12\}$ and $\{13,14\}$ have the same self-fusion multiplicities. In particular, after identifying $m_R$ and $m_B$ using Eqs.~\eqref{eq:rho_11_def} and \eqref{eq:rho_13_def}, the only freedom is given by the pair $\{\chi_9 \rightarrow m_{RB}, \chi_{10} \rightarrow f_{RB} \}$. As a result, we can identify the local order parameter corresponding to the self-fermion $f_B$ as the 2-site quantity $\widetilde{\sigma}_b \cdot \v{n}_g^{\B}$, which correspondingly transforms in the irrep $\chi_{14}$. Analogous 2-site local order parameters can be derived for the other self-fermions, while the self-boson $m_{RB}$ has an on-site local order parameter (as is the case for $m_R$ and $m_B$).

In the absence of the local order parameters in Eqs.~\eqref{eq:rho_11_def} and \eqref{eq:rho_13_def}, the freedom in labeling irreps of $G_{m_R, m_B}$ can be seen by explicitly computing the ``fusion rules'' of irreps of the group. Using the Schur orthogonality relations $\frac{1}{|G|}\sum_{g \in G} \chi_i(g) \overline{\chi}_j(g) =  \delta_{i j}$ for irreps $\chi_i, \chi_j$ of the group $G$, we compute several of the fusion rules of irreps of $G_{m_R, m_B}$ as
\begin{equation}
\label{eq:G_mRmB_fusion_rules}
    \begin{split}
        \chi_9 \otimes \chi_9 &= \chi_{10} \otimes \chi_{10} = \chi_1 \oplus \chi_2 \oplus \chi_3 \oplus \chi_4, \qquad 
        \chi_9 \otimes \chi_{10} = \chi_5 \oplus \chi_6 \oplus \chi_7 \oplus \chi_8 \\
        \chi_{11} \otimes \chi_{11} &= \chi_{12} \otimes \chi_{12} = \chi_1 \oplus \chi_2 \oplus \chi_5 \oplus \chi_6 , \qquad
        \chi_{11} \otimes \chi_{12} = \chi_3 \oplus \chi_4 \oplus \chi_7 \oplus \chi_8 \\
        \chi_{13} \otimes \chi_{13} &= \chi_{14} \otimes \chi_{14} = \chi_1 \oplus \chi_2 \oplus \chi_7 \oplus \chi_8, \qquad
        \chi_{13} \otimes \chi_{14} = \chi_3 \oplus \chi_4 \oplus \chi_5 \oplus \chi_6.
    \end{split}
\end{equation}
Under an identification of the irreps of $G_{m_R, m_B}$ with the anyons of $D(D_4)$, Eq.~\eqref{eq:G_mRmB_fusion_rules} recovers the relevant anyon fusion rules of $D(D_4)$ (shown in Table~\ref{tab:D4_fusion_rules})~\cite{Iqbal_24}. Moreover, these fusion rules are invariant under the interchanges $\chi_{9} \leftrightarrow \chi_{10}, \chi_{11} \leftrightarrow \chi_{12}, \chi_{13} \leftrightarrow \chi_{14}$, demonstrating the freedom in identifying irreps of $G_{m_R, m_B}$ with anyons of $D(D_4)$.

\begin{table}
\centering
\renewcommand{\arraystretch}{1.3}
\begin{tabular}{c||c|c|c|c|c|c|}
$\times$ & $e_R$ & $e_{RG}$ & $m_R$ & $m_{RG}$ & $m_{RB}$ & $s_{RGB}$ \\
\hline \hline
$e_R$ & $1$ & $e_G$ & $f_R$ & $f_{RG}$ & $f_{RB}$ & $\bar{s}_{RGB}$ \\
\hline
$e_G$ & $e_{RG}$ & $e_R$ & $m_R$ & $f_{RG}$ & $m_{RB}$ & $\bar{s}_{RGB}$ \\
\hline
$e_B$ & $e_{RB}$ & $e_{RGB}$ & $m_R$ & $m_{RG}$ & $f_{RB}$ & $\bar{s}_{RGB}$ \\
\hline
$e_{RG}$ & $e_G$ & $1$ & $f_R$ & $m_{RG}$ & $f_{RB}$ & $s_{RGB}$ \\
\hline
$e_{RGB}$ & $e_{GB}$ & $e_B$ & $f_R$ & $m_{RG}$ & $m_{RB}$ & $\bar{s}_{RGB}$ \\
\hline
$m_R$ & $f_R$ & $f_R$ & $1 + e_G + e_B + e_{GB}$ & $m_G + f_G$ & $m_B + f_B$ & $m_{GB} + f_{GB}$ \\
\hline
$m_G$ & $m_G$ & $f_G$ & $m_{RG} + f_{RG}$ & $m_R + f_R$ & $s_{RGB} + \bar{s}_{RGB}$ & $m_{RB} + f_{RB}$ \\
\hline
$m_B$ & $m_B$ & $m_B$ & $m_{RB} + f_{RB}$ & $s_{RGB} + \bar{s}_{RGB}$ & $m_R + f_R$ & $m_{RG} + f_{RG}$ \\
\hline
$m_{RG}$ & $f_{RG}$ & $m_{RG}$ & $m_G + f_G$ & $1 + e_{RG} + e_{RGB} + e_B$ & $m_{GB} + f_{GB}$ & $m_B + f_B$ \\
\hline
$s_{RGB}$ & $\bar{s}_{RGB}$ & $s_{RGB}$ & $m_{GB} + f_{GB}$ & $m_B + f_B$ & $m_G + f_G$ & $1 + e_{RG} + e_{GB} + e_{RB}$ \\
\hline
\end{tabular}
\caption{Fusion rules (up to permutation of colors) of the anyons in $D(D_4)$. Table adapted from Ref.~\cite{d4_hasse}.}
\label{tab:D4_fusion_rules}
\end{table}

\begin{table}[htbp]
\centering
\renewcommand{\arraystretch}{1.2}
\setlength{\tabcolsep}{3pt}
\begin{tabular}{c c|cccc|cccccccccc}
\hline
 & & $e$ & $V_B$ & $V_R V_B$ & $V_R$
 & $T_G$ & $S_B$ & $S_B V_R$ & $S_R S_B$ & $S_R S_B V_R$ & $S_B T_G$ & $S_R S_B T_G$ & $S_R$ & $S_R V_B$ & $S_R T_G$ \\

 & & $1$ & $1$ & $1$ & $1$
 & $4$ & $2$ & $2$ & $2$ & $2$
 & $4$ & $4$ & $2$ & $2$ & $4$ \\
\hline

$\chi_1$ & $1$ & 1 & 1 & 1 & 1 & 1 & 1 & 1 & 1 & 1 & 1 & 1 & 1 & 1 & 1 \\

$\chi_2$ & $e_G$ & 1 & 1 & 1 & 1 & $-1$ & 1 & 1 & 1 & 1 & $-1$ & $-1$ & 1 & 1 & $-1$ \\

$\chi_3$ & $e_{RB}$ & 1 & 1 & 1 & 1 & 1 & $-1$ & $-1$ & 1 & 1 & $-1$ & 1 & $-1$ & $-1$ & $-1$ \\

$\chi_4$ & $e_{RGB}$ & 1 & 1 & 1 & 1 & $-1$ & $-1$ & $-1$ & 1 & 1 & 1 & $-1$ & $-1$ & $-1$ & 1 \\

$\chi_5$ & $e_B$ & 1 & 1 & 1 & 1 & 1 & 1 & 1 & $-1$ & $-1$ & 1 & $-1$ & $-1$ & $-1$ & $-1$ \\

$\chi_6$ & $e_{BG}$ & 1 & 1 & 1 & 1 & $-1$ & 1 & 1 & $-1$ & $-1$ & $-1$ & 1 & $-1$ & $-1$ & 1 \\

$\chi_7$ & $e_R$ & 1 & 1 & 1 & 1 & 1 & $-1$ & $-1$ & $-1$ & $-1$ & $-1$ & $-1$ & 1 & 1 & 1 \\

$\chi_8$ & $e_{RG}$ & 1 & 1 & 1 & 1 & $-1$ & $-1$ & $-1$ & $-1$ & $-1$ & 1 & 1 & 1 & 1 & $-1$ \\

\hline

$\chi_9$ & $m_{RB}$ & 2 & $-2$ & 2 & $-2$ & 0 & 0 & 0 & 2 & $-2$ & 0 & 0 & 0 & 0 & 0 \\

$\chi_{10}$ & $f_{RB}$ & 2 & $-2$ & 2 & $-2$ & 0 & 0 & 0 & $-2$ & 2 & 0 & 0 & 0 & 0 & 0 \\

$\chi_{11}$ & $m_R$ & 2 & 2 & $-2$ & $-2$ & 0 & 2 & $-2$ & 0 & 0 & 0 & 0 & 0 & 0 & 0 \\

$\chi_{12}$ & $f_R$ & 2 & 2 & $-2$ & $-2$ & 0 & $-2$ & 2 & 0 & 0 & 0 & 0 & 0 & 0 & 0 \\

$\chi_{13}$ & $m_B$ & 2 & $-2$ & $-2$ & 2 & 0 & 0 & 0 & 0 & 0 & 0 & 0 & 2 & $-2$ & 0 \\

$\chi_{14}$ & $f_B$ & 2 & $-2$ & $-2$ & 2 & 0 & 0 & 0 & 0 & 0 & 0 & 0 & $-2$ & 2 & 0 \\

\end{tabular}
\caption{Character table of $G_{m_R, m_B} \cong \Z_2^3 \rtimes \Z_2^2$. The first header row lists a representative element from each conjugacy class, and the second header row lists conjugacy class sizes. The second column provides an identification of $\mathrm{Irr}(G_{m_R, m_B})$ with the anyons generated by fusion descendants of $m_R, m_B$. The first four classes comprise the center $Z(G_{m_R,m_B}) = \langle V_R \rangle \times \langle V_B \rangle$; each non-center coset of $Z(G_{m_R,m_B})$ that contains $T_G$ forms a single class of size 4, while the remaining cosets each split into two classes of size 2.}
\label{tab:char_tab_GRB}
\end{table}

The group $G_{m_R, m_B}$ is class-2 nilpotent since there exist finite Abelian groups $N, Q$ such that the extension $1 \longrightarrow N \longrightarrow G_{m_R, m_B} \longrightarrow Q \longrightarrow 1$ is central. Here $N = \langle V_R \rangle \times \langle V_B \rangle \cong \Z_2^2$ and $Q = G_{m_R,m_B}/N \cong \Z_2^3$. For reference, $G_{m_R, m_B}$ is indexed by the GAP~\cite{GAP4} ID $\mathrm{SmallGroup}(32, 27)$.

\section{Effective theory at the transition}

\subsection{Consistent quantum field theory}
\label{sec:field_theory}

We propose a consistent field theory that captures the phase transitions of interest. This theory consists of 2 Luttinger liquids with fields $\theta^\alpha,\phi^\alpha$ for $\alpha=\R,\B$ coupled to an Ising CFT with order parameter field $\phi_G$ and mass field $\epsilon_G$. Each Luttinger Liquid, which independently realize a disordered phase terminating on a BKT point, is associated with one of the two 2-dimensional order parameters, and the Ising CFT captures the 1-dimensional order parameter detecting condensation of $e_G$ charges.

The proposed field theory does not capture transitions in which $\theta_b^{\R}$ or $\theta_r^{\B}$ order, since we numerically do not observe such phases. As a result, to leading order we propose $\theta_g^\alpha \approx \theta^\alpha$ where $\theta_g^\alpha$ is a discrete 4-state rotor and $\theta^\alpha$ is the coarse-grained $U(1)$ rotor. The Hamiltonian density is given by:
\begin{equation}
\label{eq:ham_density}
\begin{split}
    \mathcal{H} &= \frac{1}{2 \pi}\sum_{\alpha = \R, \B} \bigg[ K_\alpha (\partial_x \theta^\alpha)^2 + \frac{1}{4 K_\alpha} (\partial_x \varphi^\alpha)^2  +  J_\alpha \cos(4 \theta^\alpha) + M_\alpha \cos(\varphi^\alpha) \bigg]  +  H_{\text{Ising}}[\phi_G, \epsilon_G] \\
    &+ \sum_{\alpha = \R, \B} g_\alpha  \phi_G \sin(2 \theta^\alpha) + \underbrace{\lambda \bigg[\cos(2 \theta^\R - 2 \theta^\B) - \cos(2 \theta^\R + 2 \theta^\B) \bigg]}_{-J(\v{n}^{\R}_g\cdot \v{n}^{\B}_g)^2} \\
    &
\end{split}
\end{equation}
The fields $\theta^\alpha, \varphi^\alpha \in [0, 2 \pi)$ are compact fields satisfying the commutation $[\partial_x \theta^\alpha(x), \varphi^\beta(y)] = i 2 \pi \delta_{\alpha \beta} \delta(x - y)$. In this normalization convention, the scaling dimensions of vertex operators are $\Delta[\cos(m \theta^\alpha)] = m^2 / (4 K_\alpha)$ and  $\Delta[\cos(n \varphi^\alpha)] = n^2 K_\alpha$. The coupling $\phi_G \sin(2 \theta^\alpha)$ can be understood as a result of the energetic constraint on the $\G$ sublattice, where $e_G$ anyons can appear to due their presence in the fusion channels $m_R \times m_R$ and $m_B \times m_B$. In fact, integrating out the $\phi_G$ field leads to the energetic coupling in the last term.

The symmetry generators of $G_{m_R, m_B}$ act on the fields as
\begin{equation}
        V_R : \begin{cases}
            \theta^\R \rightarrow \pi + \theta^\R
        \end{cases}, \
        V_B : \begin{cases}
            \theta^\B \rightarrow \pi + \theta^\B
        \end{cases}, \ 
        T_G: \begin{cases}
            \phi &\rightarrow -\phi \\
            \theta^\alpha &\rightarrow - \theta^\alpha \\
            \varphi^\alpha &\rightarrow - \varphi^\alpha
        \end{cases}, \ 
        S_R : \begin{cases}
            \theta^\R &\rightarrow \frac{\pi}{2} - \theta^\R \\
            \varphi^\R &\rightarrow - \varphi^\R
        \end{cases}, \ 
        S_B : \begin{cases}
            \theta^\B &\rightarrow \frac{\pi}{2} - \theta^\B \\
            \varphi^\B &\rightarrow - \varphi^\B
        \end{cases}
\end{equation}

From this field theory, we can then predict: i) An Ising transition where $\phi_G$ acquires a finite value by tuning $\beta^z_\G$ strength, corresponding to the condensation of $e_G$. ii) When this happens, and according to the value of $K_\alpha$, one or both $\theta^\alpha$ fields can order, corresponding to the condensation of the non-Abelian fluxes $m_\alpha$ (see Sec.~\ref{sec:emergence_TC}). Moreover, as we discussed in the main text, there is a third alternative route, even when $\beta^z_\G=0$, leading to a $G_{m_r,m_B} \rightarrow \Z_2^2$ transition that is driven by the relevant $\lambda$-coupling. Finally, for $\beta^z_{\G}=0$, we expect the boundaries of phase diagram with constant $t_\R=1$ for small $t_\B$ (and analogously for $t_\B=1$ and small $t_\R$)  to remain critical (see Sec.~\ref{sec:gapless_edge}). 

Notably, the rotors $\theta_b^{\R}, \theta_r^{\B}$ do not order despite the $U(1)$ rotors $\theta^{\R}, \theta^{\B}$ ordering. This follows from exact relations between the discrete rotors on the $b,r$ and $g$ sublattices. From the derivation in App.~\ref{sec:high_T_expansion}, $\cos(\theta_g^{\R}) = \sigma_g^{\R}/\sqrt{2}$ and $\sin(\theta_g^{\R}) = \sigma_g^{\R} \widetilde{\sigma}_g/\sqrt{2}$, while $\cos(\theta_b^{\R}) = \frac{1}{2}\sigma_b^{\R} (1 + \widetilde{\sigma}_b)$ and $\sin(\theta_b^{\R}) = \frac{1}{2}\sigma_b^{\R} (1 - \widetilde{\sigma}_b)$. These yield $e^{i \theta_g^{\R}} = \sigma_g^{\R}\, e^{i \pi \widetilde{\sigma}_g / 4}$ and $e^{i \theta_b^{\R}} = \sigma_b^{\R}\, e^{i \pi (1 - \widetilde{\sigma}_b) / 4}$, so that $e^{i (\theta_b^{\R} - \theta_g^{\R})} = \sigma_b^{\R} \sigma_g^{\R}\, e^{i
\pi/4}\, e^{-i \pi (\widetilde{\sigma}_g + \widetilde{\sigma}_b)/4}$. From this we extract the exact microscopic identities
\begin{equation}
\label{eq:theta_B_sublattice}
    \begin{split}
        \theta_b^{\R} &= \theta_g^{\R} + \frac{\pi}{2} (1 - \sigma_b^{\R} \sigma_g^{\R}) + \frac{\pi}{4} (1 - \widetilde{\sigma}_b - \widetilde{\sigma}_g).
    \end{split}
\end{equation}
We emphasize that Eq.~\eqref{eq:theta_B_sublattice} is an exact lattice identity: the variables $\sigma_g^{\R}, \widetilde{\sigma}_g$ on the right-hand side are not independent of $\theta_g^{\R}$, but parametrize it through the relations above. For this reason, the constraints among $\theta_g^{\R}, \sigma_g^{\R}, \widetilde{\sigma}_g$ reduce $\theta_b^{\R}$ to the expected $\Z_4$ rotor.

Equation~\eqref{eq:theta_B_sublattice} makes the fate of $\theta_b^{\R}$ in the ordered phase transparent. In the symmetry broken phase of the main text, $\theta_g^{\R}$ develops long-range order, whose smooth, long-wavelength component is the coarse-grained field $\theta^{\R}$; the Ising variables $\sigma_g^{\R}$ and $\widetilde{\sigma}_g$ that parametrize it order as well. The variable $\sigma_b^{\R}$, odd under the broken generator $V_R$, likewise acquires an expectation value, whereas $\widetilde{\sigma}_b$ is charged under the unbroken generator $S_R$ and remains disordered. Consequently, following an analogous derivation for the $\B$ sublattice rotors, the only fluctuating quantity on the right-hand side of Eq.~\eqref{eq:theta_B_sublattice} is $\widetilde{\sigma}_b$, so
\begin{equation}
    \begin{split}
        \theta_b^{\R} \rightarrow \mathrm{const} + \frac{\pi}{4} (1 - \widetilde{\sigma}_b), \qquad
        \theta_r^{\B} \rightarrow \mathrm{const} + \frac{\pi}{4} (1 - \widetilde{\sigma}_r),
    \end{split}
\end{equation}
where the constant collects the ordered contributions. The rotors $\theta_b^{\R}, \theta_r^{\B}$ thus reduce to Ising-like variables that remain disordered in the $\Z_2^2$-symmetric phase, justifying their absence from the field theory in Eq.~\eqref{eq:ham_density}.

\subsection{Symmetry-broken phase in $\beta_\R, \beta_\B \rightarrow \infty$ limit}
\label{sec:sb_phase}

\begin{figure}
    \centering
    \includegraphics[width=0.4\linewidth]{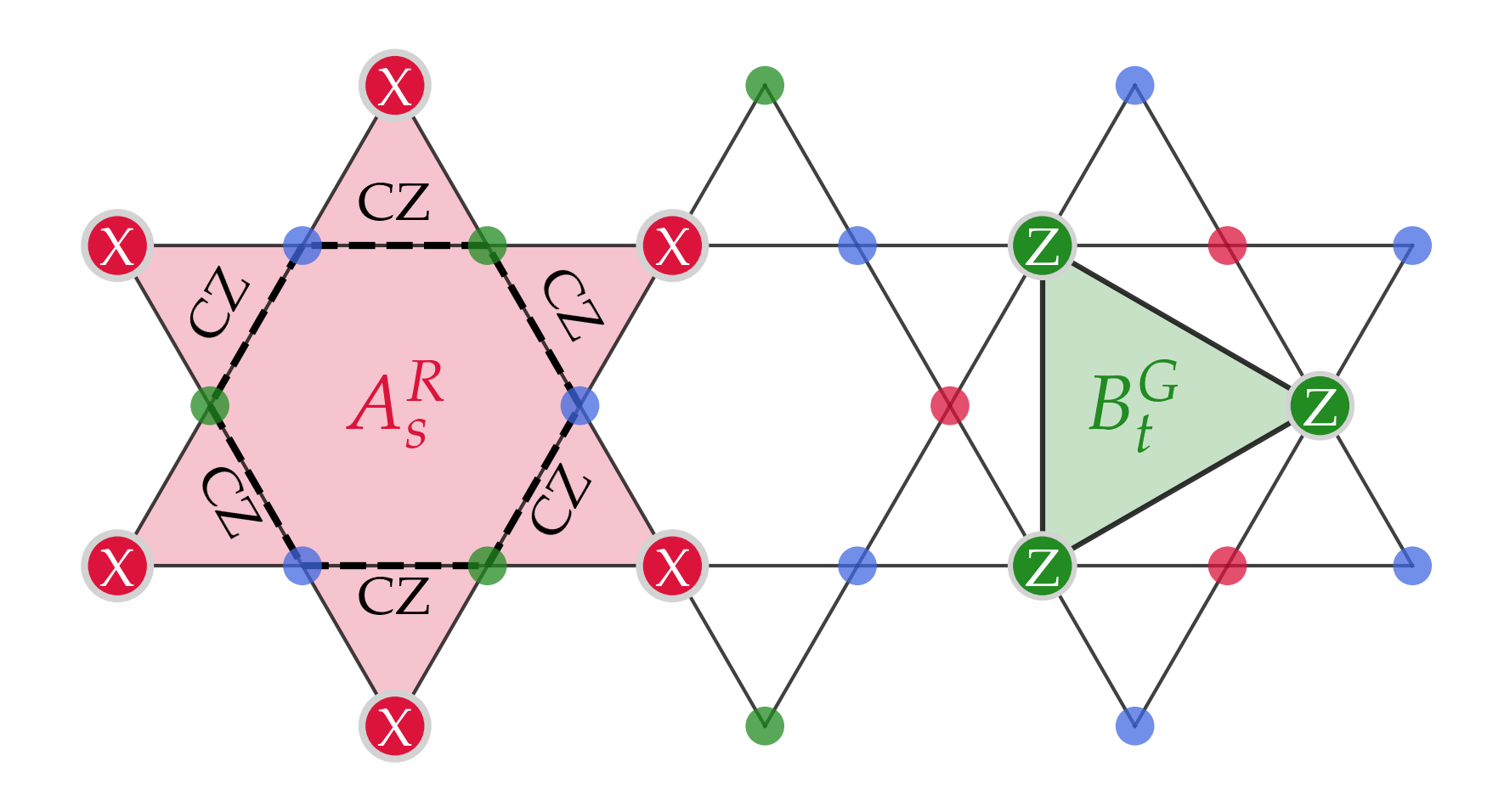}
    \caption{\textbf{Kagome lattice $D_4$ Hamiltonian terms}. 12-body star operator $A_s^{\R}$ and 3-body triangle operator $B_t^{\G}$ in the kagome lattice realization of $D_4$ given in Eq.~\eqref{eq:D4_hamiltonian}. Dotted lines denote $\CZ_{ij}$ gates.}
    \label{fig:kagome_stabilizer}
\end{figure}

In the kagome-lattice model of $D_4$ topological order presented in Ref.~\cite{Yoshida_2016, Iqbal_24}, the microscopic Hamiltonian is given by
\begin{equation}
\label{eq:D4_hamiltonian}
    H = - \sum_{\alpha \in \{\R, \G, \B\}} \bigg[ \sum_{s \in \{\davidsstar \}} A_{s}^{\alpha} - \sum_{t \in \{\triangle\}} B_t^{\alpha} \bigg] .
\end{equation}
Here the 12-body \emph{star} operators are $A_{\davidsstar}^\alpha = \prod_{i_\mathrm{in} = 1}^6 \CZ_{i_\mathrm{in}, i_\mathrm{in} + 1} \prod_{i_\mathrm{out}}^6 X^\alpha_{i_\mathrm{out}}$ such that the sites $i_\mathrm{in}, i_{\mathrm{in} + 1}, i_\mathrm{out}$ all live on different color sublattices (see $A_{\davidsstar}^{\R}$ in Fig.~\ref{fig:kagome_stabilizer}). The 3-body \emph{triangle} operators are $B_t^\alpha = \prod_{j \in t} Z_j^\alpha$ where the sites $j \in t$ are nearest-neighbor sites on sublattice $\alpha$ (see $B_t^{\G}$ in Fig.~\ref{fig:kagome_stabilizer}). Ground states of this Hamiltonian are simultaneous $+1$ eigenstates of all $A_{\davidsstar}^\alpha, B_t^\alpha $ operators.

Taking $\beta_\R =\beta_\B = \infty$ in the deformed state $\ket{\beta_\R, \beta_\B} = e^{\frac{\beta_\R}{2} \sum_{r \in {\R}} X_r^{\R}} e^{\frac{\beta_\B}{2} \sum_{b \in {\B}} X_b^{\B}} \ket{D_4}$ is equivalent to enacting the projection $X_r^{\R} = +1, X_b^{\B} = +1$ on all $r, b$ sites on the ground state $\ket{D_4}$. Under this projection, the constraint $B_g^{\G} = +1$ remains satisfied (since the $\G$ sublattice is untouched). However, $B_r^{\R}, B_b^{\B} \rightarrow 0$ since $\frac{1}{2}(1+X_j^{\alpha})Z_j^{\alpha}\frac{1}{2}(1+X_j^{\alpha})=0$. 

We can understand the orbit of $A_{\davidsstar}^{\R}$ under this projection by studying the action of a single projector $\frac{1}{2}(1 + X_b^{\B})$ on the 3-body term $\CZ_{b g} \CZ_{b g'}$:
\begin{equation}
    \begin{split}
        \frac{1}{2}(1 + X_b^{\B}) \CZ_{b g} \CZ_{b g'} \frac{1}{2} (1 + X_b^{\B}) &= \frac{1}{4} (1 + X_b^{\B}) \bigg(\CZ_{b g} \CZ_{b g'} + X_b^{\B} \frac{1}{4}[1 +  Z_g^{\G} - Z_b^{\B} + Z_b^{\B} Z_g^{\G}] [1 + Z_{g'}^{\G} - Z_b^{\B} + Z_b^{\B} Z_{g'}^{\G} ]\bigg) \\
        &= \frac{1}{2} \bigg( \frac{1}{4}[2 + 2 Z_{g}^{\G} Z_{g '}^{\G}] + \frac{1}{4}[2 + 2 Z_{g}^{\G} Z_{g '}^{\G}] \bigg) = \frac{1}{2} (1 + Z_g^{\G} Z_{g'}^{\G}).
    \end{split}
\end{equation}
where we have used the identity $\CZ_{i, j} = \frac{1}{2}(1 + Z_i + Z_j - Z_i Z_j)$. Identical algebra can be used to show that $\frac{1}{2}(1 + X_r^{\R}) \CZ_{r g} \CZ_{r g'} \frac{1}{2} (1 + X_r^{\R}) = \frac{1}{2} (1 + Z_g^{\G} Z_{g'}^{\G})$. Hence we find the orbits $A_{\davidsstar}^{\R} \rightarrow \prod_{\langle \langle g, g' \rangle \rangle \in \davidsstar} \frac{1}{2} (1 + Z_g^{\G} Z_{g'}^{\G}) = \frac{1}{4} (1 + \sum_{\langle \langle g, g' \rangle \rangle \in \davidsstar} Z_g^{\G} Z_{g'}^{\G} )$. Likewise, $A_{\davidsstar}^{\B} \rightarrow \prod_{\langle \langle g, g' \rangle \rangle \in \davidsstar} \frac{1}{2} (1 + Z_g^{\G} Z_{g'}^{\G}) = \frac{1}{4} (1 + \sum_{\langle \langle g, g' \rangle \rangle \in \davidsstar} Z_g^{\G} Z_{g'}^{\G})$. Hence, the constraints $A_{\davidsstar}^{\B}=+1$ and $A_{\davidsstar}^{\R}=+1$ become $Z_g^{\G} Z_{g'}^{\G}$ on every pair of next-nearest neighbors $g, g'$ sites on the green sublattice $\G$ of the kagome lattice.

Putting these last constraints together with $B_t^{\G}=+1$ on every green triangle, we find that $Z_g^{\G}=+1$ for all $g\in \G$.

 Finally, the orbit of $A_{\davidsstar}^{\G}$ can be understood by sequentially applying the commuting projections $X_r^{\R} = +1$ and $X_b^{\B} = +1$:
\begin{equation}
    \begin{split}
        \frac{1}{2}(1 + X_b^{\B}) \CZ_{r b} \CZ_{r' b} \frac{1}{2} (1 + X_b^{\B}) &= \frac{1}{4} (1 + X_b^{\B}) \bigg(\CZ_{r b} \CZ_{r' b} + X_b^{\B} \frac{1}{4}[1 +  Z_r^{\R} - Z_b^{\B} + Z_b^{\B} Z_r^{\R}] [1 + Z_{r'}^{\R} - Z_b^{\B} + Z_b^{\B} Z_{r'}^{\R} ]\bigg) \\
        &= \frac{1}{2} \bigg( \frac{1}{4}[2 + 2 Z_{r}^{\R} Z_{r '}^{\R}] + \frac{1}{4}[2 + 2 Z_{r}^{\R} Z_{r '}^{\R}] \bigg) = \frac{1}{2} (1 + Z_r^{\R} Z_{r'}^{\R}).
    \end{split}
\end{equation}
Thus under the projection $X_b^{\B} = +1$, $A_{\davidsstar}^{\G} \rightarrow \prod_{v \in \davidsstar} X_v^\G \frac{1}{4} (1 + \sum_{\langle \langle  r r' \rangle \rangle \in \davidsstar} Z_r^{\R} Z_{r'}^{\R})$. Moreover, applying now the projection $X_r^{\R} = +1$, implies that $Z_r^{\R} Z_{r'}^{\R} \rightarrow 0$, and hence $A_{\davidsstar}^{\G} \rightarrow \prod_{v \in \davidsstar} X_v^\G$. Imposing now that $Z_g^{\G}=+1$ for all $g\in \G$, this constraint then becomes trivial $A_{\davidsstar}^{\G} \rightarrow 0$.

All together, we find that the (projected) ground state is specified by the constraints 
\begin{equation}
     Z_g^{\G}= +1,\,\,\,\,\,\,\,\, X_r^{\R} = +1, \,\,\,\,\,\,\,\,X_b^{\B} = +1,
\end{equation}
on all sites $r, g, b$ on $\R, \G$ and $\B$ sublattices respectively. The state satisfying these constraints is given by $\ket{\beta_\R = \infty, \beta_\B = \infty} = \ket{+^\R} \ket{+^\B} \ket{\uparrow^\G}$, which explicitly breaks the $A_g^{\G} = +1$ symmetry of the ground state.

\section{Monte Carlo methods}

\subsection{Monte Carlo simulations}
\label{sec:monte_carlo_sims}

\begin{figure}
    \centering
    \includegraphics[width=\linewidth]{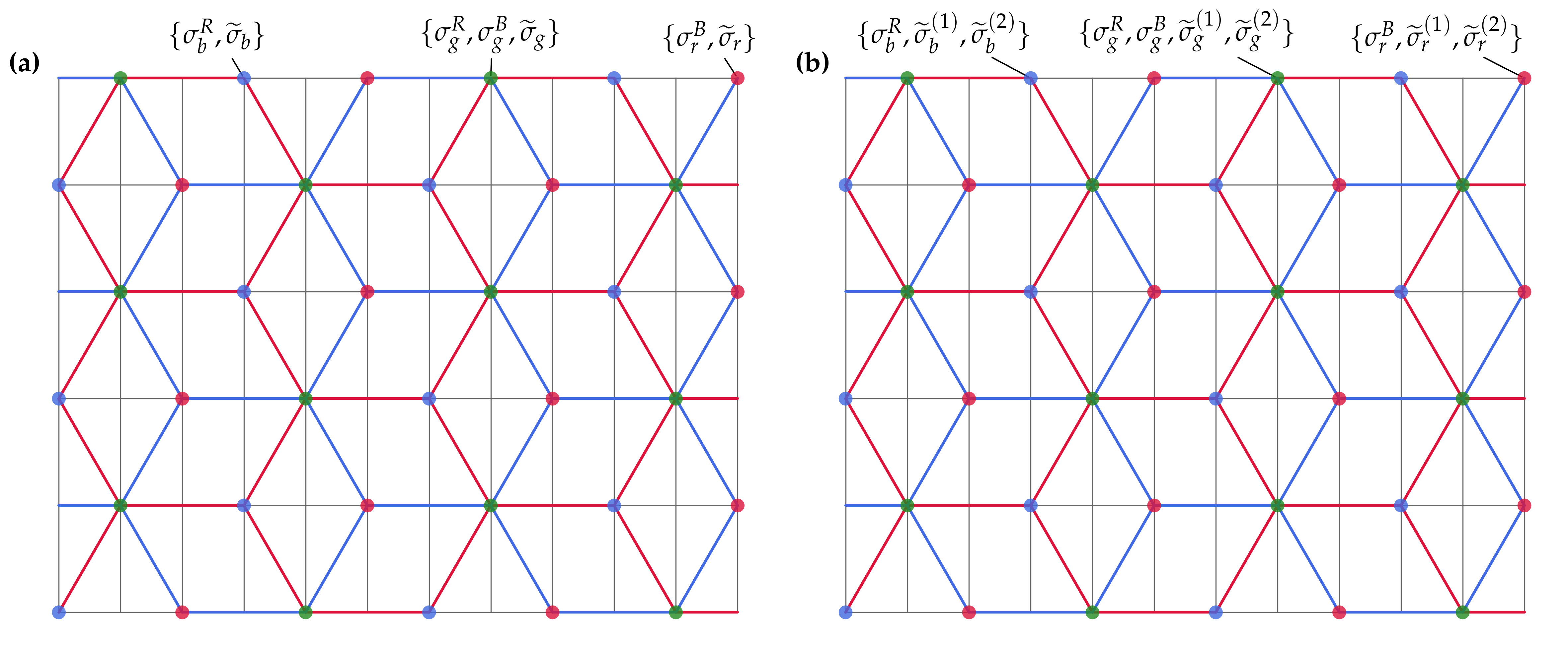}
    \caption{\textbf{Monte Carlo lattice}. Rectangular lattice tiling for $N_x = 12, N_y = 6$ with periodic boundary conditions for both honeycomb layers. Annotations indicate the Ising spins residing on each sublattice. (a) shows lattice used to simulate pure wavefunction deformation. (b) shows lattice used to simulate purity.}
    \label{fig:mcmc_lattice}
\end{figure}

This section discusses in detail the Monte Carlo simulation procedure used to sample the local stat-mech models for wavefunction deformation and the density matrix purity. To simplify the single-site Monte Carlo updates, we tile the stacked honeycomb lattices with a rectangular lattice (see Fig.~\ref{fig:mcmc_lattice}). We take system sizes of the form $(N_x, N_y) = (6N, 3N)$ to ensure periodic boundary conditions can be enforced on both sublattices. Since there is not a bijection between rectangular lattice sites and honeycomb lattice sites, we only perform updates on \textit{physical} honeycomb lattice sites (highlighted in Fig.~\ref{fig:mcmc_lattice}).

For the case of pure wavefunction deformation, we consider a 3-layer stack of the honeycomb lattices in Fig.~\ref{fig:mcmc_lattice}(a): $\sigma^{\R}$ spins lie on the first layer (the blue and green sites), $\sigma^{\B}$ spins lie on the second layer (the red and green sites), and $\widetilde{\sigma}$ spins lie on the third layer (the red, blue, and green sites). We run $10^6$ equilibration steps of Metropolis and average observables over $10^7$ Metropolis steps. 

When sampling the purity of the decohered density matrix (see Eq.~\eqref{eq:purity_ham}), we consider a 4-layer stack of the honeycomb lattices in Fig.~\ref{fig:mcmc_lattice}(b): the first two layers hosting $\{\sigma^{\R}, \sigma^{\B} \}$ and the last two hosting $\{\widetilde{\sigma}^{(1)}, \widetilde{\sigma}^{(2)} \}$. We run $5 * 10^7$ equilibration steps of Metropolis and average observables over $5 * 10^8$ Metropolis steps. Note that obtaining converged averages of observables when sampling purity requires more equilibration steps than when sampling the norm of the deformed wavefunction, as the stat-mech possesses 4 Ising spins per site and 6-spin interactions, while the Hamiltonian of the deformed wavefunction norm possesses 3 Ising spins per site and 4-spin interactions.

\subsection{Derivation of anyon condensation observables}
\label{sec:wilson_line_observables}

The form of $W_a(x, y)$ in the classical stat-mech model can be easily obtained when the deformations characterizing $\ket{\beta_\B, \beta_\R}$ commute with the operator $\hat{W}_a(x, y)$. For instance, consider the Wilson operator $\hat{W}_{e_G} (g, g') = \prod_{g'' \in \gamma_{v, v'}} Z_{g''}^{\G}$ that creates a pair of $e_G$ Abelian charges at center of stars $g$ and $g'$. Here $\gamma_{v, v'}$ is any string with support only on the $\G$ sublattice with open ends $v, v'$ lying on stars $g, g'$ respectively. One can ungauge the local symmetry $\prod_{g \in \green{\triangle}} Z_g^{\G} = +1$ (as e.g., in Ref.~\cite{sala_D4}), obtaining 
\begin{equation}
    W_{e_G}(g, g') =\frac{\langle \beta_\B, \beta_\R | \hat{W}_{e_G}(g, g') | \beta_\B , \beta_\R \rangle}{\langle \beta_\B, \beta_\R  | \beta_\B , \beta_\R \rangle}= \langle \widetilde{\sigma}_g \widetilde{\sigma}_{g'} \rangle,
\end{equation}
where the latter corresponds to a thermal average on the local stat-mech model. Hence, the relevant Monte-Carlo observable becomes $W_{e_G} = \frac{1}{|\G|^2} \sum_{(g, g') \in \G} W_{e_G}(g, g')$. The observable $W_{e_G}$ detects condensation of the anyon $e_G$. It can also be understood as a symmetry-invariant corresponding to the charged order parameter $\widetilde{\sigma}_g$, which detects breaking of the $\Z_2$ symmetry $T_G$.

When the Wilson operator $\hat{W}_a(x, y)$ and the wavefunction deformations do not commute, computation of $W_a(x, y)$ becomes significantly more challenging. In particular, this non-commutation issue arises in the computation of the expectation values of the Wilson lines creating $m_R$ and $m_B$ anyons. The Wilson line creating $m_R$ at triangles $b, g$ is defined as
\begin{equation}
    \hat{W}_{m_R}(b, g) = \prod_{r \in \gamma_{v, v'}} X_r^{\R} \prod_{b' < g' \in \gamma_{v, v'}} \CZ_{b' g'}
\end{equation}
where $\gamma_{v, v'}$ is an open string connecting sites $v, v'$ lying on triangles $b, g$ respectively. Here $b' < g' \in \gamma_{v, v'}$ means select an orientation of $\gamma_{v, v'}$ and consider all pairs $(b', g')$ where $b'$ is to the left of $g'$. The Wilson line $\hat{W}_{m_B}(r, g)$ creating $m_B$ non-Abelian anyons at triangles $r, g$ is defined analogously (with a sublattice permutation). The \emph{$\CZ$-dressing} on $\hat{W}_{m_\alpha}(t_i, t_f)$ does not commute with the Pauli $X$ deformations in $\ket{\beta_\B, \beta_\R}$, so exploiting the mapping to a $(\mathbb{Z}_2)^3$ SPT~\cite{Yoshida_2016} is not an option. As a result, we consider alternative 2-point functions whose spatial averages can also distinguish between the deconfined and condensed phases for the non-Abelian anyons $m_R, m_B$. We propose the fidelity correlator 
\begin{equation}
\label{eq:fid_corr}
    W_a(x, y) = \frac{\bra{\beta_\B, \beta_\R} e^{\frac{\beta_\R}{2} \sum_{r \in \mathcal{\R}} X_r^{\R}} e^{\frac{\beta_\B}{2} \sum_{b \in \mathcal{\B}} X_b^{\B}} \hat{W}_a(x, y) \ket{D_4}}{\langle\beta_\B, \beta_\R|\beta_\B, \beta_\R \rangle}.
\end{equation}
 Evidently when $\hat{W}_a(x, y)$ and the deformations commute, the fidelity correlator reduces to the previously studied 2-point function (as it was the case for $W_{e_G}$. One can understand this correlator as a measure of the fidelity between a deformed ground state of $D_4$ and a deformed excited state including two anyonic quasi-particles on top. When the deformed state $\ket{\beta_\B, \beta_\R}$ lies in the $D_4$ phase, the quantity $W_a(x, y) \rightarrow 0$ as $|x - y| \rightarrow \infty$, as the deformed states $\ket{\beta_\B, \beta_\R}$ and $e^{\frac{\beta_\R}{2} \sum_{r \in \mathcal{\R}} X_r^{\R}} e^{\frac{\beta_\B}{2} \sum_{b \in \mathcal{\B}} X_b^{\B}} \hat{W}_a(x, y) \ket{D_4}$ remain orthogonal. However, when the deformed state enters an \emph{a-condensed} phase, the correlator $W_a(x, y) \rightarrow \mathrm{constant}$ as $|x - y| \rightarrow \infty$, since the action of $\hat{W}_a(x, y)$ becomes trivial in the IR limit after $a$-condensation. Thus the fidelity correlators $W_a(x, y)$ are capable of distinguishing between deconfined and condensed phases with respect to an anyon $a$. We now provide the explicit expression of this quantity for $a = m_R$, as the computation for $a = m_B$ is analogous.

We expand the fidelity correlation function as 
\begin{equation}
\label{eq:W_mR_eval_1}
    \begin{split}
    W_{m_R}(b, g) &=
       \bra{\beta_\B, \beta_\R} e^{\frac{\beta_\R}{2} \sum_{r \in \mathcal{\R}} X_r^{\R}} e^{\frac{\beta_\B}{2} \sum_{b \in \mathcal{\B}} X_b^{\B}} \hat{W}_{m_R}(b, g) \ket{D_4} \\
        &= \cosh(\beta_\R)^N \cosh(\beta_\B)^N \bra{D_4} \bigg(\prod_{j \in \R} [1 + \tanh(\beta_\R) X_j^{\R}] \prod_{i \in \B} [1 + \tanh(\beta_\B) X_i^{\B}]\bigg) \hat{W}_{m_R}(b, g) \ket{D_4} \\
        &\propto \sum_{L_{\B}, L_{\R}^{b, g}} t_\R^{|L_{\R}^{b, g}|} t_\B^{|L_{\B}|} \langle D_4 | \prod_{j \in L^{b, g}_\R \cup \gamma_{x, y}} X_j^{\R} \prod_{i \in L_{\B}} X_i^{\B} \prod_{ b' < g'  \in \gamma_{b, g}} \CZ_{b' g'} |D_4 \rangle
    \end{split}
\end{equation}
where $\gamma_{x, y}$ is the string along which $\hat{W}_{m_R}(b, g)$ creates a pair of $m_R$ anyons, and $L_{\R}^{b, g}$ is the union of a closed loop configuration of $X$ operators and a string of $X$ operators creating a pair of $m_R$ anyons at triangles $b, g$. Since $L_{\R}^{b, g} \cup \gamma_{x, y}$ form closed loops on the $\R$ sublattice and $L_\B$ forms closed loops on the $\B$ sublattice, we can utilize the $\ket{D_4} \rightarrow \ket{\mathrm{SPT}}$ mapping in Ref.~\cite{Yoshida_2016}. Through this mapping, we find that
\begin{equation}
\label{eq:W_mR_topo_weight}
    \langle D_4 | \prod_{j \in L^{b, g}_\R \cup \gamma_{x, y}} X_j^{\R} \prod_{i \in L_{\B}} X_i^{\B} \prod_{ b' < g'  \in \gamma_{x, y}} \CZ_{b' g'} |D_4 \rangle = \langle + | \bigg( \prod_{\langle b' g' \rangle \in L_{\R}^{b, g}} \CZ_{b' g' } \prod_{\langle r' g' \rangle \in L_{\B}} \CZ_{r' g'} \bigg) \CZ_{b g} |+ \rangle
\end{equation}
where the sites $b', g'$ are defined at the centers of $m_R$ triangles (according to the convention in Fig.~\ref{fig:composite_figure}(a)). By writing this expectation value as a tensor network contraction (as in Ref.~\cite{sala_D4}), we can evaluate the entire quantity in Eq.~\eqref{eq:W_mR_eval_1} as
\begin{equation}
    W_{m_R}(b, g) = \frac{1}{\sqrt{2}}\sum_{L_{\B}, L_{\R}^{b, g}} \bigg( \frac{t_\B}{\sqrt{2}}\bigg)^{|L_{\B}|} \bigg( \frac{t_\R}{\sqrt{2}} \bigg)^{|L_{\R}^{b, g}|} 2^{C_{L_{\B} \cup \text{close}(L_{\R}^{b, g})}}.
\end{equation}
Here the extra factor of $1/\sqrt{2}$ comes from the definition of $\hat{W}_{m_R}(b, g)$ and $\text{close}(L_{\R}^{b, g})$ indicates a loop configuration made from closing the open-string component of $L_{\R}^{b, g}$. Note the particular choice of closure $\mathrm{close}(L_{\R}^{b, g})$ is irrelevant, as different choices correspond to string deformations $\gamma_{x, y} \rightarrow \gamma'_{x, y}$ that preserve the endpoints $x, y$. The wavefunction $\hat{W}_{m_R}(b, g) \ket{D_4}$ is invariant under these deformations, so they cannot alter $W_{m_R}(b, g)$. Equivalently, all choices of $\mathrm{close}(L_{\R}^{b, g})$ are summed over in $W_{m_R}(b, g)$, so changing $\mathrm{close}(L_{\R}^{b, g})$ amounts to permuting elements of the sum.

Now we will show that $\langle \sigma_b^{\R} \sigma_g^{\R} \CZ_{b g} \rangle \propto  W_{m_R}(b, g)$ via an exact high-temperature expansion: 
\begin{equation}
    \begin{split}
        \langle \sigma_b^{\R} \sigma_g^{\R} \CZ_{b g} \rangle &= \frac{1}{Z(\beta_\R, \beta_\B)} \sum_{\{\sigma^{\R}, \sigma^{\B}, \tilde{\sigma}\}} \bigg(\sigma_b^{\R} \sigma_{g}^{\R} \CZ_{b g} \bigg) e^{-H[\beta_\R, \beta_\B]} \\
        &\propto \sum_{\{\sigma^{\R}, \sigma^{\B}, \tilde{\sigma}\}} \bigg(\sigma_b^{\R} \sigma_{g}^{\R} \CZ_{b g} \bigg) \prod_{\langle b', g' \rangle \in \red{\hexagon}}\bigg( 1 + \sigma^{\R}_{b'} \sigma^{\R}_{g'} \CZ_{b' g'} \tanh(\beta_\R) \bigg) \prod_{\langle r', g' \rangle \in \blue{\hexagon}}\bigg( 1 + \sigma^{\B}_{r'} \sigma^{\B}_{g'} \CZ_{r' g'} \tanh(\beta_\B) \bigg) \\
        &= \sum_{L_\B} t_\B^{|L_\B|} \sum_{\{\sigma^{\R}, \tilde{\sigma} \}} \bigg(\sigma^{\R}_{b} \sigma^{\R}_{g} \CZ_{b g} \bigg) \prod_{\langle b', g'  \rangle \in \red{\hexagon}}\bigg( 1 + \sigma^{\R}_{b'} \sigma^{\R}_{g'} \CZ_{b' g'} \tanh(\beta_\R) \bigg) \prod_{\langle r', g' \rangle \in L_\B} \CZ_{r' g'} \\
        &= \sum_{L_\B, L_{\R}^{b, g}} t_\R^{|L_{\R}^{b, g}|} t_\B^{|L_{\B}|} \sum_{\{\tilde{\sigma}\}} \bigg(\prod_{\langle b', g' \rangle \in L_{\R}^{b, g}} \CZ_{b' g'} \prod_{\langle r', g' \rangle \in L_{\B}} \CZ_{r' g'} \bigg) \CZ_{b g}
    \end{split}
\end{equation}
where the last factor $ \sum_{\{\tilde{\sigma}\}} \bigg(\prod_{\langle b', g' \rangle \in L_{\R}^{b, g}} \CZ_{b' g'} \prod_{\langle r', g' \rangle \in L_{\B}} \CZ_{r' g'} \bigg) \CZ_{b g}$ is exactly the topological weight
in Eq.~\eqref{eq:W_mR_topo_weight}. Thus we can simplify this expectation to
\begin{equation}
    \langle \sigma_b^{\R} \sigma_g^{\R} \CZ_{b g} \rangle \propto \sum_{L_{\B}, L_{\R}^{b, g}} \bigg( \frac{t_\B}{\sqrt{2}}\bigg)^{|L_{\B}|} \bigg( \frac{t_\R}{\sqrt{2}} \bigg)^{|L_{\R}^{b, g}|} \frac{2^{C_{L_{\B} \cup \text{close}(L_{\R}^{b, g})}}}{\sqrt{2}},
\end{equation}
which recovers $W_{m_R}(b, g)$. Hence the relevant fidelity correlators and their spatial averages are
\begin{equation}
\label{eq:anyon_magnetization_definitions}
    \begin{split}
        W_{m_R}(b, g) &= \langle \sigma_b^{\R} \sigma_g^{\R} \CZ_{b g} \rangle \rightarrow  W_{m_R} = \frac{1}{|\R|^2} \sum_{(b, g) \in \R} W_{m_R}(b, g)\\
        W_{m_B}(r, g) &= \langle \sigma_r^{\B} \sigma_g^{\B} \CZ_{r g} \rangle \rightarrow  W_{m_B} = \frac{1}{|\B|^2} \sum_{(r, g) \in \B} W_{m_B}(r, g)\\
        W_{e_G}(g, g') &= \langle \widetilde{\sigma}_g \widetilde{\sigma}_{g'} \rangle \rightarrow W_{e_G} = \frac{1}{|\G|^2} \sum_{(g, g') \in \R \cap \B} W_{e_G}(g, g').
    \end{split}
\end{equation}
Here the spatial averaging yields observables capable of detecting condensation of anyons $m_R, m_B, e_G$ respectively.

Using identical logic, we can evaluate fidelity correlators that detect condensation of the Abelian charges $e_R$ and $e_B$. These Wilson line operators are given by $\hat{W}_{e_R} (r, r') = \prod_{r'' \in \gamma_{v, v'}} Z_{r''}^{\R}$ and $\hat{W}_{e_B} (b, b') = \prod_{b'' \in \gamma_{v, v'}} Z_{b''}^{\B}$ respectively. The corresponding fidelity correlators are $W_{e_R}(r, r') =\langle \widetilde{\sigma}_r \widetilde{\sigma}_{r'} \rangle$ and $W_{e_B}(b, b') =\langle \widetilde{\sigma}_b \widetilde{\sigma}_{b'} \rangle$, and as usual, spatial averaging yields observables capable of distinguishing $e_R$ and $e_B$ deconfined and condensed phases.

\subsection{Binder cumulants for multi-dimensional order parameters}
\label{sec:binder_derivations}

The definition of the Binder cumulant involves the ratio of the second and fourth moments of the order parameter distributions~\cite{binder_original}. Without loss of generality, consider the magnetization given by $m=\sum_j m_j$, and the symmetric observable $m^2$, which we refer to as \textit{magnetization-squared}. Suppose $m^2$ is a mean-zero Gaussian random variable with variance $\sigma^2$~\cite{binder_crit_prop}. In the disordered phase:
\begin{equation}
    \begin{split}
        \langle m^2 \rangle = \sigma^2, \qquad
        \langle m^4 \rangle = 3 \langle m^2 \rangle = 3 \sigma^2.
    \end{split}
\end{equation}
Deep in the ordered phase, the probability distribution for $m$ becomes a pair of $\delta$ function peaks: $P(m) = \frac{1}{2} [ \delta(m - N) + \delta (m + N)]$ where $N$ is the volume of the system. As a result we have
\begin{equation}
        \langle m^2 \rangle = \int dm P(m) m^2 = N^2, \qquad
        \langle m^4 \rangle = \int dm P(m) m^4 = N^4.
\end{equation}
Hence, $\langle m^4 \rangle/\langle m^2 \rangle^2=3$ in the disordered phase, and $\langle m^4 \rangle/\langle m^2 \rangle^2=1$ in the ordered phase. The Binder cumulant is defined such that it vanishes in the former and it's finite for the latter, showcasing a universal crossing point at exactly the phase transition. This can be achieved through the definition
\begin{equation}
    U_\text{1d} = \frac{3}{2} \bigg( 1 - \frac{\langle m^4 \rangle}{3 \langle m^2 \rangle^2} \bigg)
\end{equation}
where $U_\text{1d} = 1$ in the ordered phase.

Now we repeat this analysis for a two-component order parameter $\v{m}= (m_x, m_y)$, with $m_{\mu}=\sum_j m_{\mu,j}$ for $\mu \in \{x, y\}$. Suppose $\v{m}$ transforms under symmetries as $\v{m} \mapsto R \v{m}$ such that $R^\intercal R = I$. Then a symmetric observable is given by $||\v{m}|| = \v{m} \cdot \v{m} = m_x^2 + m_y^2$. Suppose now that $m_x^2, m_y^2$ are independent zero-mean Gaussian random variables both with variance $\sigma^2$ (see Ref.~\cite{binder_original,sandvik_AT}). In the disordered phase:
\begin{equation}
        \langle m_x^2 + m_y^2 \rangle = \langle m_x^2 \rangle + \langle m_y^2 \rangle = 2 \sigma^2, \quad \langle (m_x^2 + m_y^2)^2 \rangle = \langle m_x^4 \rangle + \langle m_y^2 \rangle + 2 \langle m_x^2 \rangle \langle m_y^4 \rangle = 8 \sigma^4
\end{equation}
using that $m^x, m^y$ are independent random variables. Deep in the ordered phase, the joint probability distribution for $m_x, m_y$ becomes a sum of 4 delta $\delta$ function peaks: $P(m_x, m_y) = \frac{1}{4}[\delta(m_x - N) + \delta(m_x + N)] [\delta(m_y - N) + \delta(m_y + N)]$ where $N$ is the volume. As a result we have
\begin{equation}
    \begin{split}
        \langle m_x^2 + m_y^2 \rangle &= \int dm_x dm_y P(m_x, m_y) [m_x^2 + m_y^2] = 2 N^2 \\
        \langle (m_x^2 + m_y^2)^2 \rangle &= \int d m_x d m_y P(m_x, m_y) [m_x^2 + m_y^2]^2 = 4 N^4
    \end{split}
\end{equation}
As a result, we deduce that the form of the Binder cumulant is simply
\begin{equation}
    U_\text{2d} = 2 \bigg( 1 - \frac{ \langle \|\v{m}\|^4 \rangle }{2 \langle \|\v{m}\|^2 \rangle ^2} \bigg),
\end{equation}
which is normalized to equal $1$ in the ordered phase.

We now explain and justify the particular choice of Binder cumulants $U(m_R)$ and $U(m_B)$. We define the Binder cumulants probing the condensation of non-Abelian anyons $m_R, m_B$ as
\begin{equation}
\label{eq:U(mR)_U(mB)_def}
        U(m_R) = 2 \bigg( 1  - \frac{\langle [\sum_{g, g'} \v{n}_{g}^{\R} \cdot \v{n}_{g'}^{\R}]^2 \rangle }{2 \langle \sum_{g, g'} \v{n}_g^{\R} \cdot \v{n}_{g'}^{\R} \rangle^2} \bigg),\quad
        U(m_B) = 2 \bigg( 1  - \frac{\langle [\sum_{g, g'} \v{n}_{g}^{\B} \cdot \v{n}_{g'}^{\B}]^2 \rangle}{2 \langle \sum_{g, g'} \v{n}_g^{\B} \cdot \v{n}_{g'}^{\B} \rangle^2} \bigg)
\end{equation}
where $\v{n}_g^\alpha$ is defined in Eq.~\eqref{eq:n_vec_def}. Naively, one might expect us to use all 4 2-vectors $\v{n}_g^{\R}, \v{n}_g^{\B}, \v{n}_b^{\R}, \v{n}_r^{\B}$ to form the Binder cumulants. However, observe that while both $\v{n}_g^{\R}, \v{n}_b^{\R}$ transform in the irreducible representation $\chi_{11}$ of $G_{m_R, m_B}$ (see Table~\ref{tab:char_tab_GRB}), the corresponding matrix representations differ by a relative Hadamard rotation. Similarly, both $\v{n}_g^{\B}, \v{n}_r^{\B}$ transform in the irreducible representation $\chi_{13}$ of $G_{m_R, m_B}$, though the specific matrix representations differ by an overall Hadamard rotation. As a result, the 2-vector order parameters $\v{n}_g^{\R}, \v{n}_b^{\R}$ probe the same symmetry-breaking transition, and the order parameters $\v{n}_g^{\B}, \v{n}_r^{\B}$ also probe the same symmetry-breaking transition. Hence defining the Binder cumulants using solely $\v{n}_g^{\R}, \v{n}_g^{\B}$ probes the desired phase transition while avoiding a treatment of the microscopic anisotropy between the $r, g, b$ sublattices. 

Another reason for defining Binder cumulants $U(m_R), U(m_B)$ without using vectors on both sublattices is the vanishing of the coupling between $\v{n}_g^{\R}, \v{n}_b^{\R}$ at $t_\R = 0$ and the vanishing of the coupling between $\v{n}_g^{\B}, \v{n}_r^{\B}$ at $t_\B = 0$. At $t_\R = 0$, the $\beta_\R$ coupling $\v{n}_g^{\R} \cdot \v{n}_b^{\R}$ in the Hamiltonian in Eq.~\eqref{eq:2na_ham} vanishes. As a result, correlations become $\langle \sum_{b, g} \v{n}_g^{\R} \cdot \v{n}_b^{\R} \rangle = \langle \sum_g \v{n}_g^{\R} \rangle \cdot \langle \sum_b \v{n}_b^{\R} \rangle = 0$ in the symmetric phase (likewise for $\v{n}_g^{\B}, \v{n}_r^{\B}$ when $t_\B = 0$). As a result, the Binder cumulant defined using these 2-vectors would diverge, meaning that Monte Carlo would have trouble sampling the Binder cumulants near $t_\R = 0 $ and $t_\B = 0$. 

The Binder cumulants defined in Eq.~\eqref{eq:U(mR)_U(mB)_def} do not suffer from the aforementioned divergence, as correlation in the denominator always contains a finite ``self-correlation'' term. In particular, $\langle \sum_{g, g'} \v{n}_g^{\R} \cdot \v{n}_{g'}^{\R} \rangle = \langle \sum_g |\v{n}_g^{\R}|^2 \rangle + \langle \sum_{g \neq g'} \v{n}_g^{\R} \cdot \v{n}_{g'}^{\R} \rangle > 0$ since the self-correlation $\langle \sum_g |\v{n}_g^{\R}|^2 \rangle$ is strictly positive. Hence we opt to sample the Binder cumulants $U(m_R), U(m_B)$ defined in Eq.~\eqref{eq:U(mR)_U(mB)_def}, as these remain finite in the symmetric phase and probe the symmetry-breaking transition of interest.

\section{Phase diagram analysis when proliferating two non-Abelions}
\label{app:2_NA_numerics}

\subsection{Ginzburg-Landau theory}
\label{sec:ginzburg_landau}

Here we develop a coarse-grained Ginzburg-Landau (GL) theory meant to capture potential phase transitions in $Z(\beta_\R, \beta_\B)$. This theory remains a valid description when we include a $Z^\G$ deformation, which has the effect of proliferating $e_G$ Abelian charges. Guided by our numerical results, we consider microscopic order parameters
\begin{equation}
    \Phi = \bigg( \widetilde{\sigma}_g \bigg | \sigma_g^{\R} , \sigma_g^{\R} \widetilde{\sigma}_g \bigg | \sigma^{\B}_g , \widetilde{\sigma}_g \sigma^{\B}_g \bigg).
\end{equation}
associated with condensation of $e_G$ (first entry), $m_R$ (middle block), and $m_B$ (right block) anyons. Under symmetries of $G_{m_R, m_B}$, $\widetilde{\sigma}_g$ transforms in the irrep $\chi_2$, $(\sigma_g^{\R} , \sigma_g^{\R} \widetilde{\sigma}_g)$ transforms in irrep $\chi_{11}$, and $(\sigma^{\B}_g , \widetilde{\sigma}_g \sigma^{\B}_g)$ transforms in irrep $\chi_{13}$.
The 1-dimensional irrep $\chi_2$ is a sign irrep in which only $T_G$ is represented non-trivially: $\rho_2(T_G) = -1$. The representation of generators of $G_{m_R, m_B}$ specified by $\chi_{11}$ and $\chi_{13}$ are given in Eq.~\eqref{eq:rho_11_def} and Eq.~\eqref{eq:rho_13_def} respectively. Therefore these 3 order parameters are charged under different group elements, so each of them is needed to detect the relevant symmetry-breaking patterns.
 
To write the Ginzburg-Landau functional, we introduce $3$ real order parameters $\phi_G, \v{\psi}_R, \v{\psi}_B$ that transform in the irreps $\chi_{2}, \chi_{11}, \chi_{13}$, respectively. Here $\phi_G$ is a scalar while $\v{\psi}_R, \v{\psi}_B$ are 2-vectors. From these order parameters we can construct a $G_{m_R, m_B}$-symmetric Ginzburg-Landau free energy density as
\begin{equation}
\label{eq:GL_free_energy}
\begin{split}
    F[\phi, \v{\psi}_R, \v{\psi}_B] &= z_G \phi_G^2 + z_R |\v{\psi}_R|^2 + z_B |\v{\psi}_B|^2 + g_R \phi_G (\v{\psi}_R^\intercal \sigma^x \v{\psi}_R) + g_B \phi_G (\v{\psi}_B^\intercal \sigma^x \v{\psi}_B) + \lambda (\v{\psi}_R^\intercal \sigma^x \v{\psi}_R)(\v{\psi}_B^\intercal \sigma^x \v{\psi}_B) \\
    &+ u_\phi|\phi|^4 + u_R|\v{\psi}_R|^4 + u_B |\v{\psi}_B|^4 + \cdots
\end{split}
\end{equation}
with the ellipsis denoting higher-order symmetry-invariant terms.

Consider first the symmetric line $t_\R = t_\B$, where the symmetry enlarges to $G_{m_R, m_B} \rtimes \Z_2^{\mathrm{swap}}$.  The additional swap symmetry enforces $z_R = z_B \equiv z$, $g_R = g_B \equiv g$, and $u_R = u_B$.  The simplest scenario for $m_{R,B}$ condensation involves $z$ changing sign, leading to non-zero $\v{\psi}_{R,B}$ expectation values with an orientation determined by the $\lambda$ term.  For $\lambda < 0$---assumed here---the fields order such that $\v{\psi}_R^\intercal \sigma^x \v{\psi}_R = \v{\psi}_B^\intercal \sigma^x \v{\psi}_B \neq 0$.  In turn, $\phi_g$ parasitically condenses via the $g$ coupling, even for $z_G>0$.  At the mean-field level the transition is continuous, though fluctuations could potentially drive it first order.  For $t_\R \neq t_\B$, the $z_R$ and $z_B$ parameters generically differ, and thus should not change sign simultaneously except with fine-tuning.  Again at the mean-field level, one would expect an intermediate phase to open up where (say) $\v{\psi}_R$ and $\phi_G$ condense but $\v{\psi}_B$ does not.  Such an intermediate phase---which our numerics do not resolve---can be eradicated, however, if the transition at $t_\R = t_\B$ is first order.  Other first-order scenarios are (as always) also possible, e.g., if the quartic couplings lead to a non-zero free energy minimum abruptly developing at `large' $\v{\psi}_{R,B}$ values.

\subsection{Numerical evidence for absence of intermediate phases}
\label{sec:no_intermediate}

\begin{figure}
    \centering
    \includegraphics[width=\linewidth]{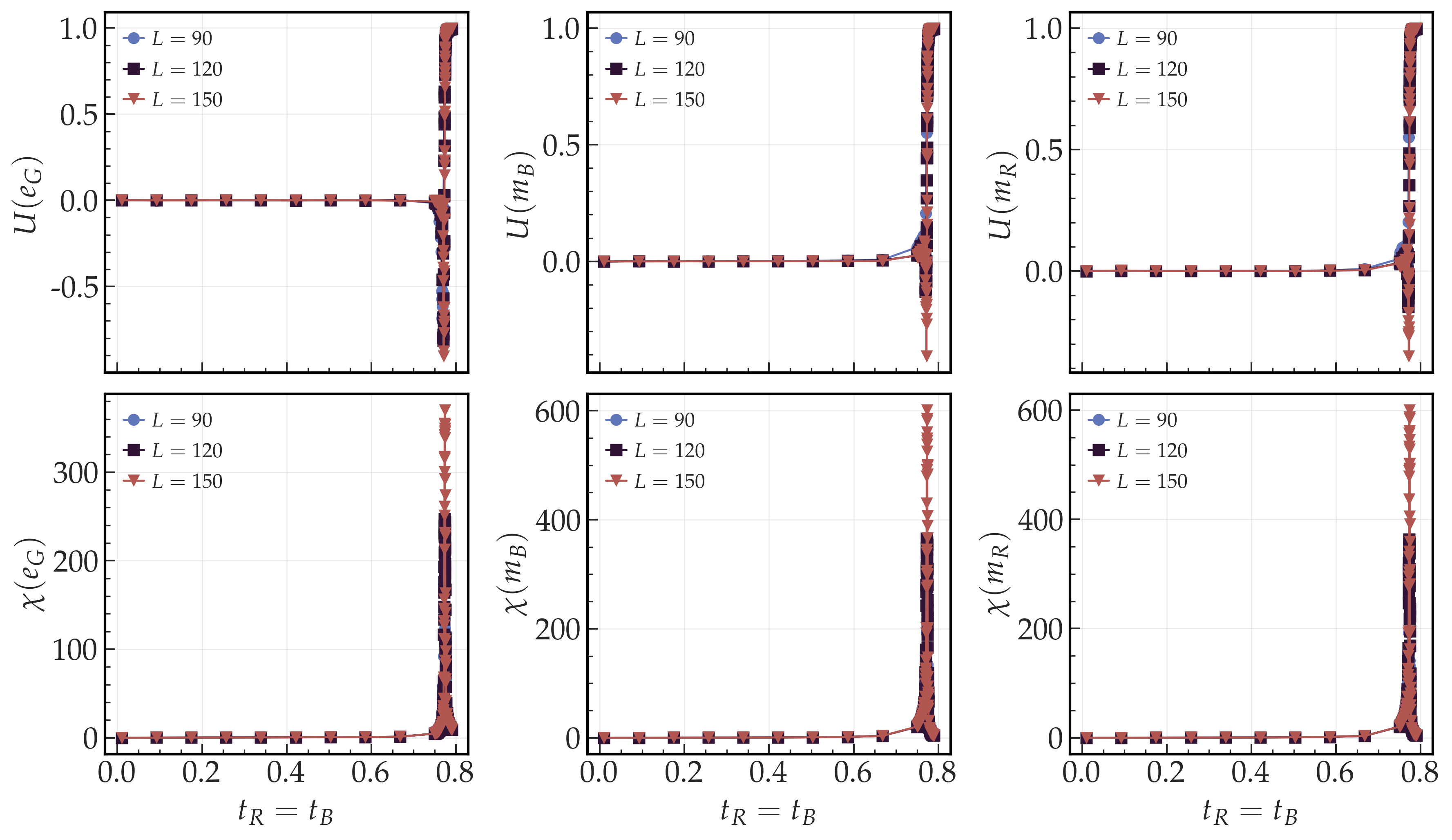}
    \caption{\textbf{Full $\mathbf{t_\R = t_\B}$ dataset.} Binder cumulants $U(e_G), U(m_B), U(m_R)$ and susceptibilities $\chi(e_G), \chi(m_B), \chi(m_R)$ along the line cut $t_\R = t_\B$. Fig.~\ref{fig:Fig_2}(b, c) resolves the critical point clearly.}
    \label{fig:diag_full_data}
\end{figure}

\begin{figure}
    \centering
    \includegraphics[width=\linewidth]{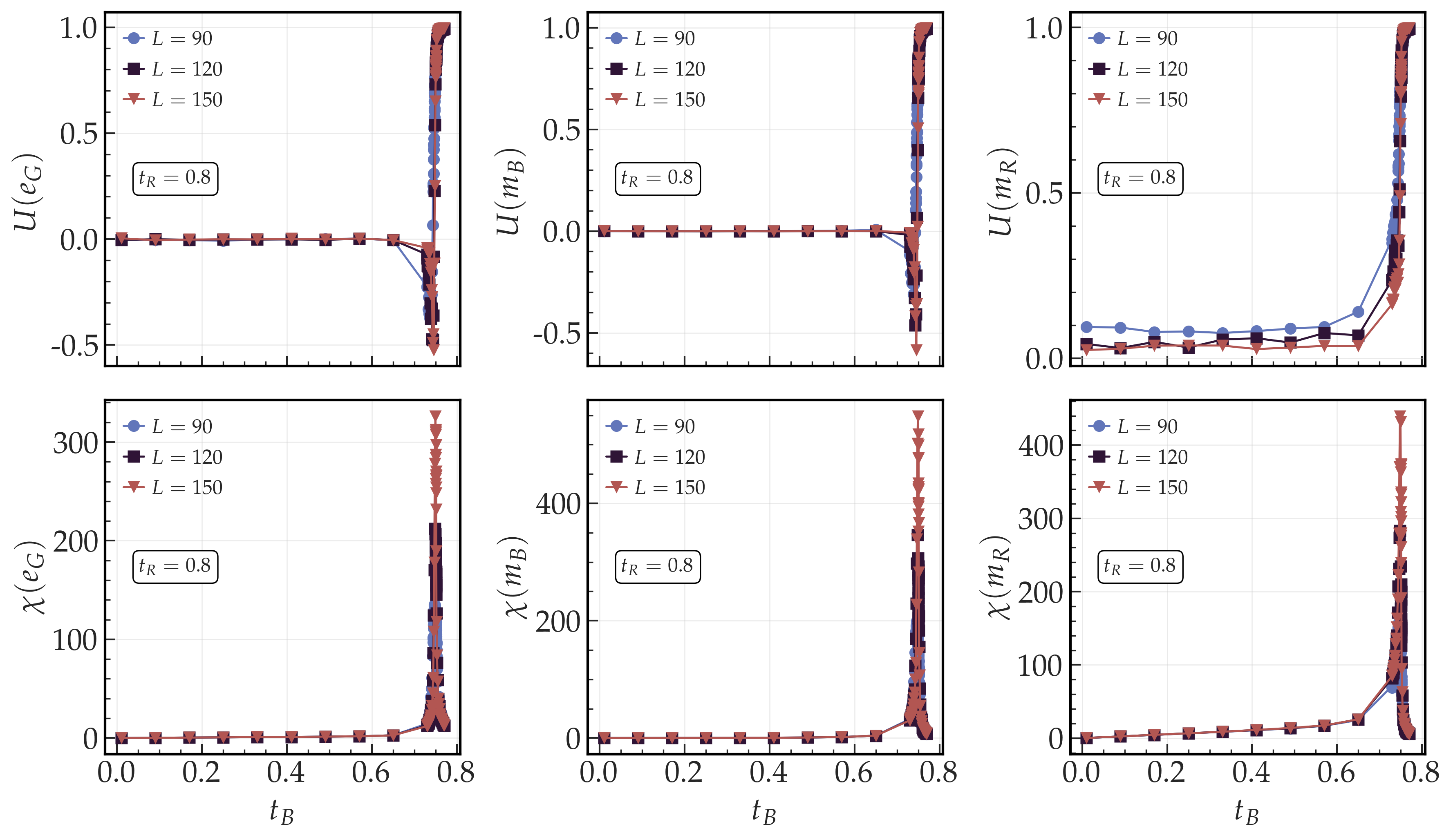}
    \caption{\textbf{Absence of intermediate phase (full dataset).} Binder cumulants $U(e_G), U(m_B), U(m_R)$ and susceptibilities $\chi(e_G), \chi(m_B), \chi(m_R)$ along the line cut $t_\R = 0.8$. Fig.~\ref{fig:vert_binders} resolves the critical point clearly.}
    \label{fig:vert_full_data}
\end{figure}

Our available numerical evidence points to the absence of an intermediate phase along the lines $t_\R = t_\B$ (see Fig.~\ref{fig:diag_full_data}) and $t_\R = 0.8$ (see Fig.~\ref{fig:vert_full_data}). That is, there does not appear to be a phase distinct from both $D_4$ and a trivial TO in which only a subset of $e_G, m_R, m_B$ are condensed. The entire phase diagram in the $(t_\R, t_\B)$ plane is reflection-symmetric across the line $t_\R = t_\B$, so we only study data along cuts $t_\R = 0.8$ and $t_\R = 0.4$.

\begin{figure}
    \centering
    \includegraphics[width=\linewidth]{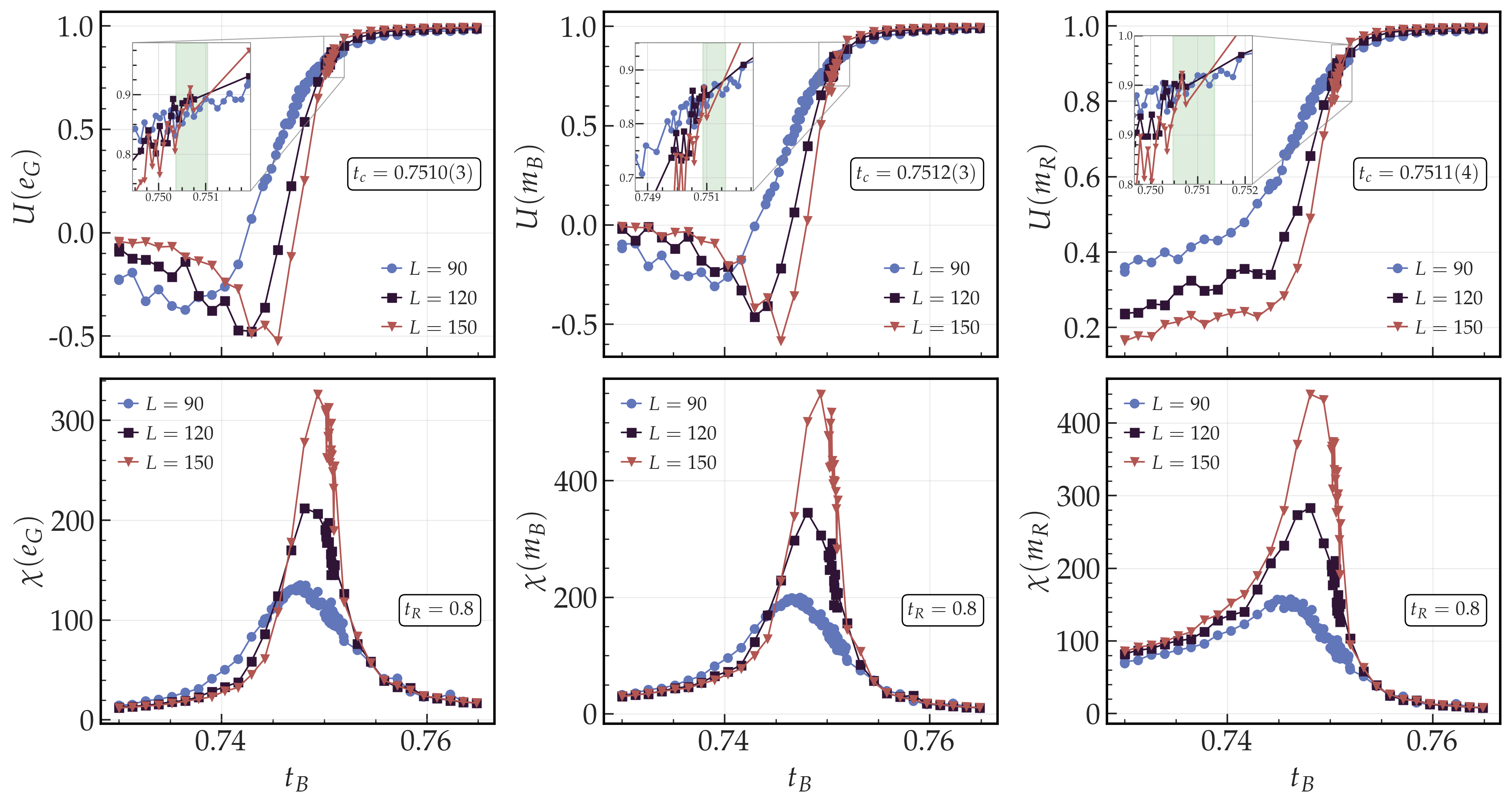}
    \caption{\textbf{Absence of intermediate phase (simultaneous transition).} Binder cumulants $U(e_G), U(m_B), U(m_R)$ and susceptibilities $\chi(e_G), \chi(m_B), \chi(m_R)$ along the line cut $t_\R = 0.8$, showing simultaneous condensation of $e_G, m_R, m_B$ anyons. \textit{Inset:} Critical tension $t_c$ given by the crossing point among different (linear) system sizes $L$.}
    \label{fig:vert_binders}
\end{figure}

Along the cut $t_\R = 0.8$, we find that the Binder cumulants for the $e_G, m_R, m_B$ diagnostics share approximately (with $1 \sigma$) identical crossing points between distinct system sizes (see Fig.~\ref{fig:vert_binders}). This observation indicates that $e_G, m_R, m_B$ condense simultaneously even when $t_\R \neq t_\B$, which suggests the existence of the extended critical curve depicted in Fig.~\ref{fig:Fig_2}(a). In $U(e_G)$ and $U(m_B)$ we observe negative extrema as also discussed in the main text. However, $U(m_R)$ does not exhibit the negative extrema shown by the other 2 Binder cumulants. 

Moreover, it appears that the Binder cumulant $U(m_R)$ is not properly normalized in the disordered phase since it does not decay to $0$. However, in fact for increasing system sizes, the Binder cumulant decreases in the disordered phase, signaling an approach to $0$ in the thermodynamic limit.  

\subsection{Energy distribution at critical point}
\label{sec:energy_hist}

\begin{figure}
    \centering
    \includegraphics[width=\linewidth]{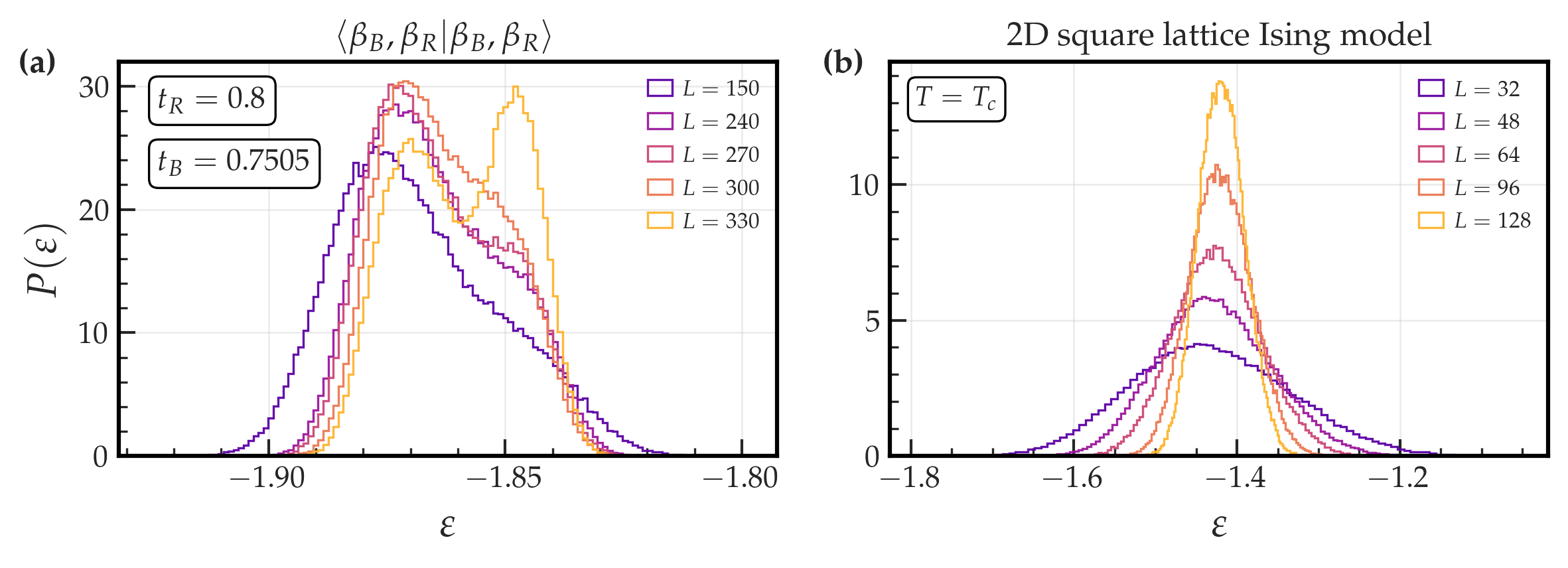}
    \caption{\textbf{Energy density near criticality}. (a) Un-normalized probability distribution $P(\varepsilon)$ of energy density $\varepsilon = H[\beta_\R, \beta_\B]/L^2$ at $t_\R = 0.8, t_\B = 0.7505$ (near the phase transition) for several linear system sizes $L$. Note that a ``shoulder'' in $P(\varepsilon)$ develops for $L \sim 240$, and bimodality manifests for $L \sim 330$. (b) Un-normalized probability distribution of energy density of 2D square lattice Ising model at the critical temperature for several system sizes. Unimodal distribution is consistent with a continuous phase transition.}
    \label{fig:off_diag_energy_hist}
\end{figure}

This section presents a histogram of the energy density $\varepsilon = H[\beta_\R, \beta_\B]/L^2$ at the critical tension. In particular, we focus on the near-critical region $t_\R = 0.8, t_\B = 0.7505$. As visible in Fig.~\ref{fig:off_diag_energy_hist}, with increasing linear system size $L$, the probability distribution of the energy near criticality appears to develop a bimodal distribution. In particular, note that a ``shoulder'' in $P(\varepsilon)$ develops near $\varepsilon = -1.85$. This feature in the energy density distribution emerges only for sufficiently larger system sizes ($L \sim 240$). More obvious bimodality manifests for $L \sim 330$.

As derived in Refs.~\cite{LeeKosterlitzBimodal, LeeKosterlitzBimodal2}, a bimodal energy distribution near criticality is characteristic of a first-order phase transition. Intuitively, one can understand this result by noting that the bimodal distribution of energy density corresponds to two coexistent but distinct phases contributing to the thermal distribution. Competing energetic minima separated by a finite-width barrier in the probability distribution of energy density is indicative of the same situation in the free energy density, which is a signature of a first-order phase transition.

Hence the distribution of energy densities near criticality in Fig.~\ref{fig:off_diag_energy_hist}(a) supports that conjecture that the corresponding phase transition (see Fig.~\ref{fig:Fig_2}(a)) is first-order.

\subsection{Numerical analysis at infinite deformation strength}
\label{sec:gapless_edge}

In this section, we show numerical evidence that no additional phase exists in the limit of infinitely strong deformations $(t_\R = 1, t_\B)$ and $(t_\R, t_\B = 1)$. Moreover, we suggest that in fact, extended critical lines for a finite interval of $t_\R$ and $t_\B$ respectively, are consistent with the data and with the `accidental' U$(1)$ 1-form symmetry exhibited by the deformed wavefunction~\cite{sohal2025obstructionergodicitylocalityu1} mentioned in the main text, which precludes the appearance of a relevant term for a finite region of the Luttinger parameter. 

\begin{figure}
    \centering
    \includegraphics[width=0.7\linewidth]{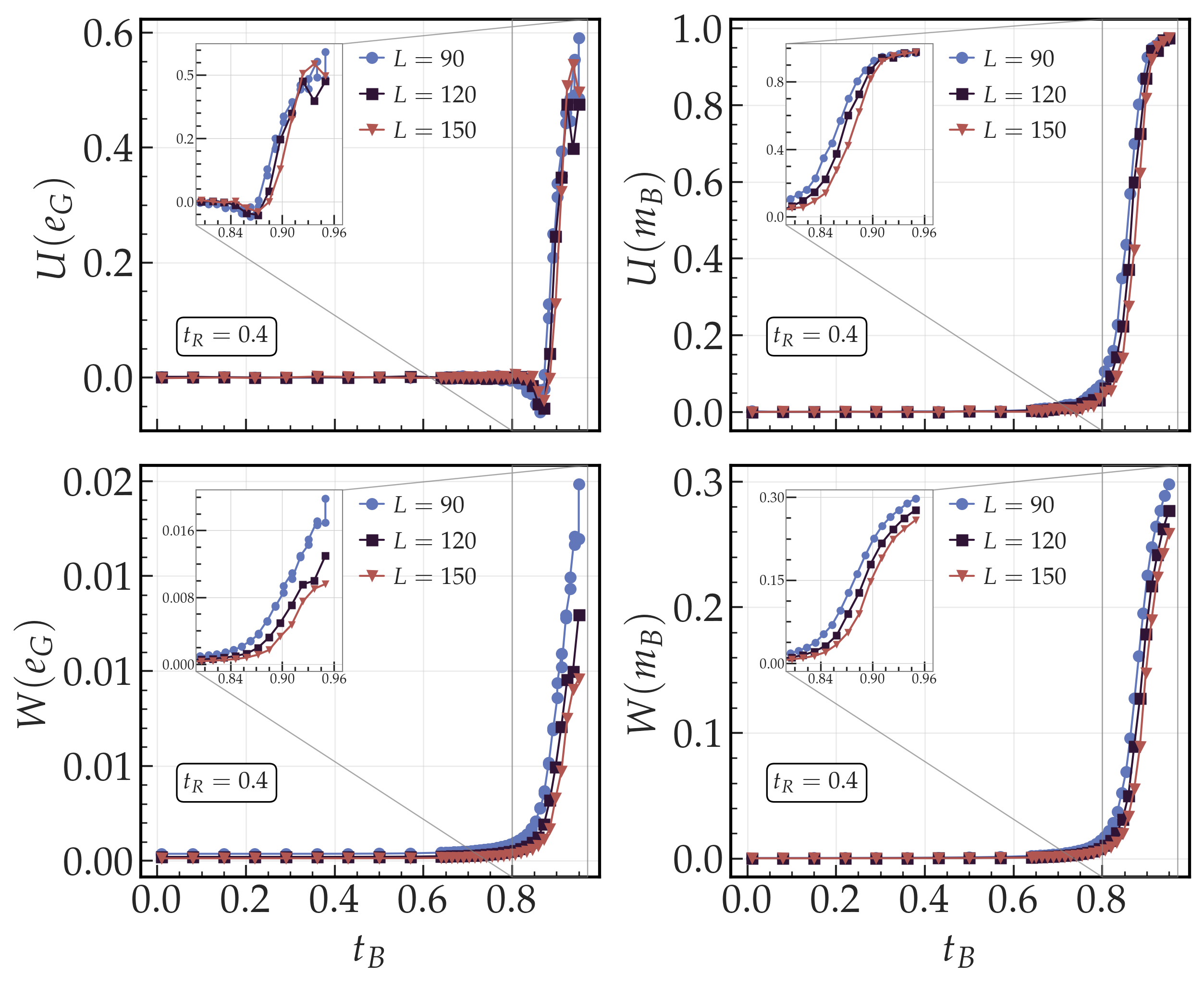}
    \caption{\textbf{No intermediate phase near $t_\B=1$.} Binder cumulants $U(e_G), U(m_B)$ and observables $W(e_G), W(m_B)$ for $t_\R = 0.4$. No crossing between $U$ and decay of $W$ for different linear system sizes $L$ indicate no $t_\B < 1$ transition. Growth of $U$ and $W$ suggests a critical point at $t_\B = 1$.}
    \label{fig:gapless_edge_binders}
\end{figure}

Recalling the reflection symmetry of the phase diagram about the diagonal $t_\R = t_\B$, we study the cut $t_\R = 0.4$ and consider $t_\B\to 1$. We study the Binder cumulants $U(e_G), U(m_B)$ along $t_\R = 0.4$ in Fig.~\ref{fig:gapless_edge_binders}. We observe an increase from $U = 0$ to $U= 1$ in both sets of Binder cumulants, and we do not identify a crossing point between $U$ for different linear system sizes $L$. In fact, for $U(m_B)$, we observe a ``merging'' of Binder cumulants towards $1$ without a crossing. On the other hand, the data for $U(e_G)$ becomes noisier as $t_\B\to 1$, displaying a negative peak that appears to shift to larger $t_\B$ for increasing system sizes. Given that negative peaks in Binder cumulants are associated with either first order phase transitions, proximity to a multi-critical point or to coupling to a fluctuating multi-component order parameter, at least one of these outcomes is consistent with proximity to a critical boundary (although we notice that $t_\R = 0.4$ is somewhat near the extended critical curve in Fig.~\ref{fig:Fig_2}(a), which could contribute to the fluctuations in order parameters driving the negative dip of $U(e_G)$). Since $W(e_G) \approx 0$ and does not converge with system size in the entire window shown in Fig.~\ref{fig:gapless_edge_binders}, we conclude that, the anyon $e_G$ has not condensed. Moreover, despite increasing monotonically, $W(m_B)$ decays with system size, so the system did not enter a long range ordered phase. However, notice that from this data alone we cannot eliminate the possibility of a critical point $t_\B < 1$, leading to a highly ``squeezed'' ordered phase.

\section{Two non-Abelian errors plus one Abelian error}
\label{sec:mR_mB_eG}

\subsection{Loop model and spin Hamiltonian}
\label{sec:mR_mB_eG_loop_model}

Consider the deformed wavefunction $ \ket{\beta_\B, \beta_\R, \beta_{\G}^z} = e^{\frac{\beta_\B}{2} \sum_{b \in \B} X_b^{\B}} e^{\frac{\beta_\R}{2} \sum_{r \in \R} X_r^{\R}} e^{\frac{\beta_{\G}^z}{2} \sum_{g \in \G} Z_g^{\G}} \ket{D_4}$.
As previously explained, the deformations $e^{\frac{\beta_\B}{2} \sum_{b \in \B} X_b^{\B}}$ and $ e^{\frac{\beta_\R}{2} \sum_{r \in \R} X_r^{\R}} $ create pairs of $m_B$ and $m_R$ non-Abelian anyons respectively, while $e^{\frac{\beta_{\G}^z}{2} \sum_{g \in \G} Z_g^{\G}}$ creates pairs of the Abelian anyon $e_G$. Importantly, as noted in Fig.~\ref{fig:composite_figure}(b), $e_G \in m_R \times m_R$ and $e_G \in m_B \times m_B$. As a result, in the loop model $\bra{\beta_\B, \beta_\R, \beta_{\G}^z} \beta_\B, \beta_\R, \beta_{\G}^z \rangle$ we will find that $Z_g^{\G}$ strings need not form closed loops. That is, $Z_g^{\G}$ strings can end on $X_r^{\R}$ and $X_b^{\B}$ loops.

We can evaluate the norm of the deformed wavefunction using the mapping between the topologically ordered ground state $\ket{D_4}$ and the gauged $\Z_2^3$ SPT state discussed in Ref.~\cite{Yoshida_2016, sala_D4}. We find that
\begin{equation}
\label{eq:Z_mR_mB_eG}
\begin{split}
    Z(\beta_\R, \beta_\B, \beta_{\G}^z) &= \langle \beta_\B, \beta_\R, \beta_{\G}^z | \beta_\B, \beta_\R, \beta_{\G}^z \rangle = \langle D_4 | e^{\beta_{\G}^z \sum_{g \in \G} Z_g^{\G}} e^{\beta_{\R} \sum_{r \in \R} X_r^{\R}} e^{\beta_{\B} \sum_{b \in \B} X_b^{\B}} | D_4 \rangle \\
    &= \sum_{L_\R, L_\B, \gamma_\G} t_{\R}^{|L_\R|} t_{\B}^{|L_\B|} {t_{\G}^z}^{|\gamma_\G|} \bra{D_4} \prod_{r \in \R} X_r^{L_\R} \prod_{b \in L_\B} X_b^{\B} \prod_{g \in \gamma_\G} Z_g^{\G} \ket{D_4} \\
    &= \sum_{L_\R, L_\B, \gamma_\G} t_{\R}^{|L_\R|} t_{\B}^{|L_\B|} {t_{\G}^z}^{|\gamma_\G|} \bra{\mathrm{SPT}} \prod_{\langle b,g \rangle \in L_\R}\CZ_{bg} \prod_{\langle r, g \rangle \in L_\B} \CZ_{r g} \prod_{\langle g, g' \rangle \in \gamma_\G} \widetilde{Z}_g \widetilde{Z}_{g'} \ket{\mathrm{SPT}} \\
    &= \sum_{L_\R, L_\B} \bigg(\frac{t_\R}{\sqrt{2}}\bigg)^{|L_\R|} \bigg(\frac{t_\B}{\sqrt{2}}\bigg)^{|L_\B|} 2^{C_{L_\R \cup L_\B}} \sum_{\gamma_\G} {t_{\G}^z}^{|\gamma_\G|}
\end{split}
\end{equation}
where $t_{\R} = \tanh(\beta_\R), t_{\B} = \tanh(\beta_\B), t_{\G}^z = \tanh(\beta_{\G}^z)$. Here $L_\R, L_\B$ are loops on the A-B stacked honeycomb lattices in Fig.~\ref{fig:composite_figure}(a), and $\gamma_\G$ is a loop on the triangular lattice dual to the green honeycomb lattice. Importantly, $\gamma_{\G}$ can consist of any closed loop configuration or open string configuration that ends on $L_\R \cup L_\B$.

It is simple to verify that the partition function of the following local spin Hamiltonian recovers (up to a multiplicative factor) the loop model $Z(\beta_\R, \beta_\B, \beta_{\G}^z)$:
\begin{equation}
\label{eq:2na_1a_ham}
    H[\beta_\R, \beta_\B, \beta_{\G}^z] = -\beta_\R \sum_{\langle b, g \rangle \in \red{\hexagon}} \sigma_b^{\R} \sigma_g^{\R} \CZ_{bg} - \beta_\B \sum_{\langle r, g \rangle \in \blue{\hexagon}} \sigma_g^{\B} \sigma_r^{\B} \CZ_{gr} - \beta_{\G}^z \sum_{\langle g, g' \rangle \in \green{\triangle}} \widetilde{\sigma}_g \widetilde{\sigma}_{g'}.
\end{equation}
Note that $H[\beta_\R, \beta_\B, \beta_{\G}^z] = H[\beta_\R, \beta_\B] - \beta_{\G}^z \sum_{\langle g, g' \rangle \in \green{\triangle}} \widetilde{\sigma}_g \widetilde{\sigma}_{g'}$ (see Eq.~\eqref{eq:2na_ham}). Moreover, the symmetry of $H[\beta_\R, \beta_\B, \beta_{\G}^z]$ is $G_{m_R, m_B}$, which is identical to that of $H[\beta_\R, \beta_\B]$. As a result, we can detect condensation of anyons in the deformed wavefunction using the same order parameters proposed previously.

\subsection{Phase boundary analysis}

\subsubsection{Emergence of toric code phase}
\label{sec:emergence_TC}

\begin{figure}
    \centering
    \includegraphics[width=\linewidth]{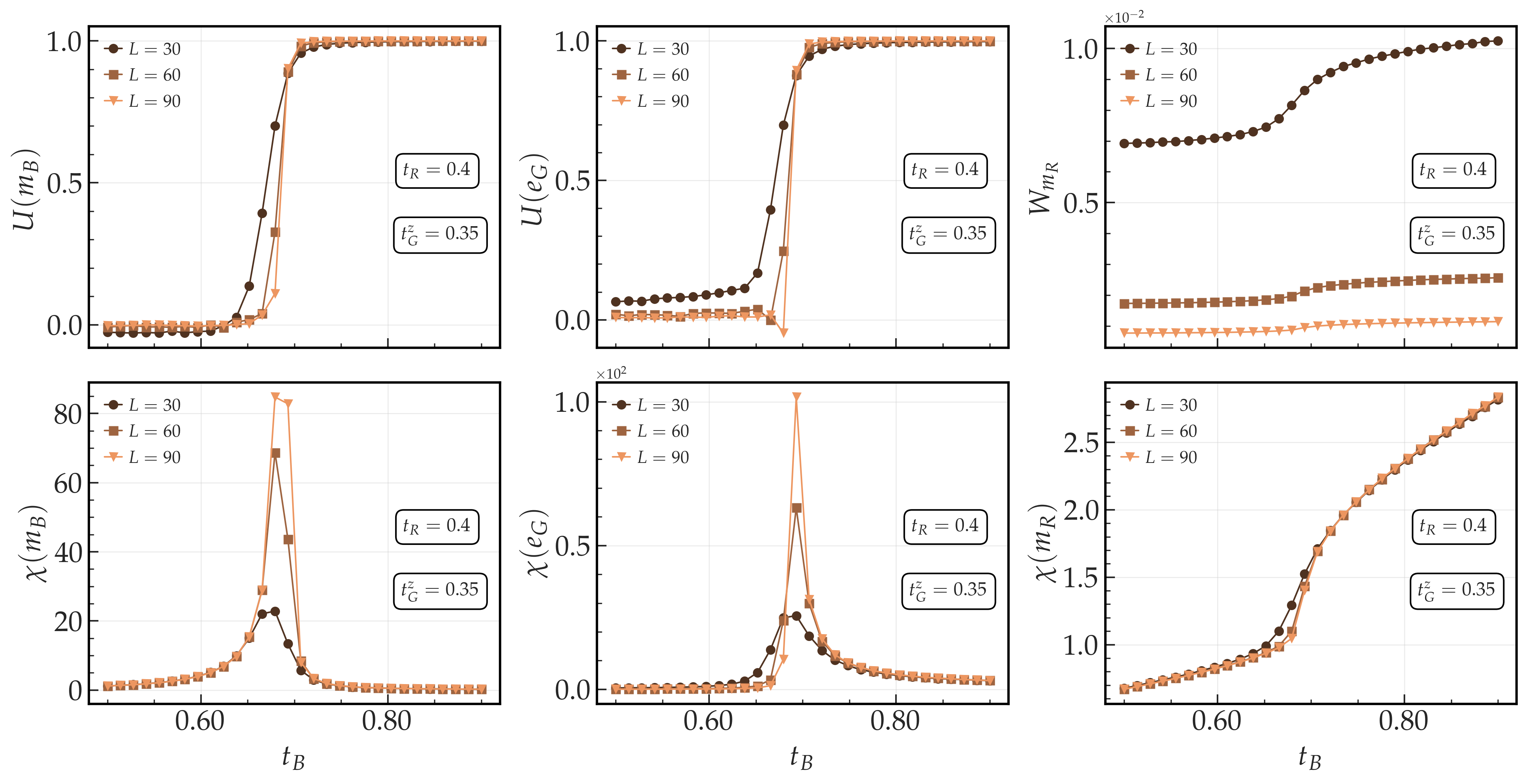}
    \caption{\textbf{Transition to toric code}. Binder cumulants $U(e_G), U(m_B)$, observable $W_{m_R}$, and susceptibilities $\chi(e_G), \chi(m_B), \chi(m_R)$ along the line cut $t_\R = 0.4, t_{\G}^z = 0.35$, showing simultaneous condensation of $e_G, m_B$ anyons without condensation of $m_R$. Non-Abelian anyon $m_R$ splits into Abelian Toric code anyons.}
    \label{fig:finite_betaG_binders}
\end{figure}

Upon adding finite $\beta_{\G}^z > 0$ to the Hamiltonian in Eq.~\eqref{eq:2na_1a_ham}, we find that a toric code phase emerges along the edge of the phase diagram in Fig.~\ref{fig:Fig_2}(a). We numerically diagnose this toric code phase by measuring the observables $W_{m_R}, W_{m_B}, W_{e_G}$ along the cut $t_\R = 0.4, t_{\G}^z = 0.35$ (see Fig.~\ref{fig:finite_betaG_binders}). We find that $W_{m_B}, W_{e_G}$ both exhibit long-range order whereas $W_{m_R}$ vanishes in the thermodynamic limit. This long-range order corresponds to a symmetry breaking pattern $G_{m_R, m_B} \cong \Z_2^3 \rtimes \Z_2^2 \rightarrow \Z_2^3$, and to the condensation of $e_G, m_B$ anyons, leaving a residual toric code phase~\cite{d4_hasse}. 

We can understand this phase transition ($D_4 \ \mathrm{TO} \rightarrow$ toric code) from the field theory proposed in Sec.~\ref{sec:field_theory}. From the Hamiltonian density in Eq.~\eqref{eq:ham_density}, this transition is induced by a relevant $\cos(4\theta^{\B})$ term (as noted in Ref.~\cite{sala_D4}). A relevant $\cos(4\theta^{\B})$ pins $\theta^{\B} = (2n+1) \pi /4$ or $\theta^{\B} = n \pi /2$ depending on the sign of $J_B$. The residual symmetry of the system is $\langle V_R \rangle \times \langle S_B \rangle \times \langle S_R \rangle$ or $\langle V_R \rangle \times \langle V_B S_B \rangle \times \langle S_R \rangle$; in either case, we find that $\theta^{\B} = (2n+1) \pi /4$.

We can also study this phase transition using the Ginzburg-Landau theory proposed in Sec.~\ref{sec:ginzburg_landau}. The $\Z_2^2$ symmetry-breaking transition that we observe numerically corresponds to $z_B$ in Eq.~\eqref{eq:GL_free_energy} changing sign. This sign change induces ordering of the 2-vector order parameter $\psi_B$ such that there is 1 of 2 possible sets of residual symmetries. In the first case, the polar angle is $\pi/4$ or $5 \pi/4$, resulting in the symmetry breaking transition $G_{m_R, m_B} \rightarrow \langle V_R \rangle \times \langle S_B \rangle \times \langle S_R \rangle \cong \Z_2^3$. In the second case, the polar angle is $3 \pi/4$ or $- \pi/4$, resulting in the symmetry breaking transition $G_{m_R, m_B} \rightarrow \langle V_R \rangle \times \langle V_B S_B \rangle \times \langle S_R \rangle \cong \Z_2^3$.

Although the $m_R$ anyon remains deconfined in the toric code condensate, this non-Abelian anyon splits into Abelian toric code anyons. Correspondingly, the 2-dimensional order parameter $\v{n}_g^{\R}$ characterizing condensation of the $m_R$ anyon splits into scalar order parameters within the symmetry-broken phase. This splitting is the cause of the ``cusp'' behavior exhibited by the susceptibility in Fig.~\ref{fig:finite_betaG_binders}, since $\chi(m_R)$ is computed assuming a vector order parameter.

In the next section we consider an infinitely strong deformation $\beta_{\G}^z = \infty$. In this case, even for small values of $t_\R$ and $t_\B$, there is a transition $D_4 \ \mathrm{TO} \rightarrow \mathrm{2 \ toric \ codes}$. In the statistical mechanical model, this transition corresponds to the symmetry-breaking pattern $G_{m_R, m_B} \rightarrow \langle V_R \rangle \times \langle V_B \rangle \times \langle S_B \rangle \times \langle S_R \rangle \cong \Z_2^4$. From the field theory, we can understand this transition as being induced by the Ising mass field ($\phi_G$). From the Ginzburg-Landau theory, this transition can be obtained by changing the sign of $m_G$, which induces ordering of the scalar order parameter $\phi_G$. In both cases we recover the same symmetry-breaking phase transition, and we observe condensation of the $e_G$ Abelian anyon, leaving a pair of decoupled toric codes behind.

\subsubsection{Infinitely strong deformation}
\label{sec:infinite_betaG}
We now consider an infinitely strong deformation $\beta_{\G}^z = \infty$. In this case, deforming $\ket{D_4}$ with $e^{\frac{\beta_{\G}^z}{2} \sum_{g \in \G} Z_g^{\G}}$ is equivalent to enacting the projection $Z_g^{\G} =+1$ on all sites $g \in \G$ in the ground state. We will study the effect of the deformations $e^{\frac{\beta_\R}{2} \sum_{r \in \R} X_r^{\R}}$ and $e^{\frac{\beta_\B}{2} \sum_{b \in \B} X_b^{\B}}$ on the projected state $\prod_{g \in \G} \frac{1}{2} (1 + Z_g^{\G}) \ket{D_4}$. 

Evidently the 3-body stabilizer $B_t^{\G} = 1$ under this projection. The other 3-body stabilizers $B_t^{\R}, B_t^{\B}$ remain invariant under this projection since they commute with $Z_g^{\G}$ for all $g \in \G$. The 12-body stabilizer $A_s^{\G} = \prod_{i_\mathrm{in} = 1}^6 \CZ_{i_\mathrm{in}, i_\mathrm{in} + 1} \prod_{i_\mathrm{out}}^6 X^{\G}_{i_\mathrm{out}} \rightarrow 0$ since $X^{\G}_g Z^{\G}_g = -Z^{\G}_g X^{\G}_g$. The 2 other 12-body stabilizers transform as
\begin{equation}
    \begin{split}
        A_s^{\R} = \prod_{i_\mathrm{in} = 1}^6 \CZ_{i_\mathrm{in}, i_\mathrm{in} + 1} \prod_{i_\mathrm{out}}^6 X^{\R}_{i_\mathrm{out}} \xrightarrow{Z^{\G}=+1} \prod_{i_\mathrm{out}}^6 X^{\R}_{i_\mathrm{out}}, \qquad
        A_s^{\B} = \prod_{i_\mathrm{in} = 1}^6 \CZ_{i_\mathrm{in}, i_\mathrm{in} + 1} \prod_{i_\mathrm{out}}^6 X^{\B}_{i_\mathrm{out}} \xrightarrow{Z^{\G}=+1} \prod_{i_\mathrm{out}}^6 X^{\B}_{i_\mathrm{out}}.
    \end{split}
\end{equation}
Observe that there are 4 non-trivial surviving stabilizers, which are exactly the stabilizers of 2 decoupled toric codes on a honeycomb lattice. As a result, the projected ground state is given by $\prod_{g \in \G} \frac{1}{2} (1 + Z_g^{\G}) \ket{D_4} = \ket{\mathrm{TC_\R}} \ket{\mathrm{TC_\B}}$ where $\ket{\mathrm{TC_\alpha}}$ denotes a toric code ground state on the $\alpha \in \{\R, \B\}$ honeycomb lattice. Thus the full deformed ground state can be written as
\begin{equation}
\begin{split}
    \ket{\beta_\R, \beta_\B, \beta_{\G}^z = \infty} &=  e^{\frac{\beta_\B}{2} \sum_{b \in \B} X_b^{\B}} e^{\frac{\beta_\R}{2} \sum_{r \in \R} X_r^{\R}} \prod_{g \in \G} \frac{1}{2} (1 + Z_g^{\G}) \ket{D_4} = \bigg(e^{\frac{\beta_\R}{2} \sum_{r \in \R} X_r^{\R}} \ket{\mathrm{TC_{\R}}} \bigg) \otimes \bigg(e^{\frac{\beta_\B}{2} \sum_{b \in \B} X_b^{\B}}\ket{\mathrm{TC}_{\B}} \bigg).
\end{split}
\end{equation}
Thus the phase diagram of this deformed state will be identical to that of a pair of decoupled toric codes subjected to Pauli $X$ noise. For each factor, the deformed wavefunction possesses a phase transition at a finite deformation strength ($\beta_\R$ and $\beta_\B$ respectively). Hence the accessible phases correspond to 2 toric codes, 1 toric code, and a topologically trivial phase.

\section{Equivalence of $e$ and $m$ condensates in deformed toric code wavefunctions}
\label{app:TC_e_m_equiv}

Here we prove an equivalence between the $e$-condensed and $m$-condensed phases of the deformed toric code wavefunction. In particular, we explicitly construct a continuous path of wavefunctions connecting the two phases using the em-duality of the toric code, which is known to be implementable using a finite-depth local unitary (FDLU) circuit~\cite{kobayashi_em_duality, potter_em_duality, tu_em_duality, shirley_em_duality}. Consider the pair of deformed toric code wavefunctions
\begin{equation}
    \ket{\beta, X} = e^{\beta \sum_v X_v} \ket{\mathrm{TC}}, \qquad \ket{\beta, Z} = e^{\beta \sum_v Z_v} \ket{\mathrm{TC}}
\end{equation}
where strings of $Z_v \ (X_v)$ operators create pairs of $e \ (m)$ anyons and $\ket{\mathrm{TC}}$ is a toric code ground state wavefunction. Any pair of FDLU operators can be connected via a continuous path, so consider an FDLU $U(\theta)$ such that $U(0) = I$ and $U(\pi)$ implements the em-duality of the toric code. The FDLU proposed in Ref.~\onlinecite{potter_em_duality} and thoroughly discussed in Ref.~\onlinecite{tu_em_duality} (see Fig.~12) acts as 
\begin{equation}
    U X_i U^\dagger = Z_j  \prod_{p \in P_j} B_p, \qquad U Z_i U^\dagger = X_j \prod_{v \in V_j} A_v.
\end{equation}
Here $A_v$ and $B_p$ are the star and plaquette operators whose excitations are e and m anyons respectively, and $V_j, P_j$ are finite-support subsets of plaquettes and stars on the lattice. Crucially, the em-duality operator maps a single $X \ (Z)$ on site $i$ to a single $Z \ (X)$ on site $j$ (where $|i - j|$ is finite) multiplied by a product of stabilizers. The action of $U = U(\pi)$ on the deformed wavefunction $\ket{\beta, X}$ is then
\begin{equation}
    U \ket{\beta, X} = U e^{\beta \sum_i X_i} \ket{\mathrm{TC}} = e^{\beta U (\sum_i X_i) U^\dagger} U \ket{\mathrm{TC}} = e^{\beta \sum_j Z_j (\prod_{p \in P_j} B_p)} \ket{\mathrm{TC}} = e^{\beta \sum_j Z_j} \ket{\mathrm{TC}} = \ket{\beta, Z}.
\end{equation}
The translational symmetry of the deformation $e^{\beta \sum_i X_i}$ renders the shift $i \rightarrow j$ enacted by $U$ unimportant. Moreover, since $\ket{\mathrm{TC}}$ is a $+1$ eigenstate of $B_p$, the decoration of $Z_j$ by stabilizers $B_p$ does not alter the deformed wavefunction. Hence we find that $U(\pi) \ket{\beta, X} = \ket{\beta, Z}$.

The two-parameter family of functions $\ket{\beta, Z, \theta} = U(\theta) e^{\beta \sum_v Z_v} \ket{\mathrm{TC}}$ can evidently be smoothly connected to the wavefunction $\ket{\beta, X}$ by tuning between $\theta =0$ and $\theta = \pi$. For sufficiently large $\beta > \beta_\mathrm{critical}$, it is known that $\ket{\beta, Z}$ and $\ket{\beta, X}$ enter an e-condensed and m-condensed phase respectively. As a result, one can continuously tune $\theta \in [0, \pi]$ to deform $\ket{\beta > \beta_\mathrm{critical}, Z, \theta} \rightarrow \ket{\beta > \beta_\mathrm{critical}, X}$, demonstrating that these two wavefunctions are in fact phase equivalent.

\section{Three non-Abelian errors}
\label{sec:3_NA_anyons}
\subsection{Loop model}

\begin{figure}
    \centering
    \includegraphics[width=0.9\linewidth]{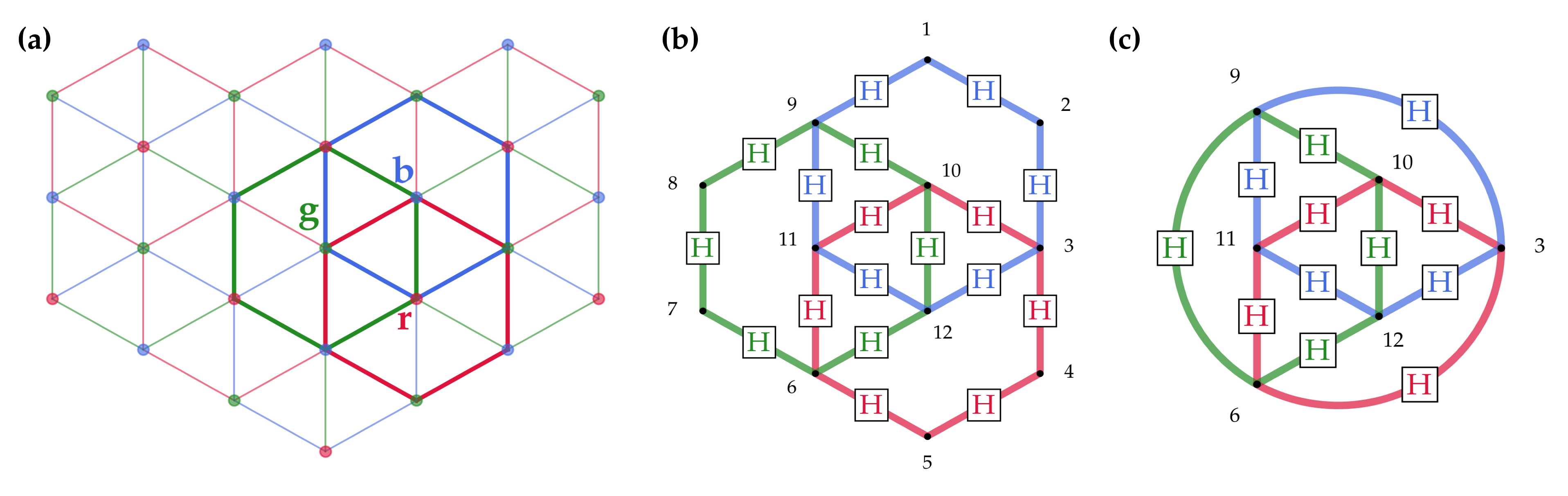}
    \caption{\textbf{Borromean ring tensor network}. (a) Borromean ring loop configuration of $L_\R, L_\B, L_\G$ in $\langle \beta_\R, \beta_\B, \beta_\G | \beta_\R, \beta_\B, \beta_\G \rangle$. Here $r, g, b$ label centers of red, green, blue hexagons respectively. (b) Computation of Borromean ring configuration as contraction of Hadamard circuit. Spins (black sites) are numerically indexed. (c) Tensor network obtained after contracting over spins $\sigma_1, \sigma_2, \sigma_4, \sigma_5, \sigma_7, \sigma_8$.}
    \label{fig:borromean_config}
\end{figure}

We can evaluate the norm of $\ket{\beta_\R, \beta_\B, \beta_\G}$ resulting from proliferating the anyons $m_R, m_B$ and $m_G$ using the same approach as in previous sections. As before, we find that
\begin{equation}
\begin{split}
    \langle \beta_\R, \beta_\B, \beta_\G | \beta_\R, \beta_\B, \beta_\G \rangle &= \langle D_4 | e^{\beta_\R \sum_{r} X_r^{\R}} e^{\beta_\B \sum_{b} X_b^{\B}}  e^{\beta_\G \sum_{g} X_g^{\G}} \ket{D_4} \\
    &\propto \sum_{L_\R, L_\B, L_\G} \tanh(\beta_\R)^{|L_\R|} \tanh(\beta_\B)^{|L_\B|} \tanh(\beta_\G)^{|L_\G|}  W_{L_\R, L_\B, L_\G}
    \end{split}
\end{equation}
where the ``topological weight'' is given by
\begin{equation}
    W_{L_\R, L_\B, L_\G} = \langle D_4 | \prod_{r \in L_\R} X_r^{\R} \prod_{b \in L_\B} X_b^{\B} \prod_{g \in L_\G} X_g^{\G} | D_4 \rangle = \langle \mathrm{SPT} | \prod_{\langle b g \rangle \in L_\R} \CZ_{b g} \prod_{\langle rg \rangle \in L_\B} \CZ_{r g} \prod_{\langle r b \rangle \in L_\G} \CZ_{r b} | \mathrm{SPT} \rangle
\end{equation}
where in the last step we used the $\ket{D_4} \rightarrow \ket{\mathrm{SPT}}$ mapping discussed in Ref.~\cite{Yoshida_2016, sala_D4}. Here $\ket{\mathrm{SPT}}$ is a $\Z_2^3$-symmetric SPT, which can be mapped to $D_4$ topological order by gauging the global symmetry~\cite{Yoshida_2016}. We can then apply the disentangling circuit $U = \prod_{\langle r, g, b \rangle \in \triangle} CCZ_{r g b}$ where $\langle r, g, b \rangle \in \triangle$ are nearest-neighbor sites forming a triangle. Under this circuit, $U\ket{ \mathrm{SPT}} \rightarrow \ket{+^\R} \ket{+^\B} \ket{+^\G}$. We can now write the expectation value $W_{L_\R, L_\B}$ as
\begin{equation}
\begin{split}
    W_{L_\R, L_\B, L_\G} &= \bra{+^\R, +^\B, +^\G} \prod_{\langle b g \rangle \in L_\R} \CZ_{b g} \prod_{\langle rg \rangle \in L_\B} \CZ_{r g} \prod_{\langle r b \rangle \in L_\G} \CZ_{r b} \ket{+^\R, +^\B, +^\G} \\ 
    &= \sum_{\{\tilde{\sigma}\}} \bigg(\prod_{\langle b, g \rangle \in L_\R} \CZ_{bg} \prod_{\langle g, r \rangle \in L_\B} \CZ_{gr} \prod_{\langle r, b \rangle \in L_\G} \CZ_{r b} \bigg) = \prod_{\ell \in (L_\R \cup L_\B \cup L_\G)} \tr(\CZ_\ell)
\end{split}
\end{equation}
where $\ell$ is a connected component of the net configuration $L_\R \cup L_\B \cup L_\G$. $\CZ$ is diagonal in the Z-basis and can be written as
\begin{center}
\begin{tikzpicture}[thick]

\draw (-1.5,0.5) node[above] {$\sigma_i$} -- (-1.5, -0.45) node[below] {$\sigma_i$};
\draw (-0.5,0.45) node[above] {$\sigma_j$} -- (-0.5, -0.45) node[below] {$\sigma_j$};

\draw (-1.5, -0.25) rectangle (-0.5, 0.25);
\node at (-1, 0.0) {CZ};

\node at (-0.05,0) {$=$};

\node at (0.55,0) {$\sqrt{2}$};

\draw (1,0.5) node[above] {$\sigma_i$} -- (1,-0.45) node[below] {$\sigma_i$};
\draw (2,0.45) node[above] {$\sigma_j$} -- (2,-0.45) node[below] {$\sigma_j$};

\draw (1,0) -- (1.25,0);
\draw (1.75,0) -- (2,0);
\draw (1.5,0) circle (0.25);
\node at (1.5,0) {$H$};

\node at (2.65,0) {$=$};

\node at (4.0,0) {$\sqrt{2} (-1)^{\sigma_i \sigma_j}$};

\end{tikzpicture}
\end{center}
combining a Hadamard gate and a delta tensor, since $\CZ_{i j} = \begin{cases}
    -1 & \sigma_i = \sigma_j = 1 \\
    +1 & \mathrm{else}
\end{cases}$
for $\sigma_j \in \{0, 1\}$. Using this correspondence, we find that $\tr(\CZ_\ell) = \sqrt{2}^{|\ell|} \tr(\prod_{e \in \ell} H_e)$, so we can simplify
\begin{equation}
\label{eq:3na_topo_weight}
    W_{L_\R, L_\B, L_\G} = \prod_{\ell \in (L_\R \cup L_\B \cup L_\G)} \frac{1}{2^{|\ell|}} \sqrt{2}^{|\ell|} \tr(H^{|\ell|}) = 2^{C_{L_\R \cup L_\B \cup L_\G}} \sqrt{2}^{-|L_\R| - |L_\B| - |L_\G|}
\end{equation}
where $\ell$ is a connected component of $L_\R \cup L_\B \cup L_\G$. Not all configurations of $\{L_\R, L_\B, L_\G\}$ yield finite $W_{L_\R, L_\B, L_\G}$: in fact we show that \emph{Borromean ring} configurations have vanishing weight. We can compute the Borromean ring contraction in Fig.~\ref{fig:borromean_config}(b), denoted by $\borromeansymbol$, by first contracting over the spins $\sigma_1, \sigma_2, \sigma_4, \sigma_5, \sigma_7, \sigma_8$ as
\begin{equation}
    \begin{split}
         \tr \bigg(\prod_{e \in \borromeansymbol} H_e \bigg) &= \sum_{\v{\sigma}} f(\v{\sigma}) \bigg[\sum_{\sigma_1, \sigma_2} (-1)^{\sigma_1 \sigma_9 + \sigma_1 \sigma_2 + \sigma_2 \sigma_3} \bigg] \bigg[\sum_{\sigma_4, \sigma_5} (-1)^{\sigma_4 \sigma_3 + \sigma_4 \sigma_5 + \sigma_5 \sigma_6} \bigg] \bigg[\sum_{\sigma_7, \sigma_8} (-1)^{\sigma_7 \sigma_6 + \sigma_7 \sigma_8 + \sigma_8 \sigma_9} \bigg] \\
        &= \sum_{\v{\sigma}} f(\v{\sigma}) \bigg[\sqrt{2} (-1)^{\sigma_3 \sigma_9} \bigg] \bigg[\sqrt{2} (-1)^{\sigma_6 \sigma_3} \bigg] \bigg[\sqrt{2} (-1)^{\sigma_6 \sigma_9} \bigg] = \tr \bigg(\prod_{e \in \trianglesymbol} H_e \bigg)
    \end{split}
\end{equation}
where $\v{\sigma}$ is the vector of spins not including $\sigma_1, \sigma_2, \sigma_4, \sigma_5, \sigma_7, \sigma_8$ and $f(\v{\sigma})$ is the network of Hadamard operators not contracted over. Here $\trianglesymbol$ is the tensor network Fig.~\ref{fig:borromean_config}(c). Now we can contract over the spins $\sigma_3, \sigma_6, \sigma_9$ as
\begin{equation}
    \begin{split}
        \tr \bigg(\prod_{e \in \trianglesymbol} H_e \bigg) &= \sum_{\v{\sigma}} g(\v{\sigma}) \sum_{\sigma_3, \sigma_6, \sigma_9} (-1)^{\sigma_6 \sigma_9 + \sigma_3 \sigma_9 + \sigma_3 \sigma_6 + \sigma_9 \sigma_{11} + \sigma_9 \sigma_{10} + \sigma_6 \sigma_{11} + \sigma_6 \sigma_{12} + \sigma_3 \sigma_{10} + \sigma_3 \sigma_{12}} \\
        &= \sum_{\v{\sigma}} g(\v{\sigma}) \bigg[1 + (-1)^{\sigma_4 + \sigma_5} - (-1)^{\sigma_4 + \sigma_5} + (-1)^{\sigma_2 + \sigma_5} - (-1)^{\sigma_2 + \sigma_5} + (-1)^{\sigma_2 + \sigma_5} - (-1)^{\sigma_2 + \sigma_5} - 1] \\
        &= 0
    \end{split}
\end{equation}
where $\v{\sigma}$ is the vector of spins not including $\sigma_3, \sigma_6, \sigma_9$ and $g(\v{\sigma})$ is the network of Hadamard operators not contracted over. Hence it is clear that when $L_\R \cup L_\B \cup L_\G$ contains a Borromean ring configuration, $W_{L_\R, L_\B, L_\G} = 0$. Thus we restrict the sum in $\bra{\beta_\R, \beta_\B, \beta_\G}\beta_\R, \beta_\B, \beta_\G \rangle $ to configurations $\{L_\R, L_\B, L_\G \}$ not containing a Borromean ring. Thus we can write the norm of the deformed wavefunction as
\begin{equation}
    \langle \beta_\R, \beta_\B, \beta_\G | \beta_\R, \beta_\B, \beta_\G \rangle = \mathop{\scalebox{2}{$\sum$}}_{(L_\R, L_\B, L_\G) \ \not\ni \borromean} \bigg(\frac{\tanh(\beta_\R)}{\sqrt{2}} \bigg)^{|L_\R|} \bigg(\frac{\tanh(\beta_\B)}{\sqrt{2}} \bigg)^{|L_\B|} \bigg(\frac{\tanh(\beta_\G)}{\sqrt{2}} \bigg)^{|L_\G|} 2^{C_{L_\R \cup L_\B \cup L_\G}}.
\end{equation}

\subsection{High-temperature expansion of spin Hamiltonian}

In this section, we propose a local spin Hamiltonian whose partition function recovers the loop model derived in the previous section. This Ising-like Hamiltonian is inspired by the Hamiltonian introduced in Sec.~\ref{sec:high_T_expansion}. Consider the Hamiltonian
\begin{equation}
\label{eq:3na_ham}
    H[\beta_\R, \beta_\B, \beta_\G] = - \beta_\R \sum_{\langle b, g \rangle \in \red{\hexagon}} \sigma_b^{\R} \sigma_g^{\R} \CZ_{b g} - \beta_\B \sum_{\langle r, g \rangle \in \blue{\hexagon}} \sigma_r^{\B} \sigma_g^{\B} \CZ_{r g} - \beta_{\G} \sum_{\langle r, b \rangle \in \green{\hexagon}} \sigma_r^{\G} \sigma_b^{\G} \CZ_{r b}.
\end{equation}
Here the honeycomb lattices $\red{\hexagon}, \green{\hexagon}, \blue{\hexagon}$ are A-B-C stacked as in Fig.~\ref{fig:borromean_config}(a). 

We now carry out the exact high-temperature expansion of the Hamiltonian:
\begin{equation}
    \begin{split}
        Z_H(\beta_\R, \beta_\B, \beta_\G) &= \sum_{\{\sigma^{\R}, \sigma^{\B}, \sigma^{\G}, \tilde{\sigma}\}} e^{-H[\beta_\R, \beta_\B, \beta_\G]} \\
        &= \sum_{\{\sigma^{\R}, \sigma^{\G}, \sigma^{\B}, \tilde{\sigma}\}} \bigg( \prod_{\langle b, g \rangle \in \red{\hexagon}} \cosh(\beta_\R) \bigg[1 + \sigma_b^{\R} \sigma_g^{\R} \CZ_{bg}\tanh(\beta_\R) \bigg] \bigg) \bigg( \prod_{\langle g, r \rangle \in \blue{\hexagon}} \cosh(\beta_\B) \bigg[1 + \sigma_g^{\B} \sigma_r^{\B} \CZ_{gr}\tanh(\beta_\B) \bigg] \bigg) \\
        &* \bigg( \prod_{\langle r, b \rangle \in \green{\hexagon}} \cosh(\beta_\G) \bigg[1 + \sigma_r^{\G} \sigma_b^{\G} \CZ_{r b}\tanh(\beta_\G) \bigg] \bigg)\\
        &\propto \sum_{L_\R, L_\B, L_\G} \tanh(\beta_\R)^{|L_\R|} \tanh(\beta_\B)^{|L_\B|} \tanh(\beta_\G)^{|L_\G|} \sum_{\{\tilde{\sigma}\}} \bigg(\prod_{\langle b, g \rangle \in L_\R} \CZ_{bg} \prod_{\langle g, r \rangle \in L_\B} \CZ_{gr} \prod_{\langle r, b \rangle \in L_\G} \CZ_{rb} \bigg) \\
        &= \sum_{L_\R, L_\B, L_\G} \tanh(\beta_\R)^{|L_\R|} \tanh(\beta_\B)^{|L_\B|} \tanh(\beta_\G)^{|L_\G|} W_{L_\R, L_\B, L_\G}
    \end{split}
\end{equation}
where $W_{L_\R, L_\B, L_\G}$ is the ``topological weight'' defined in Eq.~\eqref{eq:3na_topo_weight}. Thus the partition function evaluates to
\begin{equation}
    Z_H(\beta_\R, \beta_\B, \beta_\G) \propto \mathop{\scalebox{2}{$\sum$}}_{(L_\R, L_\B, L_\G) \ \not\ni \borromean} \bigg(\frac{\tanh(\beta_\R)}{\sqrt{2}} \bigg)^{|L_\R|} \bigg(\frac{\tanh(\beta_\B)}{\sqrt{2}} \bigg)^{|L_\B|} \bigg(\frac{\tanh(\beta_\G)}{\sqrt{2}} \bigg)^{|L_\G|} 2^{C_{L_\R \cup L_\B \cup L_\G}},
\end{equation}
which is exactly the result we obtain for $ \langle \beta_\R, \beta_\B, \beta_\G | \beta_\R, \beta_\B, \beta_\G \rangle$.

\subsection{Symmetry group $G_{m_R,m_B,m_G}$}
\label{sec:symmetry_group_3na}

The symmetry group $G_{m_R, m_B, m_G}$ of the Ising-like Hamiltonian in Eq.~\eqref{eq:3na_ham} satisfies $G_{m_R, m_B, m_G} \cong \Z_2^3 \rtimes D_4 \cong (\Z_2^2 \times D_4) \rtimes \Z_2$, which can be written in terms of the generators as 
\begin{equation} \label{eq:G_mRmBmG}
    G_{m_R, m_B, m_G} \cong \bigg ( \langle V_R \rangle \times \langle V_B \rangle \times \langle T_g \rangle \bigg) \rtimes \bigg( [\langle T_r \rangle \times \langle V_G \rangle] \rtimes \langle T_b \rangle \bigg) \cong \bigg (\langle V_R \rangle \times \langle V_B \rangle \times \bigg[(\langle T_r \rangle \times \langle V_G \rangle) \rtimes \langle T_b\rangle \bigg] \bigg) \rtimes \langle T_g \rangle.
\end{equation}
The presentation of this group is given by
\begin{equation}
G_{m_R, m_B, m_G} \;=\; \left\langle\, V_R, V_B, V_G, T_r, T_b, T_g \;\middle|\;
\begin{gathered}
V_\beta^2 = T_\alpha^2 = e, \quad T_\alpha V_\beta T_\alpha = V_\beta\\
V_R = [T_b, T_g], \quad V_B = [T_g, T_r] \\
V_G = [T_r, T_b]
\end{gathered}
\right\rangle
\end{equation}
where $[g, h] = g h g^{-1} h^{-1}$ and $\alpha = r, b, g$ and $\beta = R, B, G$. Despite the presentation including 6 generators, note that only 3 of them are independent. The group center is given by $Z(G_{m_R, m_B, m_G}) \cong \langle V_R \rangle \times \langle V_B \rangle \times \langle V_G \rangle \cong \Z_2^3$.

The action of the generators of $G_{m_R, m_B, m_G}$ on the Ising variables in Eq.~\eqref{eq:3na_ham} is given in Table~\ref{tab:mR_mB_mG_symmetries}.
\begin{table}[htbp]
\centering
\renewcommand{\arraystretch}{1.25}
\setlength{\tabcolsep}{4pt}
\begin{tabular}{c|ccccccccc}
 $G_{m_R, m_B, m_G}$ & $\widetilde{\sigma}_r$ & $\widetilde{\sigma}_b$ & $\widetilde{\sigma}_g$ & $\sigma_b^{\R}$ & $\sigma_g^{\R}$ & $\sigma_g^{\B}$ & $\sigma_r^{\B}$ & $\sigma_r^{\G}$ & $\sigma_b^{\G}$ \\ \hline
$V_R$ & & & & $-1$ & $-1$ & & & & \\
$V_B$ & & & & & & $-1$ & $-1$ & & \\
$V_G$ & & & & & & & & $-1$ & $-1$ \\
$T_r$ & $-1$ & & & & & $\widetilde{\sigma}_g$ & & & $\widetilde{\sigma}_b$ \\
$T_b$ & & $-1$ & & & $\widetilde{\sigma}_g$ & & & $\widetilde{\sigma}_r$ & \\
$T_g$ & & & $-1$ & $\widetilde{\sigma}_b$ & & & $\widetilde{\sigma}_r$ & & \\
\end{tabular}
\caption{Symmetry generators of $H[\beta_\R, \beta_\B, \beta_\G]$. Table entries denote multiplication. Blank entries are to be interpreted as $+1$, i.e., not transforming.}
\label{tab:mR_mB_mG_symmetries}
\end{table}

When $\beta_\R = \beta_\B$, the symmetry is enlarged and includes an extra swap factor ($S_{RB}:R \leftrightarrow B$), which exchanges $R$ and $B$ degrees of freedom, such that the symmetry $\langle V_R \rangle \times \langle V_B\rangle$ is enlarged to $(\langle V_R \rangle \times \langle V_B \rangle ) \rtimes S_{RB}$. An analogous swap symmetry emerges when $\beta_\R = \beta_\G$ or $\beta_\B = \beta_\G$ as well. When $\beta_\R = \beta_\B = \beta_\G$, there is an additional permutation $S_3$ symmetry of the three colors such that $G_{m_R, m_B, m_G} \rightarrow G_{m_R, m_B, m_G} \rtimes S_3$. The character table of $G_{m_R, m_B, m_G}$ is given in Table~\ref{tab:char_tab_GRBG} with the corresponding conjugacy class representatives in Table~\ref{tab:class_reps_GRBG}.

\begin{table}[htbp]
\centering
\renewcommand{\arraystretch}{1.2}
\setlength{\tabcolsep}{2.5pt}
\begin{tabular}{c c|cccccccc|cccccccccccccc}
\hline
 &  & $C_{1}$ & $C_{2}$ & $C_{3}$ & $C_{4}$ & $C_{5}$ & $C_{6}$ & $C_{7}$ & $C_{8}$
 & $C_{9}$ & $C_{10}$ & $C_{11}$ & $C_{12}$ & $C_{13}$ & $C_{14}$
 & $C_{15}$ & $C_{16}$ & $C_{17}$ & $C_{18}$ & $C_{19}$ & $C_{20}$
 & $C_{21}$ & $C_{22}$ \\

 &  & $1$ & $1$ & $1$ & $1$ & $1$ & $1$ & $1$ & $1$
 & $4$ & $4$ & $4$ & $4$ & $4$ & $4$
 & $4$ & $4$ & $4$ & $4$ & $4$ & $4$
 & $4$ & $4$ \\

\hline
$\chi_1$    & $1$ & 1 & 1 & 1 & 1 & 1 & 1 & 1 & 1 & 1 & 1 & 1 & 1 & 1 & 1 & 1 & 1 & 1 & 1 & 1 & 1 & 1 & 1 \\
$\chi_2$    & $1$ & 1 & 1 & 1 & 1 & 1 & 1 & 1 & 1 & $-1$ & $-1$ & 1 & 1 & 1 & 1 & $-1$ & $-1$ & $-1$ & $-1$ & 1 & 1 & $-1$ & $-1$ \\
$\chi_3$    & $1$ & 1 & 1 & 1 & 1 & 1 & 1 & 1 & 1 & 1 & 1 & $-1$ & $-1$ & 1 & 1 & $-1$ & $-1$ & 1 & 1 & $-1$ & $-1$ & $-1$ & $-1$ \\
$\chi_4$    & $1$ & 1 & 1 & 1 & 1 & 1 & 1 & 1 & 1 & $-1$ & $-1$ & $-1$ & $-1$ & 1 & 1 & 1 & 1 & $-1$ & $-1$ & $-1$ & $-1$ & 1 & 1 \\
$\chi_5$    & $1$ & 1 & 1 & 1 & 1 & 1 & 1 & 1 & 1 & 1 & 1 & 1 & 1 & $-1$ & $-1$ & 1 & 1 & $-1$ & $-1$ & $-1$ & $-1$ & $-1$ & $-1$ \\
$\chi_6$    & $1$ & 1 & 1 & 1 & 1 & 1 & 1 & 1 & 1 & $-1$ & $-1$ & 1 & 1 & $-1$ & $-1$ & $-1$ & $-1$ & 1 & 1 & $-1$ & $-1$ & 1 & 1 \\
$\chi_7$    & $1$ & 1 & 1 & 1 & 1 & 1 & 1 & 1 & 1 & 1 & 1 & $-1$ & $-1$ & $-1$ & $-1$ & $-1$ & $-1$ & $-1$ & $-1$ & 1 & 1 & 1 & 1 \\
$\chi_8$    & $1$ & 1 & 1 & 1 & 1 & 1 & 1 & 1 & 1 & $-1$ & $-1$ & $-1$ & $-1$ & $-1$ & $-1$ & 1 & 1 & 1 & 1 & 1 & 1 & $-1$ & $-1$ \\
\hline
$\chi_9$    & $2$ & 2 & $-2$ & 2 & 2 & $-2$ & $-2$ & 2 & $-2$ & 0 & 0 & 0 & 0 & 2 & $-2$ & 0 & 0 & 0 & 0 & 0 & 0 & 0 & 0 \\
$\chi_{10}$ & $2$ & 2 & $-2$ & 2 & 2 & $-2$ & $-2$ & 2 & $-2$ & 0 & 0 & 0 & 0 & $-2$ & 2 & 0 & 0 & 0 & 0 & 0 & 0 & 0 & 0 \\
$\chi_{11}$ & $2$ & 2 & 2 & $-2$ & 2 & $-2$ & 2 & $-2$ & $-2$ & 0 & 0 & 2 & $-2$ & 0 & 0 & 0 & 0 & 0 & 0 & 0 & 0 & 0 & 0 \\
$\chi_{12}$ & $2$ & 2 & 2 & $-2$ & 2 & $-2$ & 2 & $-2$ & $-2$ & 0 & 0 & $-2$ & 2 & 0 & 0 & 0 & 0 & 0 & 0 & 0 & 0 & 0 & 0 \\
$\chi_{13}$ & $2$ & 2 & $-2$ & $-2$ & 2 & 2 & $-2$ & $-2$ & 2 & 0 & 0 & 0 & 0 & 0 & 0 & 0 & 0 & 0 & 0 & 2 & $-2$ & 0 & 0 \\
$\chi_{14}$ & $2$ & 2 & $-2$ & $-2$ & 2 & 2 & $-2$ & $-2$ & 2 & 0 & 0 & 0 & 0 & 0 & 0 & 0 & 0 & 0 & 0 & $-2$ & 2 & 0 & 0 \\
$\chi_{15}$ & $2$ & 2 & 2 & 2 & $-2$ & 2 & $-2$ & $-2$ & $-2$ & 2 & $-2$ & 0 & 0 & 0 & 0 & 0 & 0 & 0 & 0 & 0 & 0 & 0 & 0 \\
$\chi_{16}$ & $2$ & 2 & 2 & 2 & $-2$ & 2 & $-2$ & $-2$ & $-2$ & $-2$ & 2 & 0 & 0 & 0 & 0 & 0 & 0 & 0 & 0 & 0 & 0 & 0 & 0 \\
$\chi_{17}$ & $2$ & 2 & $-2$ & 2 & $-2$ & $-2$ & 2 & $-2$ & 2 & 0 & 0 & 0 & 0 & 0 & 0 & 0 & 0 & 2 & $-2$ & 0 & 0 & 0 & 0 \\
$\chi_{18}$ & $2$ & 2 & $-2$ & 2 & $-2$ & $-2$ & 2 & $-2$ & 2 & 0 & 0 & 0 & 0 & 0 & 0 & 0 & 0 & $-2$ & 2 & 0 & 0 & 0 & 0 \\
$\chi_{19}$ & $2$ & 2 & 2 & $-2$ & $-2$ & $-2$ & $-2$ & 2 & 2 & 0 & 0 & 0 & 0 & 0 & 0 & 2 & $-2$ & 0 & 0 & 0 & 0 & 0 & 0 \\
$\chi_{20}$ & $2$ & 2 & 2 & $-2$ & $-2$ & $-2$ & $-2$ & 2 & 2 & 0 & 0 & 0 & 0 & 0 & 0 & $-2$ & 2 & 0 & 0 & 0 & 0 & 0 & 0 \\
$\chi_{21}$ & $2$ & 2 & $-2$ & $-2$ & $-2$ & 2 & 2 & 2 & $-2$ & 0 & 0 & 0 & 0 & 0 & 0 & 0 & 0 & 0 & 0 & 0 & 0 & $2i$ & $-2i$ \\
$\chi_{22}$ & $2$ & 2 & $-2$ & $-2$ & $-2$ & 2 & 2 & 2 & $-2$ & 0 & 0 & 0 & 0 & 0 & 0 & 0 & 0 & 0 & 0 & 0 & 0 & $-2i$ & $2i$ \\
\end{tabular}
\caption{Character table of $G_{m_R, m_B, m_G} \cong \Z_2^3 \rtimes D_4$. The second column lists the irrep dimensions $\dim(\chi_j)$, and the second header row lists conjugacy class sizes $|C_j|$. Classes $C_1$--$C_8$ comprise the center $Z(G_{m_R,m_B,m_G}) = \langle V_R \rangle \times \langle V_B \rangle \times \langle V_G \rangle$; each non-center coset of $Z(G_{m_R,m_B,m_G})$ splits into exactly two conjugacy classes of size 4.}
\label{tab:char_tab_GRBG}
\end{table}

\begin{table}[t]
\centering
\renewcommand{\arraystretch}{1.2}
\setlength{\tabcolsep}{6pt}
\begin{tabular}{cl@{\quad}cl@{\quad}cl}
\hline
class & rep. & class & rep. & class & rep. \\
\hline
$C_{1}$ & $e$         & $C_{9}$  & $T_r$    & $C_{16}$ & $T_rT_gV_R$    \\
$C_{2}$ & $V_B$       & $C_{10}$ & $T_rV_R$ & $C_{17}$ & $T_rT_b$       \\
$C_{3}$ & $V_G$       & $C_{11}$ & $T_g$    & $C_{18}$ & $T_rT_bV_R$    \\
$C_{4}$ & $V_R$       & $C_{12}$ & $T_gV_G$ & $C_{19}$ & $T_gT_b$       \\
$C_{5}$ & $V_BV_G$    & $C_{13}$ & $T_b$    & $C_{20}$ & $T_gT_bV_B$    \\
$C_{6}$ & $V_RV_B$    & $C_{14}$ & $T_bV_B$ & $C_{21}$ & $T_rT_gT_b$    \\
$C_{7}$ & $V_RV_G$    & $C_{15}$ & $T_rT_g$ & $C_{22}$ & $T_rT_gT_bV_R$ \\
$C_{8}$ & $V_RV_BV_G$ &          &          &          &                \\
\hline\hline
\end{tabular}
\caption{Conjugacy class representatives for the character table of
$G_{m_R,m_B,m_G}$.}
\label{tab:class_reps_GRBG}
\end{table}

The group $G_{m_R, m_B, m_G}$ is a class-2 nilpotent since there exist finite Abelian groups $N, Q$ such that the extension $1 \longrightarrow N \longrightarrow G_{m_R, m_B, m_G} \longrightarrow Q \longrightarrow 1$ is central. Here $N = \langle V_R \rangle \times \langle V_B \rangle \times \langle V_G \rangle \cong Z(G_{m_R, m_B, m_G}) \cong \Z_2^3$ and $Q = G_{m_R, m_B, m_G}/N \cong \Z_2^3$. 

Analogous to the group $G_{m_R, m_B}$, the 22 irreducible representations of $G_{m_R, m_B, m_G}$ can be related to the 22 anyons of twisted $\Z_2^3$ quantum double $D^\omega(\Z_2^3)$ or equivalently, those of the $D_4$ quantum double $D(D_4)$ (see Ref.~\cite{Iqbal_24} for the 1-to-1 mapping among anyons in the two formulations). As before, this identification is not unique since the 2-dimensional fermions and bosons of the same color $f_\alpha, m_\alpha$ can have their corresponding irreps interchanged given that these anyons possess identical fusion descendants and appear with the same multiplicity in all fusion products (see Table~\ref{tab:D4_fusion_rules}). Moreover, irreps $21$ and $22$, unlike the semions $\bar{s}_{RGB}$ and $s_{RGB}$, are not self-dual, i.e., while $\mathbf{1} \in s_{RGB} \otimes s_{RGB}$, $\chi_1 \notin \rho_{21} \otimes \rho_{21}$ and $\chi_1 \notin \rho_{22} \otimes \rho_{22}$ (although $\chi_1 \in \rho_{21} \otimes \rho_{22}$). Hence, the conventional tensor product on $\mathrm{Rep}(G_{m_R, m_B, m_G})$ cannot reproduce semion fusion. For the remaining anyons, we provide an anyon-to-irrep correspondence in Table~\ref{tab:irrep_anyon_map}.  A convention is typically fixed by mapping the fidelity correlators in Eq.~\eqref{eq:fid_corr} to the 2-point correlators in Eq.~\eqref{eq:anyon_magnetization_definitions}. In implementing such a mapping, we demand that order parameters transforming in a certain irrep of the symmetry group be identified with the anyon corresponding to the relevant fidelity correlator. 

For reference, $G_{m_R, m_B, m_G}$ is indexed by the GAP~\cite{GAP4} ID $\mathrm{SmallGroup}(64, 73)$.

\begin{table}[htbp]
\centering
\resizebox{\textwidth}{!}{
\begin{tabular}{c|*{22}{c}|}
\hline
irrep & $\rho_1$ & $\rho_2$ & $\rho_3$ & $\rho_4$ & $\rho_5$ & $\rho_6$ & $\rho_7$ & $\rho_8$ & $\rho_9$ & $\rho_{10}$ & $\rho_{11}$ & $\rho_{12}$ & $\rho_{13}$ & $\rho_{14}$ & $\rho_{15}$ & $\rho_{16}$ & $\rho_{17}$ & $\rho_{18}$ & $\rho_{19}$ & $\rho_{20}$ \\
\hline
anyon & $\mathbf{1}$ & $e_R$ & $e_G$ & $e_{RG}$ & $e_B$ & $e_{RB}$ & $e_{GB}$ & $e_{RGB}$ & $m_B$ & $f_B$ & $m_G$ & $f_G$ & $m_{GB}$ & $f_{GB}$ & $m_R$ & $f_R$ & $m_{RB}$ & $f_{RB}$ & $m_{RG}$ & $f_{RG}$ \\
\hline
\end{tabular}
}
\caption{Identification of $20$ out of $22$ elements of $\mathrm{Irr}(G_{m_R, m_B, m_G})$ with the non-semionic anyons of $D(D_4)$.}
\label{tab:irrep_anyon_map}
\end{table}

\section{Three non-Abelian errors plus three Abelian errors}
\label{sec:3_NA_A_anyons}

\subsection{Loop model}

\begin{figure}
    \centering
    \includegraphics[width=\linewidth]{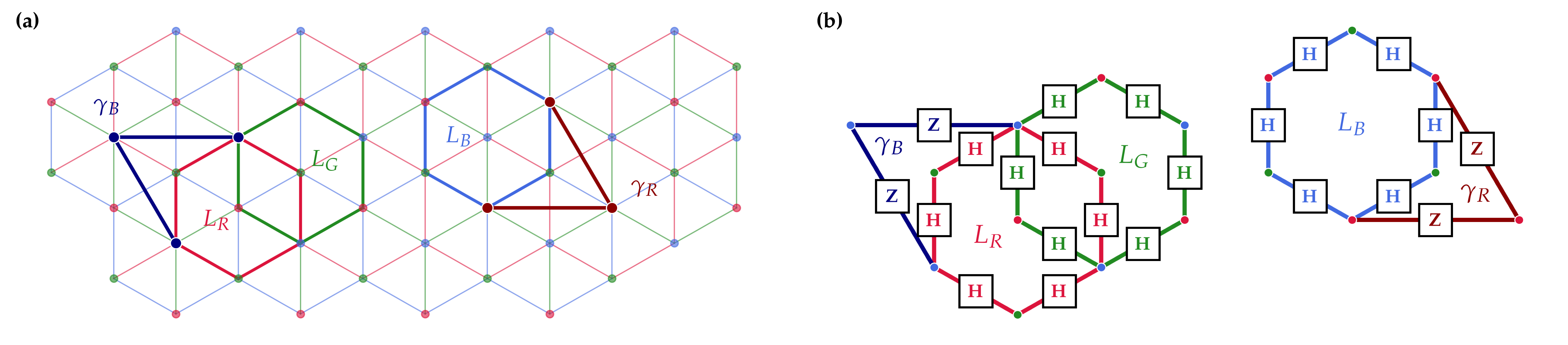}
    \caption{\textbf{Abelian and non-Abelian loop network.} Here $L_\alpha$ indicate closed loops of non-Abelian anyon worldlines, and $\gamma_\mu$ signify strings of Abelian anyon worldlines for $\{m_R, m_B, m_G, e_R, e_B, e_G\}$-nucleating deformations. (a) Example loop and string configuration in $Z(\v{\beta}, \v{\beta}^z)$; in this case $\gamma_\G$ is trivial. (b) Tensor-network whose contraction gives weight $W_{\v{L}, \v{\gamma}}$ associated with configuration $\v{L}, \v{\gamma}$ in (a).}
    \label{fig:general_loops}
\end{figure}

We can evaluate the norm of $\ket{\v{\beta}, \v{\beta}^z}$ resulting from proliferating the anyons $m_R, m_B, m_G, e_R, e_B, e_G$ using similar techniques as in previous sections. Here $\v{\beta} = (\beta_\R, \beta_\B, \beta_\G)$ and $\v{\beta}^z = (\beta_{\R}^z, \beta_{\B}^z, \beta_{\G}^z)$. When evaluating the norm of $\ket{\v{\beta}, \v{\beta}^z}$, there are 3 non-trivial observations that must be kept in mind. First, as in $Z(\beta_{\R}, \beta_{\B}, \beta_{\G})$ in App.~\ref{sec:3_NA_anyons}, the topological weight of the loop configuration will be given by $2^{C_{L_\R \cup L_\B \cup L_\G}}$, which couples together the loops of non-Abelian anyon wordlines. Second, as in $Z(\beta_\R, \beta_\B, \beta_{\G}^z)$ in App.~\ref{sec:mR_mB_eG}, strings of Abelian charge worldlines can terminate on loops of non-Abelian anyon worldlines so long as the Abelian charge is contained within the self-fusion channel of the non-Abelian anyon. Third, the anticommutation of $X$ and $Z$ deformations on the same sublattice forces that loops of $X^\alpha$ and $Z^\alpha$ operators are constrained to not cross. This is a consequence of the fermionic exchange statistics between $e_\alpha$ and $m_\alpha$ anyons. As a consequence of these factors, we find that 
\begin{equation}
\begin{split}
    \langle \v{\beta}, \v{\beta}^z | \v{\beta}, \v{\beta}^z \rangle 
    \propto \sum_{\v{L}} t_{\R}^{|L_\R|} t_{\B}^{|L_\B|} t_{\G}^{|L_\G|} \sum_{\v{\gamma}} {t_{\R}^z}^{|\gamma_{\R}|} {t_{\B}^z}^{|\gamma_{\B}|} {t_{\G}^z}^{|\gamma_{\G}|} W_{\v{L}, \v{\gamma}}
    \end{split}
\end{equation}
where $\v{L} = (L_\R, L_\B, L_\G)$ are closed loops, $\v{\gamma} = (\gamma_\R, \gamma_\B, \gamma_\G)$ are constrained string configurations, and $t_\alpha = \tanh(\beta_\alpha)$ and $t_\alpha^z = \tanh(\beta_\alpha^z) / \cosh(\beta_\alpha)$. These sums are constrained such that $L_\alpha \cap \gamma_\alpha = \varnothing$, and $\gamma_\R$ lines can end on the global $L_\B \cup L_\G$ configuration (analogous constraints hold on $\gamma_\B, \gamma_\G$). An example configuration is given in Fig.~\ref{fig:general_loops}(a). The origin of the suppression factor $1/\cosh(\beta_\alpha)$ within the tension $t_\alpha^z$ is a consequence of the anti-commutation between $X^\alpha_j$ and $Z^\alpha_j$ deformations, and the choice of first applying $X$ deformations and then $Z$ deformations (this suppression factor is absent when instead considering local deformations $e^{\beta_\alpha X_j^\alpha + \beta_\alpha^z Z_j^\alpha}$). In the following calculation, $\gamma_\alpha$ are closed loop configurations, as $\gamma_\alpha$ configurations cannot end on loops $L_\alpha$ of the same color:
\begin{equation}
\label{eq:Z_e_m_def}
\begin{split}
Z(\beta_\alpha, \beta_\alpha^z) &= \langle \Dfour | e^{\frac{\beta_\alpha}{2}\sum_j X_j^\alpha} e^{\beta_\alpha^z \sum_i Z_i^\alpha} e^{\frac{\beta_\alpha}{2}\sum_j X_j^\alpha} | \Dfour \rangle \\
&= \cosh(\beta_\alpha^z)^V \sum_{\gamma_\alpha} \tanh(\beta_\alpha^z)^{|\gamma_\alpha|} \langle \Dfour | e^{\frac{\beta_\alpha}{2}\sum_j X_j^\alpha} \prod_{j\in \gamma_\alpha} Z_j^\alpha e^{\frac{\beta_\alpha}{2}\sum_j X_j^\alpha} | \Dfour \rangle \\ 
&= \cosh(\beta_\alpha^z)^V \sum_{\gamma_\alpha} \tanh(\beta_\alpha^z)^{|\gamma_\alpha|} \langle \Dfour | e^{\beta_\alpha \sum_{j\notin \gamma_\alpha} X_j^\alpha} \prod_{j\in \gamma_\alpha} Z_j^\alpha | \Dfour \rangle \\
&= \cosh(\beta_\alpha^z)^V \sum_{\gamma_\alpha} \tanh(\beta_\alpha^z)^{|\gamma_\alpha|} \cosh(\beta_\alpha)^{V-|\gamma_\alpha|} \sum_{L_\alpha \cap \gamma_\alpha=\varnothing} \tanh(\beta_\alpha)^{|L_\alpha|} \times \langle \Dfour | \prod_{j\in \gamma_\alpha} Z_j^\alpha \prod_{i\in L_\alpha} X_i^\alpha | \Dfour \rangle \\
&= \bigl[\cosh(\beta_\alpha^z)\cosh(\beta_\alpha)\bigr]^V \sum_{\substack{\gamma_\alpha,L_\alpha:\\ \gamma_\alpha \cap L_\alpha=\varnothing}} \tanh(\beta_\alpha)^{|L_\alpha|} \left( \frac{\tanh(\beta_\alpha^z)}{\cosh(\beta_\alpha)} \right)^{|\gamma_\alpha|} \frac{2^{C_{L_\alpha}}}{\sqrt{2}^{\,|L_\alpha|}}.
\end{split}
\end{equation}

All together, the topological weight $W_{\v{L}, \v{\gamma}}$ is then given by
\begin{equation}
\label{eq:topo_weight_def}
\begin{split}
    W_{\v{L}, \v{\gamma}} &= \bra{D_4} \prod_{r \in L_\R} X_r^{\R} \prod_{b \in L_\B} X_b^{\B} \prod_{g \in L_\G} X_g^{\G} \prod_{r' \in \gamma_\R} Z_{r'}^{\R} \prod_{b' \in \gamma_\B} Z_{b'}^{\B} \prod_{g' \in \gamma_\G} Z_{g'}^{\G}
    \ket{D_4} \\
    &= \bra{\mathrm{SPT}} \prod_{\langle b,g \rangle \in L_{\R}} \CZ_{b g} \prod_{\langle r, g \rangle \in L_\B} \CZ_{r g} \prod_{\langle r, b \rangle \in L_\G} \CZ_{r, b} \prod_{\langle r, r' \rangle \in \gamma_\R} Z_{r}^{\R} Z_{r'}^{\R} \prod_{\langle b, b' \rangle \in \gamma_\B} Z_{b}^{\B} Z_{b'}^{\B} \prod_{\langle g, g' \rangle \in \gamma_\G} Z_{g}^{\G} Z_{g'}^{\G} \ket{\mathrm{SPT}}.
\end{split}
\end{equation}
In the second line, we used the $\ket{D_4} \rightarrow \ket{\mathrm{SPT}}$ mapping mentioned in Ref.~\cite{Yoshida_2016, sala_D4}. As in App.~\ref{sec:3_NA_anyons}, we then apply the disentangling circuit $U = \prod_{\langle r, g, b \rangle \in \triangle} CCZ_{r g b}$ where $\langle r, g, b \rangle \in \triangle$ are nearest-neighbor sites forming a triangle. Under this circuit, $U\ket{ \mathrm{SPT}} \rightarrow \ket{+^\R} \ket{+^\B} \ket{+^\G}$. 

Since $\CZ$ is diagonal in the $Z$-basis, we can rewrite the expectation value $W_{\v{L}, \v{\gamma}}$ as a contraction over a tensor-network of $H$ (Hadamard) and $Z$ gates (see App.~\ref{sec:3_NA_anyons}). An example of this tensor-network is given in Fig.~\ref{fig:general_loops}(b). When evaluating this contraction, it is important to note that Abelian strings $\gamma_\alpha$ can end on non-Abelian loops $L_\alpha$. As a result, the contraction will contain $Z$ insertions between Hadamard gates, with non-vanishing contributions if and only if $\gamma_\R$ intersects each connected component of $L_\B \cup L_\G$ an even number of times (with analogous constraints for $\gamma_\B, \gamma_\G$)~\cite{sala_D4}. Following the analyses in App. C.2.a of Ref.~\cite{sala_D4} and App.~\ref{sec:3_NA_anyons}, we evaluate the topological weight as
\begin{equation}
\begin{split}
    W_{\v{L}, \v{\gamma}} 
    &= \tr \bigg(\prod_{\langle b,g \rangle \in L_{\R}} \CZ_{b g} \prod_{\langle r, g \rangle \in L_\B} \CZ_{r g} \prod_{\langle r, b \rangle \in L_\G} \CZ_{r, b} \prod_{\langle r, r' \rangle \in \gamma_\R} \widetilde{Z}_{r}^{\R} \widetilde{Z}_{r'}^{\R} \prod_{\langle b, b' \rangle \in \gamma_\B} \widetilde{Z}_{b}^{\B} \widetilde{Z}_{b'}^{\B} \prod_{\langle g, g' \rangle \in \gamma_\G} \widetilde{Z}_{g}^{\G} \widetilde{Z}_{g'}^{\G} \bigg ) \\
    &= \frac{2^{C_{L_\R \cup L_\B \cup L_\G}}}{\sqrt{2}^{|L_\R| + |L_\B| + |L_\G|}} \sigma_{\v{L}}(\v{\gamma}).
\end{split}
\end{equation}
Here the sign $\sigma_{\v{L}}(\v{\gamma})$ is simply defined as the sign of the topological weight: $\sigma_{\v{L}}(\v{\gamma}) = \mathrm{sign}(W_{\v{L}, \v{\gamma}})$. As noted in Ref.~\cite{sala_D4} and as we will later show, a closed-form expression for $\sigma_{\v{L}}(\v{\gamma})$ is unnecessary for a characterization of the phase diagram of the deformed wavefunction. Thus the loop model characterizing the norm of the deformed wavefunction is given by
\begin{equation}
\label{eq:general_loop_model}
\begin{split}
    Z(\v{\beta}, \v{\beta}^z) = \langle \v{\beta}, \v{\beta}^z | \v{\beta}, \v{\beta}^z \rangle &= \sum_{\v{L}} \bigg(\frac{t_\R}{\sqrt{2}} \bigg)^{|L_\R|} \bigg(\frac{t_\B}{\sqrt{2}} \bigg)^{|L_\B|} \bigg(\frac{t_\G}{\sqrt{2}} \bigg)^{|L_\G|} 2^{C_{L_\R \cup L_\B \cup L_\G}} \bigg[\sum_{\v{\gamma}} {t_{\R}^z}^{|\gamma_\R|} {t_{\B}^z}^{|\gamma_\B|} {t_{\G}^z}^{|\gamma_\G|}  \sigma_{\v{L}}(\v{\gamma}) \bigg].
\end{split}
\end{equation}
Here $L_\alpha \cap \gamma_\alpha = \varnothing$, since loops of the same color cannot cross. Additionally, the configurations $\gamma_\R$ can have endpoints on $L_\B \cup L_\G$, $\gamma_\B$ can have endpoints on $L_\G \cup L_\R$, and $\gamma_\G$ can have endpoints on $L_\R \cup L_\B$.

\subsection{Local spin formulation}

We now propose a stat-mech model, a set of local weights, whose partition function recovers the loop model $Z(\v{\beta}, \v{\beta}^z)$ in Eq.~\eqref{eq:general_loop_model}. From the analysis in App.~\ref{sec:high_T_expansion} and App.~\ref{sec:3_NA_anyons}, we know that the $\CZ$ decoration on the nearest-neighbor Ising spins in the stat-mech model can recover the appropriate topological weight. Moreover, from the analysis in App.~\ref{sec:mR_mB_eG} we know that using the same $\widetilde{\sigma}$ degrees of freedom for $\beta_\alpha$ and $\beta_\alpha^{z}$ deformations reproduces the feature that Abelian anyon worldlines $\gamma_\mu$ can end on certain non-Abelian anyon closed loops $L_\nu$. Now we prove that using the following stat-mech model, we can recover the no-crossing conditions between loops $L_\alpha$ and strings $\gamma_\alpha$ of the same color:
\begin{equation}
\label{eq:Z_e_m_stat_mech_model}
    \begin{split}
        Z(\beta_\alpha, \beta_\alpha^z) &= \bigl[\cosh(\beta_\alpha)\cosh(\beta_\alpha^z)\bigr]^V \sum_{ \{\sigma^\alpha, \tsigma \} } \prod_{\langle i, j\rangle \in \hexagon_\alpha} \left[1 + t_\alpha \,\sigma_i^\alpha \sigma_j^\alpha \CZ_{ij} + t_\alpha^z \,\tsigma_{i'} \tsigma_{j'} \right] \\
        &= \bigl[\cosh(\beta_\alpha)\cosh(\beta_\alpha^z)\bigr]^V \sum_{\substack{\gamma_\alpha,L_\alpha:\\ \gamma_\alpha \cap L_\alpha=\varnothing}} t_\alpha^{|L_\alpha|} {t_\alpha^z}^{|\gamma_\alpha|} \frac{2^{C_{L_\alpha}}}{\sqrt{2}^{\,|L_\alpha|}} = \langle \beta_\alpha, \beta_\alpha^z | \beta_\alpha, \beta_\alpha^z \rangle.
    \end{split}
\end{equation}
Here $t_\alpha = \tanh(\beta_\alpha)$, $t_\alpha^z = \tanh(\beta_\alpha^z) / \cosh(\beta_\alpha)$, $\langle i, j \rangle$ is a bond on the honeycomb lattice, and $\langle i', j' \rangle$ is the bond on the dual triangular lattice intersecting $\langle i, j \rangle$. Notice that the correspondence between bonds $\langle i, j \rangle \leftrightarrow \langle i', j' \rangle$ is one-to-one with periodic boundary conditions.  From the form of the local weight, we immediately note that each bond $\langle i, j \rangle$ can only carry either a $t_\alpha$ factor or a $t_\alpha^z$ factor, which enforces the no-crossing condition. Moreover, the suppression factor found in $t_\alpha^z$ within the loop $Z(\v{\beta}, \v{\beta}^z)$ is already accounted for in this definition of $t_\alpha^z$. Thus we have reproduced the no-crossing constraint of the loop model $Z(\beta_\alpha, \beta_\alpha^z)$ in Eq.~\eqref{eq:Z_e_m_def} using the stat-mech model in Eq.~\eqref{eq:Z_e_m_stat_mech_model}. We can combine all parts of the stat-mech model to exactly reproduce the loop model $Z(\v{\beta}, \v{\beta}^z)$. This general stat-mech model takes the form
\begin{equation}
\label{eq:generaL_stat_mech_model}
    \begin{split}
        Z(\v{\beta}, \v{\beta^z)} &= \bigg( \prod_{\alpha = \R, \B, \G} \cosh(\beta_\alpha)\cosh(\beta_\alpha^z) \bigg)^V \sum_{\{\sigma^{\R}, \sigma^{\B}, \sigma^{\G}, \tsigma \}} \prod_{\alpha = \R, \B, \G} \prod_{\langle i, j\rangle \in \hexagon_\alpha} \left[1 + t_\alpha \,\sigma_i^\alpha \sigma_j^\alpha \CZ_{ij} + t_\alpha^z \,\tsigma_{i'} \tsigma_{j'} \right].
    \end{split}
\end{equation}
Not only does this stat-mech model exactly reproduce the desired loop model, but it is also manifestly $G_{m_R, m_B, m_G}$-symmetric. Thus we conclude that adding finite $\beta_\alpha^z$ does not change the symmetry of the underlying stat-mech model.

Ideally we would like to identify a parameter regime for $t_\alpha, t_\alpha^z$ in which the stat-mech model can be written as the partition function of some $G_{m_R, m_B, m_G}$-symmetric local spin Hamiltonian. To this end, we take the following link Hamiltonian ansatz and derive the conditions on the tensions such that the stat-mech model in Eq.~\eqref{eq:generaL_stat_mech_model} can be recovered:
\begin{equation}
H_\mathrm{test}[i, j] = -\bigg[
c_0 + c_1\,\sigma_i\sigma_j\CZ_{ij} + c_2\,\tsigma_{i'}\tsigma_{j'} + c_3\,\sigma_i\sigma_j\CZ_{ij}\tsigma_{i'}\tsigma_{j'} \bigg],
\end{equation}
where $c_k \in \mathbb{R}$ are constants. Defining $x_k = \tanh(c_k)\in [-1,1]$, we can compute the Boltzmann weight associated with $H_\mathrm{test}$ as
\begin{equation}
    \begin{split}
        e^{-H_\mathrm{test}} = e^{c_0} \Biggl[1 + \frac{1}{x_1x_2x_3}(x_1+x_2x_3)\,\sigma_i\sigma_j\CZ_{ij} + \frac{1}{x_1x_2x_3}(x_2+x_1x_3)\,\tsigma_{i'}\tsigma_{j'}+ \frac{1}{x_1x_2x_3}(x_3+x_1x_2)\, \sigma_i\sigma_j\CZ_{ij}\tsigma_{i'}\tsigma_{j'} \Biggr].
    \end{split}
\end{equation}
Thus the consistency equations for the tension mandate that
\begin{equation}
    \begin{split}
    t_\alpha &= \frac{1}{x_1x_2x_3}(x_1+x_2x_3), \qquad t_\alpha = \cosh(\beta_\alpha)\frac{1}{x_1x_2x_3}(x_2+x_1x_3), \qquad x_3 = -x_1x_2.
    \end{split}
\end{equation}
Whenever these conditions are satisfied, there exists a $G_{m_R, m_B, m_G}$-symmetric local spin Hamiltonian corresponding to the stat-mech model in Eq.~\eqref{eq:generaL_stat_mech_model}. Such a Hamiltonian does not exists, for example, when $\beta_\alpha \to\infty$. In this case, $\cosh(\beta_\alpha)\to\infty$ and $t_\alpha \to 1$, so $x_1x_2x_3 = x_1+x_2x_3$ and $x_1x_2x_3 = \cosh(\beta_\alpha)(x_2+x_1x_3) \to \infty$, which is incompatible with $x_k \in [-1,1]$.

\subsection{Distinguishing maximal condensations using symmetry-breaking patterns}
\label{app:distinguishing_maximal_condensation}

\subsubsection{Charge condensate}
Let us begin with the simpler scenario in which all Abelian charges $e$ condense, which can be achieved by taking all $\beta_\alpha^z$ in Eq.~\eqref{eq:generaL_stat_mech_model} sufficiently large. One then finds the symmetry-breaking pattern $G_{m_R,m_B,m_G}\to\mathbb{Z}_2^3$, where the residual symmetry is generated by $\mathbb{Z}_2^3\cong\langle V_R\rangle\times\langle V_B\rangle\times\langle V_G\rangle$. At the level of the quantum wavefunction, this corresponds to a condensation transition between $D_4$ TO and a trivial TO~\cite{d4_hasse}. The remaining symmetry has a natural interpretation: after condensation, un-condensed anyons that can be rotated into each other through fusion with a condensed anyon are identified~\cite{Burnell_2018}. Here this gives $m_\alpha\equiv f_\alpha$ and $s_{RGB}\equiv\bar s_{RGB}$, so the $14$ un-condensed anyons form pairs and reduce to seven sectors with representatives $\{m_R,m_G,m_B,m_{RG},m_{RB},m_{GB},s_{RGB}\}$. Together with the new vacuum, each of these eight sectors is labeled by a one-dimensional irrep of the residual $\mathbb{Z}_2^3$.

The non-Abelian fluxes $m_\alpha$ are themselves rendered Abelian by this condensation. Each flux fuses with itself into the vacuum and three nontrivial Abelian charges, all of which condense: $m_\alpha\times m_\alpha = 1 + e + e' + ee'$. With every charge channel identified with the vacuum, the self-fusion collapses to $m_\alpha\times m_\alpha = 1$; since the anyons of $D(D_4)$ are self-dual, a self-fusion containing only the vacuum forces unit quantum dimension, so $m_\alpha$. We therefore identify $m_R,m_G,m_B$ with the sign irreps of $\langle V_R\rangle,\langle V_G\rangle,\langle V_B\rangle$, while the remaining anyons $m_{RG},m_{RB},m_{GB},s_{RGB}$ carry the four remaining nontrivial one-dimensional irreps.

\subsubsection{Dyon condensate}

The stat-mech model also admits a spontaneous symmetry breaking transition driven by the condensation of $m_R$, $m_B$, and $e_G$, obtained by tuning $\beta_\R,\beta_\B$ while keeping $\beta_\G=\beta_\alpha^z$ small enough. The corresponding symmetry breaking pattern is $G_{m_R,m_B,m_G}\to D_4$, with the residual symmetry generated via $D_4\cong(T_r\times V_G)\rtimes T_b$ as defined in
Sec.~\ref{sec:symmetry_group_3na}, and it again corresponds to a condensation transition between $D_4$ TO and a trivial TO. Here the condensate corresponds to the algebra $\mathcal{L}= 1 + e_G +  m_R +  m_B + m_{RB}$~\cite{d4_hasse} (with $m_{RB}\in m_R\times m_B$ condensing automatically), all of which are identified with the vacuum.

According to Sec.~V.C of Ref.~\cite{Bais09}, $D_\mathcal{A}/D_\mathcal{T} = D_\mathcal{T}/D_\mathcal{U}$, where $D_\mathcal{A}, D_\mathcal{T}, D_\mathcal{U}$ are the total quantum dimensions of the original, broken, and deconfined theories. Here $D_\mathcal{A} = \sqrt{\dim(D(D_4))} = \sqrt{\sum_a d_a^2}=8$, and $D_\mathcal{U} = 1$ since condensing $\mathcal{L}$ yields a topologically trivial phase. Therefore, the broken theory has total quantum dimension $\mathcal{D}_{\mathcal{T}}^2=\sum_i d_{a_i}^2=8$. Hence, either all $8$ condensate sectors have $d_{a_i}=1$ (as in the charge condensate), or there are $4$ non-trivial sectors with only one of them corresponding to a non-Abelian anyon of quantum dimension $d_{a_i} = 2$ (i.e., $\mathcal{D}_{\mathcal{T}}^2=2^2 +1+1+1+1=8$).

Because the condensate now contains non-Abelian bosons, we read off the un-condensed content directly from the condensation prescription detailed in Sec.~IV.A of Ref.~\cite{Bais09}: each anyon $a$ of $D(D_4)$ branches into $a \to \sum_i n_i^a a_i$ (Eq.~(14) of Ref.~\cite{Bais09}), where the $a_i$ are the condensate sectors and $n_i^a$ the multiplicity of anyon $a_i$ in the branching of anyon $a$. This is also referred to as \emph{the restriction of $a$ into $\mathcal{L}\times a$}~\cite{Kong2014}, which preserves the quantum dimension. By Eq.~(17) in Ref.~\cite{Bais09}, conjugation gives $\bar a \to \sum_i n_i^a \bar a_i$. To fix the multiplicities $n_i^a$, we compute the branching of the product $a\times\bar a= \sum_c N_{a\bar a}^c c$ and extract the coefficient of the new vacuum. By Eq.~(16) in Ref.~\cite{Bais09}, 
\begin{equation}
\label{eq:branch-aabar}
  \Big(\sum_j n_j^a a_j\Big)\times\Big(\sum_i n_i^a \bar a_i\Big) = \sum_{c,k} N_{a\bar a}^c \, n_k^c\, c_k.
\end{equation}
We first extract the multiplicity of the new vacuum from both sides of Eq.~\eqref{eq:branch-aabar}. Expanding the left-hand side, the new vacuum is collected from each pairwise product $a_j\times\bar a_i$ with multiplicity $N^{1}_{a_j\bar a_i}$, so it appears in total with multiplicity $\sum_{i,j} n_i^a n_j^a\,N^{1}_{a_j\bar a_i}$. An anyon fuses to the vacuum exactly once and only with its own antiparticle, so $N^{1}_{a_j\bar a_i}=\delta_{ij}$~\cite{Kitaev_2006}. Hence
\begin{equation}
  \sum_{i,j} n_i^a n_j^a\,N^{1}_{a_j\bar a_i}
  = \sum_{i,j} n_i^a n_j^a\,\delta_{ij}
  = \sum_i (n_i^a)^2.
\end{equation}
On the right-hand side of Eq.~\eqref{eq:branch-aabar}, the new vacuum is the condensate itself, so the multiplicity of an anyon $c$ (belonging to the original theory) in $\mathcal{L}$ equals the multiplicity $n_1^c$ with which $c$ branches to the vacuum (Eq.~(15) of Ref.~\cite{Bais09}),
\begin{equation}
  \mathcal{L} = \sum_c n_1^c\, c .
\end{equation}
Each condensed anyon here contains the vacuum once, $n_1^c\in\{0,1\}$, recovering $\mathcal{L}=1+e_G+m_R+m_B+m_{RB}$. The new vacuum coefficient on the right-hand side of Eq.~\eqref{eq:branch-aabar} is then
\begin{equation}
  \sum_c N_{a\bar a}^{c}\, n_1^c = \sum_c n_1^c\, N_{ca}^{a} \equiv N_a ,
\end{equation}
where $N_a \equiv N^{a}_{\mathcal{L}a}=\sum_c n_1^c\, N_{ca}^{a}$ is the multiplicity of $a$ in $\mathcal{L}\times a$, and we used the fusion symmetry $N_{a\bar a}^{c}=N_{ca}^{a}$~\cite{Kitaev_2006}. Equating the two new-vacuum coefficients in Eq.~\eqref{eq:branch-aabar} and combining with Eq.~(19) of Ref.~\cite{Bais09} yields
\begin{equation}
  \label{eq:cond-rules}
  \sum_i ({n_i^a})^2 = N_a, \qquad \sum_i n_i^a\, d_{a_i} = d_a,
\end{equation}

Restricting the charge $e_R$,
\begin{equation}
  \mathcal{L}\times e_R = e_R + e_{RG} + f_R + m_B + f_{RB},
\end{equation}
gives $N_{e_R}=1$, so it forms a single sector of dimension $d_{e_R}=1$. A similar calculation shows that $N_{e_B} = N_{e_{RB}} = 1$, so they form single sectors of dimension $d_{e_B} = d_{e_{RB}} =1$ respectively. Since $e_{RG}=e_G\times e_R$ also appears in $\mathcal{L} \times e_R$, the charges identify pairwise, $e_R\equiv e_{RG}$, $e_B\equiv e_{GB}$, $e_{RB}\equiv e_{RGB}$, leaving the three nontrivial Abelian sectors $e_R,e_B,e_{RB}$ (with $e_R\times e_B=e_{RB}$). 

Restricting the anyon $m_G$,
\begin{equation}
  \mathcal{L}\times m_G = m_G + f_G + m_{RG} + f_{RG}
  + m_{GB} + f_{GB} + s_{RGB} + \bar s_{RGB},
\end{equation}
gives $N_{m_G}=1$, so it forms a single sector that retains $d_{m_G}=2$. The total quantum dimension of the set $\{1, e_R, e_{B}, e_{RB}, m_G\}$ precisely matches that of the broken theory, and once again we can identify the corresponding anyons with the five irreps of $D_4$: $m_G\times m_G = 1 + e_R + e_B + e_{RB}$. The remaining un-condensed non-Abelian anyons are all identified with an anyon in this set. 

Unlike the charge condensate, where the analogous three charges all condensed and collapsed $m_G$ to the vacuum, here only $e_G$ condenses and the non-Abelian $m_G$ survives as the two-dimensional irrep, rendering the residual symmetry non Abelian. In both cases, the symmetry of the resulting state is the minimal one consistent with the fusion multiplicities of the un-condensed anyons.

\section{Decohered density matrix}
\label{sec:dec_density_mat}

\subsection{Numerical evidence for purity instability}
\label{sec:numerical_purity_instability}

In order to study the purity of the decohered density matrix, we follow the same reasoning as in Sec.~\ref{sec:high_T_expansion} to derive a spin Hamiltonian whose partition function recovers $\tr(\rho^2)$:
\begin{equation}
\label{eq:purity_ham}
H_2[\mu_\R, \mu_\B] = -2 \mu_\R \sum_{\langle b, g \rangle \in \red{\hexagon}} \sigma_b^{\R} \sigma_g^{\R} \CZ^{(1)}_{b g} \CZ^{(2)}_{b g} - 2 \mu_\B \sum_{\langle r, g \rangle \in \blue{\hexagon}} \sigma_r^{\B} \sigma_g^{\B} \CZ^{(1)}_{r g} \CZ^{(2)}_{r g}
\end{equation}
where $\tanh(\mu_\alpha) = p_\alpha / (1 - p_\alpha)$ for $\alpha \in \{\R, \B\}$. The Hamiltonian $H_2[\mu_\R, \mu_\B]$ possesses a symmetry $G_{m_R, m_B}^{(2)} \cong (G_{m_R, m_B} \times G_{m_R, m_B}) \rtimes \Z_2$. The two copies of $G_{m_R, m_B}$ in the purity symmetry group $G_{m_R, m_B}^{(2)}$ act on $\sigma_i^{\alpha} \sigma_j^{\alpha} \CZ^{(1)}_{i j}$ and $\sigma_i^{\alpha} \sigma_j^{\alpha} \CZ^{(2)}_{i j}$ respectively. Moreover, the outer $\Z_2$ acts by conjugation as a SWAP symmetry on the two factors of $G_{m_R, m_B}$. Through a similar calculation, one can identify the Hamiltonian $H_n[\mu_\R, \mu_\B]$ whose partition function captures the $n^\mathrm{th}$ R\'{e}nyi moment $\tr(\rho^n)$ (see e.g., Ref.~\cite{sala_D4} for a similar analysis).

Using the spin Hamiltonian in Eq.~\eqref{eq:purity_ham}, we performed Monte Carlo simulations to sample the susceptibility diagnostics discussed in the main text (see Fig.~\ref{fig:Fig_2}(d)). Defining the magnetization as
\begin{equation}
    m(e_G^{12}) = \frac{1}{|\R \cap \B|}\sum_{g \in \R \cap \B} \widetilde{\sigma}_g^{(1)} \widetilde{\sigma}_g^{(2)} ,
\end{equation}
we define the corresponding susceptibility via
\begin{equation}
   \chi(e_G^1 \times e_G^2) = |\R \cap \B|\frac{p}{(1 - p)} [ \langle m(e_G^{12})^2 \rangle - \langle |m(e_G^{12})| \rangle^2].
\end{equation}
Here $p = p_R = p_B$ and $|\R \cap \B|$ denotes the total number of vertices on the triangular lattice containing the $\widetilde{\sigma}_g$ degrees of freedom.

\begin{figure}
    \centering
    \includegraphics[width=\linewidth]{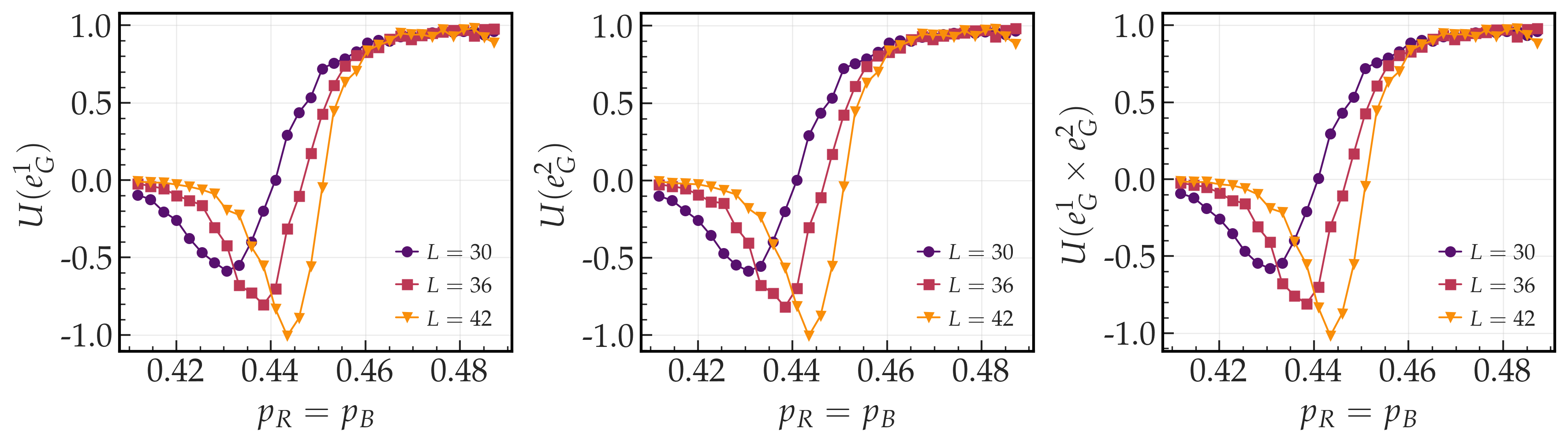}
    \caption{\textbf{Binder cumulants for $e_G^1, e_G^2, e_G^{1} \times e_G^2$}. $U$ is inconclusive in determining if a phase transition exists for $p_R = p_B \leq 1/2$. However, merging and saturation to 1 of $U$ without crossing for different system sizes $L$ strongly suggests the existence of a transition.}
    \label{fig:purity_U_data}
\end{figure}

As shown in Fig.~\ref{fig:Fig_2}(d) sharpening of susceptibility peaks with increasing linear system size $L$ indicate condensation of $e_G^1 \times e_G^2$ for $p_R = p_B \leq 1/2$; however, their drift with increasing system size $L$ prevents determination of a precise numerical value. 
From Fig.~\ref{fig:purity_U_data}, we find that $e_G^1$ and $e_G^2$ condense simultaneously (via susceptibility peaks) with $e_G^1 \times e_G^2$.

Looking at the Binder cumulant diagnostics $U$ (Fig.~\ref{fig:purity_U_data}), we cannot conclusively determine whether or not a transition exists for $p_R = p_B < 1/2$ (see Fig.~\ref{fig:purity_U_data}). However, several features in these plots strongly suggest that a transition exists for $p_R = p_B \leq 1/2$. The saturation of Binder cumulants in Fig.~\ref{fig:purity_U_data} to 1 indicates the existence of an ordered phase. However, the data in the ``ordered'' region is too noisy to resolve crossings between Binder cumulants of different linear system sizes. The Binder cumulants also possess negative extrema that increase in depth with system size $L$. As mentioned in the main text, these negative dips can be indicative of a first or weakly first order phase transition, proximity to a multi-critical point, or fluctuations along a projection of a vector order parameter; however, we cannot make any definite claim given the available data.

\subsection{Eigenvalues of maximally decohered density matrix}
\label{sec:decoherence_eigenvalues}

Following the analysis in~\cite{sala_D4}, we consider the stability of the decohered density matrix at maximal decoherence $p_\R = p_\B = 1/2$. Since the error channels $\mathcal{E}_j^{\R}$ and $\mathcal{E}_j^{\B}$ commute, we simultaneously diagonalize them to obtain the maximally decohered density matrix $\rho_{1/2} = \frac{1}{4^{|\R| + |\B|}} \sum_\eta \ket{\eta} \bra{\eta}$. Here the orthogonal states $\ket{\eta}$ are given by $\ket{\eta} = \prod_{b \in \B} \prod_{r \in \R} (1 + \eta_b X_b^{\B}) (1 + \eta_r X_r^{\R}) \ket{D_4}$
with $\eta_j = \pm 1$. The eigenvalues of $\rho_{1/2}$ are $P(\eta) = \frac{1}{4^{|\R| + |\B|}} \braket{\eta | \eta}$, which we can be expanded as
\begin{equation}
    P(\eta) = \frac{1}{4^{|\R| + |\B|}} \sum_{L_\R, L_\B} \frac{2^{C_{L_\R \cup L_\B}}}{\sqrt{2}^{|L_\R| + |L_\B|}} \prod_{\red{\ell} \in L_\R} \eta_\red{\ell} \prod_{ \blue{\ell} \in L_\B} \eta_\blue{\ell}
\end{equation}
since only closed loops of $X^{\R}$ and $X^{\B}$ have finite expectation value in $\ket{D_4}$. From here, it's then easy to obtain the infinite moment of the maximally decohered density matrix, since this is (proportional to) the maximum eigenvalue: $\lim_{n \rightarrow \infty} \tr(\rho^n)^{1/n} \propto \max_{\eta} P(\eta)$, given by the configuration with $\prod_{\red{\ell} \in L_\R} \eta_\red{\ell} \prod_{ \blue{\ell} \in L_\B} \eta_\blue{\ell}=1$ for all $L_\R, L_\B$.
This maximum eigenvalue exactly corresponds to the upper-right corner of the phase diagram in Fig.~\ref{fig:Fig_2}(a), which (as we showed) possesses long-range correlations in the observables $W_{m_R}, W_{m_B}, W_{e_G}$. Hence, according to the infinite moment of the density matrix, the threshold $p_c^{(\infty)}$ is strictly smaller than $1/2$, unlike the result previously found for a noise channel which proliferates a single type of non-Abelian anyon~\cite{sala_D4}. Moreover, we notice that all eigenvalues at maximal error rate have the symmetry $G_{m_R, m_B}$ that we identified when considering wavefunction deformations.

\end{document}